\documentclass[10pt,journal,compsoc]{IEEEtran}

\usepackage{newlfont}
\usepackage{color}
\definecolor{shadecolor}{rgb}{0.92,0.92,0.92}
\usepackage{colortbl}
\usepackage[table]{xcolor}
\usepackage[color,matrix,arrow,all]{xy}

\usepackage{graphicx}
\usepackage{float}
\usepackage{subfigure}
\usepackage{multirow}
\usepackage{multicol}
\usepackage{marginnote}
\usepackage{CJKutf8}
\usepackage{makecell}
\usepackage{pifont}

\usepackage{tabularx}

\usepackage{graphicx,epsfig}
\usepackage{balance}
\usepackage{amsmath,amssymb,amsfonts,bm}
\usepackage{framed}
\usepackage{lipsum}
\usepackage{color}
\usepackage{hyperref}
\usepackage{balance}  
\usepackage{listings,upquote,soul,framed}
\usepackage{hyperref}
\usepackage{xspace}
\usepackage{epstopdf}
\usepackage{enumitem}
\usepackage{graphicx}
\usepackage[mathscr]{eucal}
\usepackage{graphicx}
\usepackage{bbding}
\usepackage{algorithmic}
\usepackage{amssymb}
\usepackage{epstopdf}
\usepackage{epsfig,endnotes}
\usepackage{url}
\usepackage{dsfont}
\usepackage{array}
\usepackage{booktabs}
\usepackage{threeparttable}
\usepackage[lined,boxed,vlined,ruled,linesnumbered]{algorithm2e}
\usepackage[justification=centering]{caption}
\colorlet{shadecolor}{gray!20}

\usepackage{forest}
\usetikzlibrary{arrows.meta,shapes,positioning,shadows,trees}

\usepackage[utf8]{inputenc}
\usepackage{amsmath}
\usepackage{amssymb}
\usepackage{tikz,lipsum,lmodern}
\usepackage[most]{tcolorbox}
\usepackage{graphicx}
\usepackage{afterpage}

\newcommand{\zxh}[1]{\textcolor{blue}{#1}}
\newcommand{\hjx}[1]{\textcolor{red}{HJX: #1}}
\newcommand{\whr}[1]{\textcolor{brown}{#1}}

\newcommand{\zw}[1]{\textcolor{purple}{#1}}
\definecolor{zhycolor}{RGB}{0,84,0}
\newcommand{\zhy}[1]{\textcolor{zhycolor}{#1}}

\newcommand{\llm}{\textsc{LLM}\xspace}
\newcommand{\llms}{\textsc{LLMs}\xspace}

\newcommand{\hi}[1]{\vspace{.25em} \noindent {\bf #1}\xspace}
\newcommand{\bfit}[1]{\textbf{\textit{#1}}}

\newcommand{\datallm}{\textsc{DATA4LLM}\xspace}
\newcommand{\llmdata}{\textsc{LLM4DATA}\xspace}

\newcommand{\dataconcept}{\textbf{IaaS}\xspace}

\newcommand{\insertfig}{\includegraphics[width=.95\linewidth]{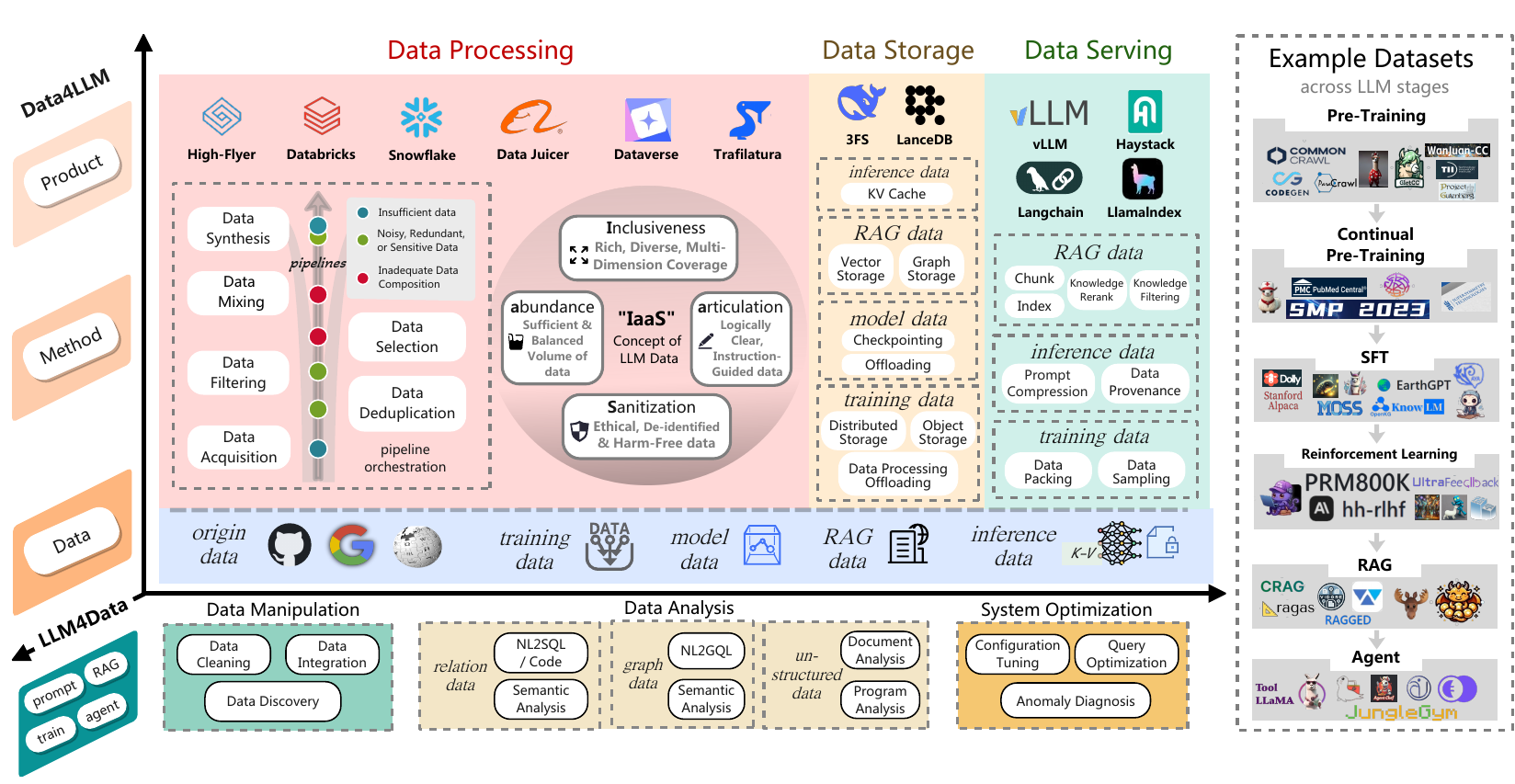}\vspace{-.25cm}\captionof*{figure}{Fig. 1: Overview of LLM $\times$ DATA (with ``\dataconcept'' Concept).}\vspace{-.5cm}\label{fig:fig1-overview}}

\makeatletter
\apptocmd{\@maketitle}{\centering\insertfig}{}{}
\makeatother

\begin{document}

\pagestyle{plain}
\pagenumbering{roman}

\twocolumn

\pagestyle{plain}
\title{Data $\times$ \llm: From Principles to Practices}
\title{Advances and Challenges in Data$\times$\llm}
\title{A Survey of LLM $\times$ DATA}

\pagestyle{plain}
\pagenumbering{arabic}

\renewcommand\thesection{\arabic{section}}
\setcounter{section}{0}

\pagenumbering{arabic}
\setcounter{page}{1}
\setcounter{figure}{1}
\setcounter{table}{0}



\author{
    \IEEEauthorblockN{Xuanhe Zhou\IEEEauthorrefmark{1}\IEEEauthorrefmark{5}, 
    Junxuan He\IEEEauthorrefmark{1}\IEEEauthorrefmark{5}, 
    Wei Zhou\IEEEauthorrefmark{1}\IEEEauthorrefmark{5}, 
    Haodong Chen\IEEEauthorrefmark{1}\IEEEauthorrefmark{5},
    Zirui Tang\IEEEauthorrefmark{1}\IEEEauthorrefmark{5},
    Haoyu Zhao\IEEEauthorrefmark{1}\IEEEauthorrefmark{5},
    Xin Tong\IEEEauthorrefmark{1},
    Guoliang Li\IEEEauthorrefmark{2}, 
    Youmin Chen\IEEEauthorrefmark{1},
    Jun Zhou\IEEEauthorrefmark{1},
    Zhaojun Sun\IEEEauthorrefmark{1},
    Binyuan Hui\IEEEauthorrefmark{3},
    Shuo Wang\IEEEauthorrefmark{2},
    Conghui He\IEEEauthorrefmark{4},
    \\
    Zhiyuan Liu\IEEEauthorrefmark{2},
    Jingren Zhou\IEEEauthorrefmark{3},
    Fan Wu\IEEEauthorrefmark{1}}
    \\
    \IEEEauthorblockA{\IEEEauthorrefmark{1}Shanghai Jiao Tong University}
    \IEEEauthorblockA{\IEEEauthorrefmark{2}Tsinghua University}
    \IEEEauthorblockA{\IEEEauthorrefmark{3}Alibaba Group}
    \IEEEauthorblockA{\IEEEauthorrefmark{4}Shanghai AI Laboratory}

    \textcolor{blue}{\href{https://github.com/weAIDB/awesome-data-llm}{https://github.com/weAIDB/awesome-data-llm}}
    
\IEEEcompsocitemizethanks{
\protect \item
\IEEEauthorrefmark{5} Co-first authors with equal contributions.
}
}

\IEEEtitleabstractindextext{
{\leftskip=0pt \rightskip=0pt plus 0cm

\begin{abstract}
The integration of large language model (LLM) and data management (DATA) is rapidly redefining both domains. In this survey, we comprehensively review the bidirectional relationships. On the one hand, \textbf{\datallm}, spanning large-scale data processing, storage, and serving, feeds LLMs with diversity, redundant, high quality, and sanitized data (following the ``\dataconcept'' concept) required for stages like pre-training, post-training, retrieval-augmented generation, and agentic workflows: $(i)$ Data processing for LLMs includes scalable acquisition, deduplication, filtering, selection, domain mixing, and synthetic augmentation; $(ii)$ Data Storage for LLMs focuses on efficient data and model formats, distributed and heterogeneous storage hierarchies, KV-cache management, and fault-tolerant checkpointing; $(iii)$ Data serving for LLMs tackles challenges in RAG (e.g., knowledge post-processing), LLM inference (e.g., prompt compression, data provenance), and training strategies (e.g., data packing and shuffling). On the other hand, in \textbf{\llmdata}, LLMs are emerging as general-purpose engines for data management. We review recent advances in $(i)$ data manipulation, including automatic data cleaning, integration, discovery; $(ii)$ data analysis, covering reasoning over structured, semi-structured, and unstructured data, and $(iii)$ system optimization (e.g., configuration tuning, query rewriting, anomaly diagnosis), powered by LLM techniques like retrieval-augmented prompting, task-specialized fine-tuning, and multi-agent collaboration.



\end{abstract}

}
}

\maketitle


\section{INTRODUCTION}
\label{sec:introducion}





\IEEEPARstart{L}{arge} language models (LLMs\footnote{We use LLMs to refer to billion-scale language models capable of supporting general NLP tasks~\cite{zhao2023survey} or multimodal tasks~\cite{mllmsurvey,datajuicersurvey}.}) have made remarkable progress in both {{\it general domain applications}} (e.g., open-domain question answering~\cite{opendomainqa}, cross-modal video summarization~\cite{crossdomainsummary}, general-purpose code generation~\cite{codesurvey}) and {{\it specific domain applications}} (e.g., biomedical literature analysis~\cite{mdedicinesurvey}, legal document review~\cite{lai2024large}, SQL generation for business intelligence~\cite{chatbi2024}). As shown in Figure 1, apart from technical advances in \llms~\cite{minaee2024large,moe2021,zhang2023instruction,feedback2022,quantization2024,speculative2023}, data management has emerged as a critical factor in unlocking \llms' full potential in these applications (\datallm). It includes efficient and scalable solutions for data processing, storage, and serving across the \llm lifecycle, as evidenced in recent academic studies~\cite{goyal2024scaling,marion2023less,lin2024data} and industry reports~\cite{radford2019language,yang2023baichuan,bai2023qwen,llamaindex}. Conversely, \llm-powered techniques are increasingly being adopted to enhance data management tasks, such as data manipulation, analysis, and system optimization (\llmdata).

\hi{\datallm.} 
Effective data management is fundamental to the scalable development and deployment of \llms.
To illustrate this, we highlight representative scenarios where \llms depend on specialized techniques for data processing, storage, and serving across various stages of the \llm lifecycle.



\noindent\bfit{Example-\textcircled{1} Data Processing for LLMs.}
Processing a large-scale training dataset (e.g., $\sim$4 TB multi-modal tokens utilized in Qwen2.5-VL pretraining~\cite{qwen25vl}) poses several challenges. 
First, acquiring diverse raw data (e.g., over 10,000 object categories for visual grounding) demands substantial efforts in data collection (Section~\ref{subsubsec:acquisition}) and, in many cases, data synthesis (Section~\ref{subsubsec:synthesis}).
Second, preparing high-quality training samples requires robust pre-processing, including rigorous data filtering (Section~\ref{subsubsec:filtering}), along with dedicated evaluation approaches.
Third, the overall performance of \llms depends heavily on an end-to-end pipeline that effectively schedules and coordinates these processing tasks, especially for the pretraining stage (Section~\ref{subsubsec:pipelines}).

\noindent\bfit{Example-\textcircled{2} Data Storage for LLMs.} 
Managing storage for \llms, spanning both training datasets (see Example-\textcircled{1}) and massive model parameters (e.g., DeepSeek-R1 with 671B parameters~\cite{guo2025deepseek}), poses significant challenges.
First, large-scale datasets must be partitioned and distributed across multiple storage nodes, introducing challenges in data placement and consistency management (Section~\ref{DataDistribution}). Second, to support efficient \llm training and inference, these storage nodes must deliver high I/O throughput for timely data transfer to compute nodes (Section~\ref{subsubsec:DataTransmission}). Third, the massive size of model parameters increases the risk of training interruptions, necessitating robust fault tolerance mechanisms to recover and resume training from intermediate states (Section~\ref{datafault}).

\noindent\bfit{Example--\textcircled{3} Data Serving for LLMs.} Data serving plays a critical role in selecting and preparing input data (e.g., the task-specific prompts), directly affecting the quality of \llm's responses. Taking retrieval-augmented generation (RAG) as an example, EyeLevel.ai~\cite{eyelevelai} observed that when relying solely on vector similarity, RAG accuracy declines notably with 10,000-page documents, and the performance degradation can reach up to 12\% with 100,000 pages (still fewer than enterprise-scale datasets). Several challenges arise in this context. First, the retrieved knowledge is typically noisy and must be filtered and re-ranked to ensure relevance and factual accuracy (Section~\ref{sec:DataSelection}). Second, the retrieved content is often lengthy and exceeds the input capacity or comprehension of \llms, necessitating effective compression techniques to preserve utility while improving performance (Section~\ref{sec:DataCompression}).

\vspace{.5em}

\hi{\llmdata.} 
Conversely, various \llm-based techniques can be leveraged to enhance core data management tasks, including data manipulation, data analysis, and system-level optimization.
The following examples illustrate how \llms can be applied to improve these tasks in practice.



\vspace{.5em}

\noindent\bfit{Example-\textcircled{1} LLM-based Data Manipulation.}
Data manipulation, including cleaning, integration, and discovery, is critical for ensuring high-quality datasets.
Traditional methods depend on rigid rules and domain-specific configurations, requiring extensive manual efforts and struggling with complex data samples~\cite{LLMBench, LLMClean, LLMErrorBench}.
For instance, standardizing date formats (e.g., ``Fri Jan 1st 10:36:28 2021'' vs. ``1996.07.10 AD at 15:08:56'') or resolving textual inconsistencies (e.g., ``Monticello VA, Jasper'' vs. ``Monticello VAA'') typically requires intricate programming scripts or handcrafted constraints~\cite{CleanAgent, GIDCL}.
These approaches also struggle with cross-row error detection, such as mismatched city-state-zip entries.
In contrast, \llms can infer semantic similarities and autonomously generate cleaning workflows to resolve such inconsistencies without requiring explicit rule definitions~\cite{AutoDCWorkflow, GIDCL, Jellyfish}.
This semantic understanding enables \llms to adapt flexibly to diverse data issues and support more scalable and context-aware data manipulation (Section~\ref{subsec:manipulation}).


\noindent\bfit{Example-\textcircled{2} LLM-based Data Analysis.} 
Data analysis over heterogeneous sources, such as medical records and transactional data, is essential in many real-world applications.
Traditional deep learning models, while effective at performing specific semantic-level analysis, struggle to generalize across diverse data formats and task types.
For instance, tasks such as table extraction and table-based question answering across heterogeneous sources (e.g., relational tables and knowledge graphs) often require the development of separate, specialized models.
This process is both resource-intensive and difficult to scale.
In contrast, \llms offer a unified reasoning framework that leverages broad semantic understanding, enabling them to support a wide range of analytical tasks across various data modalities with greater flexibility and reduced efforts for task-specific engineering (Section~\ref{subsec:analysis}). 

\noindent\bfit{Example-\textcircled{3} LLM-based System Optimization.}
System optimization entails configuring parameters (e.g., memory settings) and monitoring runtime status (e.g., resource utilization) to ensure optimal system performance.
Traditional approaches, such as manual tuning or deep learning-based methods, are time-consuming and inefficient~\cite{TuningSurvey}.
For instance, methods of Bayesian Optimization (BO) or Reinforcement Learning (RL) require numerous workload replays over 20 hours to identify promising configurations for a single TPC-H workload~\cite{E2ETune}.
Moreover, root cause analysis over anomalies can be error-prone, particularly in multi-cause scenarios where metrics are highly interdependent~\cite{DBot}.
In contrast, \llms offer a new paradigm by integrating domain knowledge (e.g., tuning manuals) and applying advanced reasoning to instruct optimization.
By leveraging retrieval-augmented prompts, \llms can efficiently identify root causes or recommend precise configurations, enabling faster and more accurate optimization in complex environments~\cite{LLMasDBA, LLMR2, GPTuner} (Section~\ref{subsec:optimization}).


\subsection{Techniques of \datallm}


\hi{Characteristics of \llm Datasets (\S~\ref{subsec:data-characters}).} As shown in Figure 1, datasets (following the ``\dataconcept'' concept) play a critical role in enabling the desired capabilities at each \llm stage,  including (1) pre-training, (2) continual pre-training, (3) fine-tuning, (4) reinforcement learning, (5) retrieval-augmented generation (RAG), (6) LLM agents, and (7) evaluation. For each stage, we separately analyze the characters of required data (e.g., preferred formats and emphasized aspects within \dataconcept) and the corresponding data techniques (see Table~\ref{tab:overview}). 




\hi{Data Processing for \llms (\S~\ref{sec:processing}).} 
We introduce techniques to prepare high-quality datasets for \llms based on a series of processing steps. 

\noindent$\bullet$ \underline{{\it Data Acquisition.}} Data acquisition aims to (1) extract relevant data (e.g., text and images) from noisy data sources with certain structures (e.g., dynamically rendered web pages)~\cite{trafilatura,bet,bevendorff2023contentextraction,trafilatura,beautifulsoup,selenium,playwright,puppeteer}, and (2) extract data from complicated data sources (e.g., scanned or handwritten documents) with techniques such as complex layout analysis~\cite{Tesseract,PaddleOCR,wang2024mineru,huang2022layoutlmv3,wang2024yolo,wei2024got,liu2024fox,radford2021clip,wei2023vary}. 


\noindent$\bullet$ \underline{{\it Data Deduplication.}} Data deduplication aims to identify duplicates in large-scale {textual} or multi-modal data, including exact string matching ~\cite{dong2024baichuanseed, nystrom2022deduplicating}, hash identification \cite{SimHash, MinHash, dong2024baichuanseed, nystrom2022deduplicating, shen2023slimpajama,silcock2022noise, khan2024lshbloom, DotHash}, sample reweighing \cite{he2024softdedup} and embedding-based clustering \cite{abbas2023semdedup, tirumala2023d4, slyman2024fairdedup}. 

\noindent$\bullet$ \underline{{\it Data Filtering.}} We review data filtering methods at two primary levels: (1) Sample-level filtering selects high-quality and diverse samples using strategies like perplexity measuring~\cite{thrush2024improving, ankner2024perplexed, mekala2024smaller}, 
influence assessment~\cite{lin2024data, he2024shed}, clustering methods~\cite{abbas2024effective, SmallToLarge}, prompt-based scoring~\cite{wettig2024qurating, liu2023makes, shen2024seal}, or mixes of these strategies \cite{marion2023less, cao2023instruction, du2023mods}; (2) Content-level filtering aims to remove undesirable or harmful content from large-scale datasets, such as toxic language, personal identifiable information (PII), biased statements~\cite{liu2023deid, lukas2023analyzing}, and improper images and videos~\cite{CogVideoX,Hunyuanvideo,Wan}.

\noindent$\bullet$ \underline{{\it Data Selection.}} Data selection aims to select sub-datasets and evaluate their ability to accurately represent the target distribution, especially when handling diverse datasets or domains. There are methods like similarity-based data selection~\cite{xie2024efficient,xie2023data,qin2024enabling,brandfonbrener2024color}, optimization-based data selection~\cite{Dsdm,LESS,liu2024tsds}, and model-based data selection~\cite{zhang2024autonomous}.


\noindent$\bullet$ \underline{{\it Data Mixing.}} Data mixing aims to effectively integrate datasets from diverse domains without degrading quality or destabilizing \llm performance. Key techniques include:
(1) \emph{Heuristic optimization}, which empirically tunes data ratios to enhance downstream performance. Examples include two-stage mixing~\cite{feng2024maximize}, source rebalancing~\cite{shen2023slimpajama}, and entropy-based weighting~\cite{ge2025bimixbivariatedatamixing}; 
(2) \emph{Bilevel optimization}, which formulates data weighting as a nested optimization problem to jointly balance training and validation objectives~\cite{pan2024scalebio,fan2023doge}; 
(3) \emph{Distributionally robust optimization}, which enhances resilience to worst-case domain shifts by emphasizing underperforming or rare data domains~\cite{xie2023doremi,ma2024task}; 
(4) \emph{Model-based optimization}, which builds predictive models to map data mixing ratios to loss and task performance. Approaches include linear predictive modeling (e.g., REGMIX~\cite{liu2024regmix}), nonlinear function fitting~\cite{ge2025bimixbivariatedatamixing, ye2024data, gu2024cmr}, scaling law-based estimation~\cite{que2024d}, and latent source attribution~\cite{liang2024data}.

\noindent$\bullet$ \underline{{\it Data Synthesis.}} We introduce data synthesis techniques designed to address the following key challenges:  
(1) \emph{Mitigating harmful characteristics} such as toxicity or bias, which can be inherited or amplified in synthetic data (e.g., program-aided verification~\cite{zhu2024padprogramaideddistillationteach}, semantic scoring~\cite{hou2025advancing}, and multi-agent consistency filtering~\cite{shen2025satori});   
(2) \emph{Balancing data utility and privacy}, through privacy-preserving synthetic rewriting and key-entity obfuscation methods during the RAG stage~\cite{zeng2024mitigating};  
(3) \emph{Generating diverse and logically consistent reasoning data} using approaches like formal proof-based validation~\cite{huang2024mustard}, Chain-of-Thought (CoT) branching and error correction~\cite{hou2025advancing}, and high-quality problem synthesis guided by structure and complexity constraints~\cite{liu2024augmenting, ye2025limo};  
(4) \emph{Automating human-like evaluation and feedback generation} with \llm-based preference modeling~\cite{bai2022training}, judge models for response ranking~\cite{zheng2023judging}, and clustering-based diversity quantification~\cite{chen2024diversity}.

\noindent$\bullet$ \underline{{\it Data Pipelines.}} 
We first introduce \emph{frameworks} that integrate basic data processing operators and interfaces, serving as the general foundation for building data pipelines~\cite{chen2024data,park2024dataverse,sun2024integrated}. Then we showcase \emph{typical pipelines} with heuristic mechanisms that properly arrange these operators (mainly for \llm pretraining)~\cite{penedo2023refinedwebdatasetfalconllm,li2024datacomp,penedo2024fineweb}. 
Finally, we discuss strategies that go beyond heuristic designs to further optimize these data processing pipelines~\cite{chen2024data2}.

\hi{Data Storage for \llms (\S~\ref{sec:storage}).} We review data storage techniques for \llms from the following main aspects.


\noindent$\bullet$ \underline{{\it Data Formats.}}
We review commonly-used dataset and model data formats for \llms. Dataset formats include  \textit{TFRecord}~\cite{tfrecord}, \textit{MindRecord}~\cite{mindspore} for multimodal data, and \textit{tf.data.Dataset} {that can be directly fed into LLMs}~\cite{tfdata}.
For model data storage, there are formats like \textit{Pickle}~\cite{pickle} and \textit{ONNX}~\cite{onnx}.

\noindent$\bullet$ \underline{{\it LLM Data Distribution.}} LLM data distribution aims to store data across multiple storage nodes in a cluster, which mainly serves for storing large-scale \llm training data. Key approaches include (1) distributed storage systems like JuiceFS~\cite{juicefs} and 3FS~\cite{3FS}; and (2) heterogeneous storage systems for model data  (e.g., {across GPUs and CPUs})~\cite{rajbhandari2020zero,rajbhandari2021zero,rhu2016vdnn,ZeROf-Offload,yang2024protrain}.

\noindent$\bullet$ \underline{{\it LLM Data {Organization}.}} LLM data organization aims to {transform data into a format suitable for storage and retrieval} (mainly for the RAG stage) in heterogeneous forms. First, for vector RAG, relevant techniques include content formatting~\cite{chen2023dense,hosseini2024scalable,an2024thread,che2024hierarchical}, chunking~\cite{zhong2024mix}, embedding~\cite{chen2024bge,stella,li2023towards}, compression~\cite{aguerrebere2023similaritysearchblinkeye,tepper2024leanvecsearchingvectorsfaster,tepper2024gleanvecacceleratingvectorsearch,tepper2024gleanvecacceleratingvectorsearch}. Second, for graph RAG, we discuss indexing techniques such as generating textual summary for quick retrieval~\cite{edge2024local,guo2024lightrag,fan2025minirag}. We also introduce the systems that integrate these techniques, including vector search engines~\cite{faiss,milvus,weaviate,lancedb} and graph storage platforms~\cite{propertygraph,AmazonNeptune,arangodb}.

\noindent$\bullet$ \underline{{\it LLM Data Movement.}} LLM data movement aims to improve the speed of data movement across storage and compute nodes.
Relevant techniques include (1) caching data~\cite{kumar2020quiver, gu2022fluid, TectonicShift}; (2) offloading data/operator to multiple devices (e.g., across CPUs) ~\cite{graur2022cachew,audibert2023tf,graur2024pecan,zhao2024cedar}; and (3) overlapping of storage and computing in training stage~\cite{zhao2023silod,zhong2025optimizing}.

\noindent$\bullet$ \underline{{\it LLM Model Data Fault Tolerance.}} LLM model data fault tolerance aims to enhance the ability to recover from system failures during model training.
Relevant techniques include (1) checkpointing~\cite{CheckFreq,bytehdfs, wang2023gemini,ByteCheckpoint}, which stores checkpoints across a hierarchical storage system;
and (2) redundant computation, {which leverages redundant states of \llm in parallel training (e.g., pipeline parallelism~\cite{bamboo}, hybrid parallelism~\cite{oobleck,recycle}) to support rapid fault recovery.}




\noindent$\bullet$ \underline{{\it KV Cache in \llms.}} KV caching in \llms is essential for enabling fast and efficient inference by managing key-value memory usage. Existing techniques include: 
(1) \emph{Memory layout and allocation}, which optimize the physical organization of KV memory for high performance and scalability~\cite{vllm,vtensor};
(2) \emph{Storage offloading}, which places KV data on suitable storage media to balance speed and capacity~\cite{jin2024ragcache,gao2024cost};
(3) \emph{KV compression}, which reduces memory footprint through techniques like encoding compression~\cite{liu2024cachegen,liu2024minicache,gao2024faststaterestorationllm};
(4) \emph{Efficient indexing}, which accelerates KV access via specialized retrieval structures~\cite{ye2024chunkattention,zheng2024batchllm}.

\hi{Data Serving for \llms (\S~\ref{sec:serving}).} We provide an overview of data serving techniques tailored for LLMs from four aspects.


\noindent$\bullet$ \underline{{\it LLM Data Shuffling.}} \llm data shuffling aims to determine the appropriate order of data application during stages like \llm training and RAG. In the training stage, we discuss data pruning techniques (e.g., sample-scoring-based approaches~\cite{fayyaz2022bert,attendu2023nlu}, model-state-based approaches~\cite{tan2024data,albalak2023efficient,wu2024mixture,luo2024velocitune}) and data-centric training strategies~\cite{dong2023abilities}. In the RAG stage, we discuss RAG knowledge filtering~\cite{Ma_2023,DBLP:journals/corr/abs-2306-16092,chang2024mainragmultiagentfilteringretrievalaugmented} and re-ranking~\cite{eibich2024aragog,cohere,pradeep2023rankvicuna,abdallah2025asrank}.

\noindent$\bullet$ \underline{{\it LLM Data Compression.}} \llm data compression aims to compress the model's input data to stay within the context window limit or to facilitate model understanding. 
Relevant techniques include: (1) RAG knowledge compression (e.g., rule-based~\cite{xu2023recomp, shi2024compressing, jung2024familiarity} and model-based method~\cite{cheng2024xrag,rau2024context}); 
and (2) prompt compression (e.g., metric-based~\cite{jiang2023llmlingua,jiang2023longllmlingua} and model-based method~\cite{pan2024llmlingua,mu2024learning,chevalier2023adapting}).

\noindent$\bullet$ \underline{{\it LLM Training Data Packing.}} \llm training data packing aims to ensure uniform sequence lengths in training inputs. 
Relevant techniques include: (1) short sequence insertion~\cite{ding2024fewer,liu2024bucket};
(2) optimizing sequence combination~\cite{krell2021efficient,pouransari2024dataset};
and (3) semantic-cased packing~\cite{staniszewski2023structured,shi2023context}).

\noindent$\bullet$ \underline{{\it LLM Inference Data Provenance.}} 
\llm inference data provenance aims to ensure the factual consistency of \llm-generated content.
Relevant techniques include: (1) embedding markers~\cite{zhou2024bileve,christ2024undetectable,liu2023unforgeable};
and (2) statistical provenance~\cite{kirchenbauer2023watermark}).

\subsection{Techniques of \llmdata}


\hi{LLM for Data Manipulation (\S~\ref{subsec:manipulation}).} 
\llms have been increasingly applied to data manipulation tasks, with the goal of preparing high-quality datasets for non-\llm applications and enhancing data quality for downstream usage.
Key areas include data cleaning, data integration, and data discovery.

\noindent$\bullet$ \underline{{\it Data Cleaning.}}
This task involves standardizing and refining datasets through a series of operations.
We highlight three major subtasks:
(1) Data Standardization, which reformats data samples using handcrafted standardization prompts~\cite{LLMGDO, Evaporate} or agents that generate cleaning operations or pipelines~\cite{CleanAgent, AutoDCWorkflow};
(2) Data Error Processing, which identifies and corrects noisy data via direct \llm prompting~\cite{MultiNews, Cocoon, GIDCL}, context-enrichment techniques~\cite{LLMClean, LLMErrorBench}, or task-specific fine-tuning for error handling~\cite{GIDCL};
(3) Data Imputation, which fills in missing values using explicit imputation instructions and retrieval-augmented generation (RAG) methods~\cite{RetClean}.

\noindent$\bullet$ \underline{{\it Data Integration.}} 
This task focuses on identifying and reconciling semantically related datasets across heterogeneous sources. 
We review two core subtasks:
(1) Entity Matching, which aligns data entries referring to the same real-world entity using structured prompts~\cite{MatchGPT, BATCHER}, sometimes augmented with predefined code-based reasoning strategies~\cite{KcMF};
(2) Schema Matching, which establishes correspondences between schema elements using direct prompting~\cite{LLMSchemaBench}, RAG techniques incorporating multiple models~\cite{Magneto}, knowledge graph-based methods~\cite{KGRAG4SM}, and agent-based workflow generation~\cite{AgentOM, Harmonia}.

\noindent$\bullet$ \underline{{\it Data Discovery.}}
This task aims to extract informative insights from a dataset.
We cover two key subtasks:
(1) Data Profiling, which generates descriptive metadata and summaries using task-specific prompts~\cite{AutoDDG, LEDD}, and enhanced with context via RAG techniques~\cite{Pneuma};
(2) Data Annotation, which assigns semantic labels or types through various prompting strategies~\cite{CHORUS, kayali2025goby, LLMCTA}, supported by classical retrieval-based~\cite{RACOON} and \llm-generated context~\cite{Birdie}.


\hi{LLM for Data Analysis (\S~\ref{subsec:analysis}).} \llms significantly improve the analytical capabilities across structured, semi-structured, and unstructured data.

\noindent$\bullet$ \underline{{\it Structured Data Analysis.}} For relational data analysis, natural language interfaces allow users to write high-level questions instead of SQL/Python code~\cite{zhang2024finsql}. Multi-step QA frameworks (e.g., TAPERA~\cite{zhao2024tapera} and ReAcTable~\cite{zhang2023reactable}) decompose complex queries, while some end-to-end solutions fine-tune \llms specifically for tabular tasks (e.g., TableGPT~\cite{li2023tablegpt}), apply content retrieval (e.g., CABINET~\cite{patnaik2024cabinet}) or convert tables into images for analysis (e.g., Table-LLaVA~\cite{zheng2024multimodal}). For graph data, \llms facilitate semantic queries with GQL generation (e.g., $R^3$-NL2GQL~\cite{zhou2024r3}) and knowledge-aware QA by retrieving or reasoning over relevant subgraphs~\cite{xiong2024interactivekbqa}. 

\noindent$\bullet$ \underline{{\it Semi-Structured Data Analysis.}} Meanwhile, handling semi-structured data (e.g., JSON and spreadsheets) remains challenging. Recent benchmarks (e.g., TEMPTABQA~\cite{gupta2023temptabqa} and SPREADSHEETBENCH~\cite{ma2024spreadsheet}) reveal substantial performance gaps.

\noindent$\bullet$ \underline{{\it Unstructured Data Analysis.}} Finally, unstructured data analysis leverages \llms to address document and program analysis tasks. For document analysis, OCR-dependent approaches involve performing OCR on document images followed by the integration of textual, layout, and visual features for reasoning (e.g., UDOP~\cite{tang2023unifying} and DocFormerV2~\cite{appalaraju2023docformerv2}). OCR-free methods directly generate the answer with end-to-end multimodal \llms (e.g., Pix2Struct~\cite{lee2023pix2struct} and DUBLIN~\cite{aggarwal2023dublin}). For program analysis, \llms could serve as vulnerability detection tools using program analysis based training (e.g., PDBER~\cite{liu2024pdbert}) or case-driven prompt engineering (e.g., VUL-GPT~\cite{liu2023gptvulnerability}). For program related analysis, \llms could summarize repositories (e.g., SCLA~\cite{mao2024scla}) or serve as a repository-level code completer (e.g., RepoFusion~\cite{repofusion}) using their powerful semantic reasoning abilities.

\begin{figure*}[!t]
    \centering \includegraphics[width=1\textwidth]{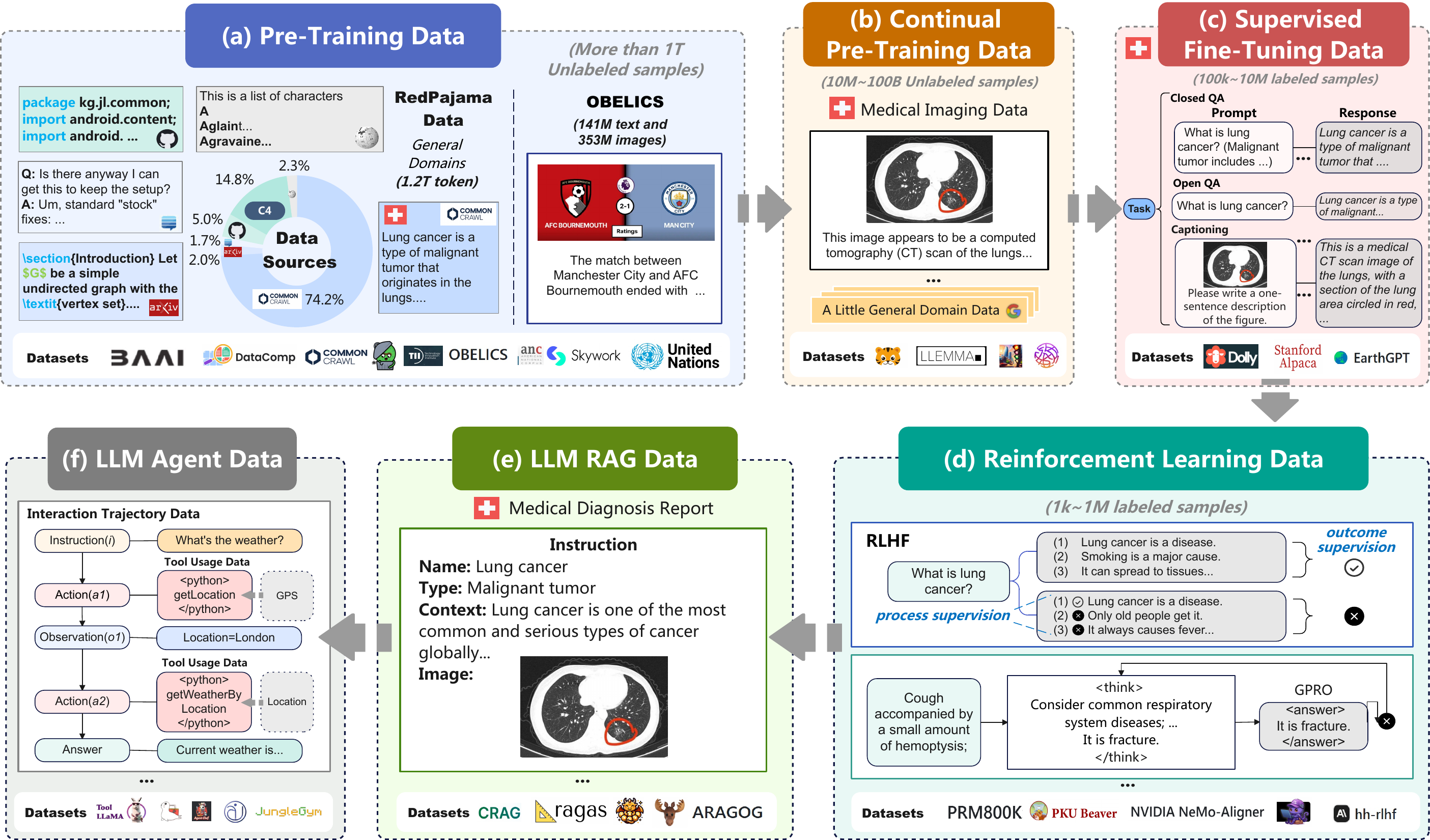}
    \caption{\textbf{Example Data Characteristics across \llm Stages} - (a) Pretraining data~\cite{together2023redpajama,laurenccon2024obelics}, (b) Continual pre-training~\cite{CL}, (c) SFT~\cite{yue2023disc}, (d) Reinforcement learning~\cite{MedicalGPT,guo2025deepseek,lightman2023let}, (e) RAG~\cite{wu2024medical}, (f) Agent~\cite{wang2025stecastepleveltrajectorycalibration,shi2025tool}.}
    \label{fig:characteristics}
    \vspace{-1em}
\end{figure*}

\hi{LLM for Data System Optimization (\S~\ref{subsec:optimization}).}
\llms equipped with advanced reasoning and code generation capabilities have been increasingly adopted in core system optimization tasks.
These include: (1) configuration tuning (identifying optimal system settings); (2) query optimization (rewriting or refining input queries for performance gains); and (3) anomaly diagnosis (analyzing system issues to ensure performance reliability).

\noindent$\bullet$ \underline{{\it Configuration Tuning.}}
This task leverages \llms to determine effective configuration parameters for improved system performance through:
(1) Prompt engineering tailored to tuning tasks, using both manually crafted~\cite{LLMBench, LATuner, lambdaTune} and automatically generated prompts~\cite{DBGPT, LLMIdxAdvis};
(2) Retrieval-augmented generation (RAG), which incorporates prior tuning experiences during offline knowledge base preparation~\cite{GPTuner} and online knowledge retrieval~\cite{Andromeda};
(3) Objective-aligned tuning, which is enhanced through targeted training techniques~\cite{DBGPT, E2ETune}.

\noindent$\bullet$ \underline{{\it Query Optimization.}}
This task utilizes \llms to rewrite queries or improve execution plans by:
(1) Designing optimization-oriented prompts that include explicit guidance~\cite{LITHE, DBGPT, LLMOpt} and in-context examples~\cite{LLMR2};
(2) Enriching optimization knowledge using RAG techniques, including \llm-generated and hybrid retrieval strategies~\cite{RBot};
(3) Enhancing optimization performance through task-specific training~\cite{LLMSteer, LLMQO, LLMOpt}.

\noindent$\bullet$ \underline{{\it Anomaly Diagnosis.}}
This task involves identifying the root causes of anomalies and suggesting effective solutions via:
(1) Direct \llm prompting based on detailed diagnosis context~\cite{DBGPTDebugger};
(2) RAG-based enrichment using relevant historical diagnosis experience~\cite{DBot, ByteHTAP};
(3) Multi-agent collaboration mechanisms for comprehensive diagnosis~\cite{DBot, Panda}.



\subsection{Comparison with Existing Surveys}

Different from existing \llm and data management surveys~\cite{wang2023data,dataselectionsurvey,DBLP:journals/tkde/ChaiWLNL23,wang2024survey,long2024llms,tabledataanalysissurevey,DBLP:conf/emnlp/TanLWBJBKL0024,2020aidb}, our survey offers a comprehensive and detailed overview of the key intersections between \llms and data management, highlighting how they can mutually benefit from each other. We uniquely position our work at the intersection of data for LLMs (e.g., how to acquire, process, store, and serve \llm data) and LLMs for data (e.g., how LLMs can be leveraged to enhance data management tasks).


\noindent$\bullet$ {\it We propose the \dataconcept concept as a principled lens to assess \llm dataset quality.} The IaaS concept identifies four essential dimensions, including inclusiveness,  abundance, articulation, and sanitization. This concept is promising to offers an evaluative criteria for guiding data management and understanding its impact across the \llm development lifecycle (see Section~\ref{subsec:data-concept}).


\noindent$\bullet$ {\it We investigate the unique characteristics of data across different LLM development stages} (Figure~\ref{fig:characteristics}), and provide a systematic overview of the associated challenges and techniques in data processing, storage, and serving (Table~\ref{tab:overview}). In contrast, prior surveys~\cite{wang2023data,dataselectionsurvey,DBLP:journals/tkde/ChaiWLNL23} primarily center on the pre-training stage without covering the full \llm lifecycle like supervised fine-tuning (SFT), retrieval-augmented generation (RAG), and agent-based applications.

\noindent$\bullet$ {\it We provide a lifecycle-based taxonomy of \datallm, introducing key tasks in data processing, storage, and serving.} For each task, we summarize representative methodologies, discuss their design principles, and analyze their strengths and limitations. In comparison, \cite{wang2023data} focuses on deduplication and filtering, \cite{dataselectionsurvey} emphasizes data selection, and \cite{annotationsurvey} reviews data annotation strategies, none of which offer a systematic perspective across the data management pipeline.

\noindent$\bullet$ We {\it introduce recent advances in \llmdata, outlining key components of \llm-driven data optimization.} While earlier work~\cite{2020aidb} has investigated the application of classical machine learning in data management, it largely neglects the distinctive strengths and limitations of \llms, particularly in manipulating data for non-LLM tasks, processing semi-structured and unstructured data, and enabling system-level optimizations.


\noindent$\bullet$ {\it We highlight open challenges and future directions} from both ends: (1) improving data management techniques to meet practical LLM training and deployment needs (e.g., efficient data evaluation, scalable multi-modal storage), and (2) enhancing LLMs’ ability (e.g., private knowledge understanding, informative representation for non-sequential and non-textual data) to perform complex data management tasks across diverse real-world scenarios.

\section{Data Management for \llm (\datallm)}





\subsection{``\dataconcept'' Concept of \llm Data}
\label{subsec:data-concept}

Based on our investigation of over 400 papers~\footnote{https://github.com/weAIDB/awesome-data-llm}, we introduce the \dataconcept concept for evaluating the quality of \llm datasets.

\begin{table*}[]
\caption{\textbf{Technique Comparison} - Data Processing, Storage, and Serving Techniques for Different \llm Stages. {``N/A'' indicates that no relevant work has been reported yet, although the corresponding techniques could potentially be applied.}}
\resizebox{\linewidth}{!}{%
\begin{tabular}{|cc|c|c|c|c|c|c|}
\hline
\multicolumn{2}{|c|}{\textbf{Stage}} &
  {\color[HTML]{222222} \textbf{\begin{tabular}[c]{@{}c@{}}Pre-training /\\ Incremental Pre-training\end{tabular}}} &
  {\color[HTML]{222222} \textbf{\begin{tabular}[c]{@{}c@{}}Supervised \\ Fine-Tuning\end{tabular}}} &
  {\color[HTML]{222222} \textbf{\begin{tabular}[c]{@{}c@{}}Reinforcement \\ Learning\end{tabular}}} &
  {\color[HTML]{222222} \textbf{Inference}} &
  {\color[HTML]{222222} \textbf{RAG}} &
  {\color[HTML]{222222} \textbf{Evaluation}} \\ \hline
\multicolumn{1}{|c|}{} &
  {\color[HTML]{222222} \textbf{Acquisition}} &
  {\color[HTML]{222222} $\checkmark$} &
  {\color[HTML]{222222} $\checkmark$} &
  {\color[HTML]{222222} $\checkmark$} &
  {\color[HTML]{222222} N/A} &
  {\color[HTML]{222222} $\checkmark$} &
  {\color[HTML]{222222} $\checkmark$} \\ \cline{2-8} 
\multicolumn{1}{|c|}{} &
  {\color[HTML]{222222} \textbf{De-duplication}} &
  {\color[HTML]{222222} $\checkmark$} &
  {\color[HTML]{222222} $\checkmark$} &
  {\color[HTML]{222222} N/A} &
  {\color[HTML]{222222} N/A} &
  {\color[HTML]{222222} N/A} &
  {\color[HTML]{222222} N/A} \\ \cline{2-8} 
\multicolumn{1}{|c|}{} &
  {\color[HTML]{222222} \textbf{Filtering}} &
  {\color[HTML]{222222} $\checkmark$} &
  {\color[HTML]{222222} $\checkmark$} &
  {\color[HTML]{222222} N/A} &
  {\color[HTML]{222222} N/A} &
  {\color[HTML]{222222} ×} &
  {\color[HTML]{222222} N/A} \\ \cline{2-8} 
\multicolumn{1}{|c|}{} &
  {\color[HTML]{222222} \textbf{Selection}} &
  {\color[HTML]{222222} $\checkmark$} &
  {\color[HTML]{222222} $\checkmark$} &
  {\color[HTML]{222222} N/A} &
  {\color[HTML]{222222} N/A} &
  {\color[HTML]{222222} N/A} &
  {\color[HTML]{222222} N/A} \\ \cline{2-8} 
\multicolumn{1}{|c|}{} &
  {\color[HTML]{222222} \textbf{Mixing}} &
  {\color[HTML]{222222} $\checkmark$} &
  {\color[HTML]{222222} $\checkmark$} &
  {\color[HTML]{222222} ×} &
  {\color[HTML]{222222} N/A} &
  {\color[HTML]{222222} ×} &
  {\color[HTML]{222222} ×} \\ \cline{2-8} 
\multicolumn{1}{|c|}{\multirow{-6}{*}{\textbf{\begin{tabular}[c]{@{}c@{}}Data \\ Processing\end{tabular}}}} &
  {\color[HTML]{222222} \textbf{Synthesis}} &
  {\color[HTML]{222222} $\checkmark$} &
  {\color[HTML]{222222} $\checkmark$} &
  {\color[HTML]{222222} $\checkmark$} &
  {\color[HTML]{222222} N/A} &
  {\color[HTML]{222222} $\checkmark$} &
  {\color[HTML]{222222} $\checkmark$} \\ \hline
\multicolumn{1}{|c|}{} &
  {\color[HTML]{222222} \textbf{Distribution}} &
  {\color[HTML]{222222} \begin{tabular}[c]{@{}c@{}}Distributed File System\\ Model Offload (GPUs, CPUs)\end{tabular}} &
  {\color[HTML]{222222} \begin{tabular}[c]{@{}c@{}}Model Offload\\  (GPUs, CPUs)\end{tabular}} &
  {\color[HTML]{222222} \begin{tabular}[c]{@{}c@{}}Model Offload \\ (GPUs, CPUs)\end{tabular}} &
  {\color[HTML]{222222} \begin{tabular}[c]{@{}c@{}}Model Offload\\  (GPUs, CPUs)\end{tabular}} &
  {\color[HTML]{222222} \begin{tabular}[c]{@{}c@{}}Model Offload \\ (GPUs, CPUs)\end{tabular}} &
  {\color[HTML]{222222} \begin{tabular}[c]{@{}c@{}}Model Offload \\ (GPUs, CPUs)\end{tabular}} \\ \cline{2-8} 
\multicolumn{1}{|c|}{} &
  {\color[HTML]{222222} } &
  {\color[HTML]{222222} } &
  {\color[HTML]{222222} } &
  {\color[HTML]{222222} } &
  {\color[HTML]{222222} } &
  {\color[HTML]{222222} } &
  {\color[HTML]{222222} } \\
\multicolumn{1}{|c|}{} &
  {\color[HTML]{222222} } &
  {\color[HTML]{222222} } &
  {\color[HTML]{222222} } &
  {\color[HTML]{222222} } &
  {\color[HTML]{222222} } &
  {\color[HTML]{222222} } &
  {\color[HTML]{222222} } \\
\multicolumn{1}{|c|}{} &
  \multirow{-3}{*}{{\color[HTML]{222222} \textbf{Transmission}}} &
  \multirow{-3}{*}{{\color[HTML]{222222} \begin{tabular}[c]{@{}c@{}}Caching Data Placement\\ Parallelized Pipeline\\ Data/Operator Offloading (CPUs)\end{tabular}}} &
  \multirow{-3}{*}{{\color[HTML]{222222} \begin{tabular}[c]{@{}c@{}}Parallelized Pipeline\\ Data/Operator Offloading (CPUs)\end{tabular}}} &
  \multirow{-3}{*}{{\color[HTML]{222222} \begin{tabular}[c]{@{}c@{}}Parallelized Pipeline\\ Data/Operator Offloading (CPUs)\end{tabular}}} &
  \multirow{-3}{*}{{\color[HTML]{222222} ×}} &
  \multirow{-3}{*}{{\color[HTML]{222222} N/A}} &
  \multirow{-3}{*}{{\color[HTML]{222222} N/A}} \\ \cline{2-8} 
\multicolumn{1}{|c|}{} &
  {\color[HTML]{222222} \textbf{Fault Tolerance}} &
  {\color[HTML]{222222} $\checkmark$} &
  {\color[HTML]{222222} $\checkmark$} &
  {\color[HTML]{222222} $\checkmark$} &
  {\color[HTML]{222222} ×} &
  {\color[HTML]{222222} ×} &
  {\color[HTML]{222222} ×} \\ \cline{2-8} 
\multicolumn{1}{|c|}{} &
  {\color[HTML]{222222} } &
  {\color[HTML]{222222} } &
  {\color[HTML]{222222} } &
  {\color[HTML]{222222} } &
  {\color[HTML]{222222} } &
  {\color[HTML]{222222} } &
  {\color[HTML]{222222} } \\
\multicolumn{1}{|c|}{} &
  {\color[HTML]{222222} } &
  {\color[HTML]{222222} } &
  {\color[HTML]{222222} } &
  {\color[HTML]{222222} } &
  {\color[HTML]{222222} } &
  {\color[HTML]{222222} } &
  {\color[HTML]{222222} } \\
\multicolumn{1}{|c|}{} &
  {\color[HTML]{222222} } &
  {\color[HTML]{222222} } &
  {\color[HTML]{222222} } &
  {\color[HTML]{222222} } &
  {\color[HTML]{222222} } &
  {\color[HTML]{222222} } &
  {\color[HTML]{222222} } \\
\multicolumn{1}{|c|}{\multirow{-9}{*}{\textbf{\begin{tabular}[c]{@{}c@{}}Data \\ Storage\end{tabular}}}} &
  \multirow{-4}{*}{{\color[HTML]{222222} \textbf{KV Cache}}} &
  \multirow{-4}{*}{{\color[HTML]{222222} N/A}} &
  \multirow{-4}{*}{{\color[HTML]{222222} N/A}} &
  \multirow{-4}{*}{{\color[HTML]{222222} N/A}} &
  \multirow{-4}{*}{{\color[HTML]{222222} \begin{tabular}[c]{@{}c@{}}Cache Space Management\\ KV Indexing\\ KV Placement\\ KV Shrinking\end{tabular}}} &
  \multirow{-4}{*}{{\color[HTML]{222222} \begin{tabular}[c]{@{}c@{}}KV Placement\\ KV Shrinking\end{tabular}}} &
  \multirow{-4}{*}{{\color[HTML]{222222} N/A}} \\ \hline
\multicolumn{1}{|c|}{} &
  {\color[HTML]{222222} } &
  {\color[HTML]{222222} } &
  {\color[HTML]{222222} } &
  {\color[HTML]{222222} } &
  {\color[HTML]{222222} } &
  {\color[HTML]{222222} } &
  {\color[HTML]{222222} } \\
\multicolumn{1}{|c|}{} &
  {\color[HTML]{222222} } &
  {\color[HTML]{222222} } &
  {\color[HTML]{222222} } &
  {\color[HTML]{222222} } &
  {\color[HTML]{222222} } &
  {\color[HTML]{222222} } &
  {\color[HTML]{222222} } \\
\multicolumn{1}{|c|}{} &
  {\color[HTML]{222222} } &
  {\color[HTML]{222222} } &
  {\color[HTML]{222222} } &
  {\color[HTML]{222222} } &
  {\color[HTML]{222222} } &
  {\color[HTML]{222222} } &
  {\color[HTML]{222222} } \\
\multicolumn{1}{|c|}{} &
  \multirow{-4}{*}{{\color[HTML]{222222} \textbf{Selection}}} &
  \multirow{-4}{*}{{\color[HTML]{222222} \begin{tabular}[c]{@{}c@{}}Sample-Scoring-Based\\ Model-State-Based\end{tabular}}} &
  \multirow{-4}{*}{{\color[HTML]{222222} \begin{tabular}[c]{@{}c@{}}Model-State-Based\\ Experience-Based\end{tabular}}} &
  \multirow{-4}{*}{{\color[HTML]{222222} N/A}} &
  \multirow{-4}{*}{{\color[HTML]{222222} ×}} &
  \multirow{-4}{*}{{\color[HTML]{222222} \begin{tabular}[c]{@{}c@{}}SLM-Based Filtering\\ LLM-Based Filtering\\ Metric-Based Re-ranking\\ LLM-Based Re-ranking\end{tabular}}} &
  \multirow{-4}{*}{{\color[HTML]{222222} ×}} \\ \cline{2-8} 
\multicolumn{1}{|c|}{} &
  {\color[HTML]{222222} \textbf{Compression}} &
  {\color[HTML]{222222} N/A} &
  {\color[HTML]{222222} N/A} &
  {\color[HTML]{222222} N/A} &
  {\color[HTML]{222222} $\checkmark$} &
  {\color[HTML]{222222} $\checkmark$} &
  {\color[HTML]{222222} N/A} \\ \cline{2-8} 
\multicolumn{1}{|c|}{} &
  {\color[HTML]{222222} \textbf{Packing}} &
  {\color[HTML]{222222} $\checkmark$} &
  {\color[HTML]{222222} $\checkmark$} &
  {\color[HTML]{222222} $\checkmark$} &
  {\color[HTML]{222222} ×} &
  {\color[HTML]{222222} ×} &
  {\color[HTML]{222222} ×} \\ \cline{2-8} 
\multicolumn{1}{|c|}{\multirow{-7}{*}{\textbf{\begin{tabular}[c]{@{}c@{}}Data \\ Serving\end{tabular}}}} &
  {\color[HTML]{222222} \textbf{Provenance}} &
  {\color[HTML]{222222} ×} &
  {\color[HTML]{222222} ×} &
  {\color[HTML]{222222} ×} &
  {\color[HTML]{222222} $\checkmark$} &
  {\color[HTML]{222222} N/A} &
  {\color[HTML]{222222} ×} \\ \hline
\end{tabular}%
}
\label{tab:overview}
\end{table*}

\begin{figure*}[!t]
    \centering 
    \includegraphics[width=.95\textwidth]{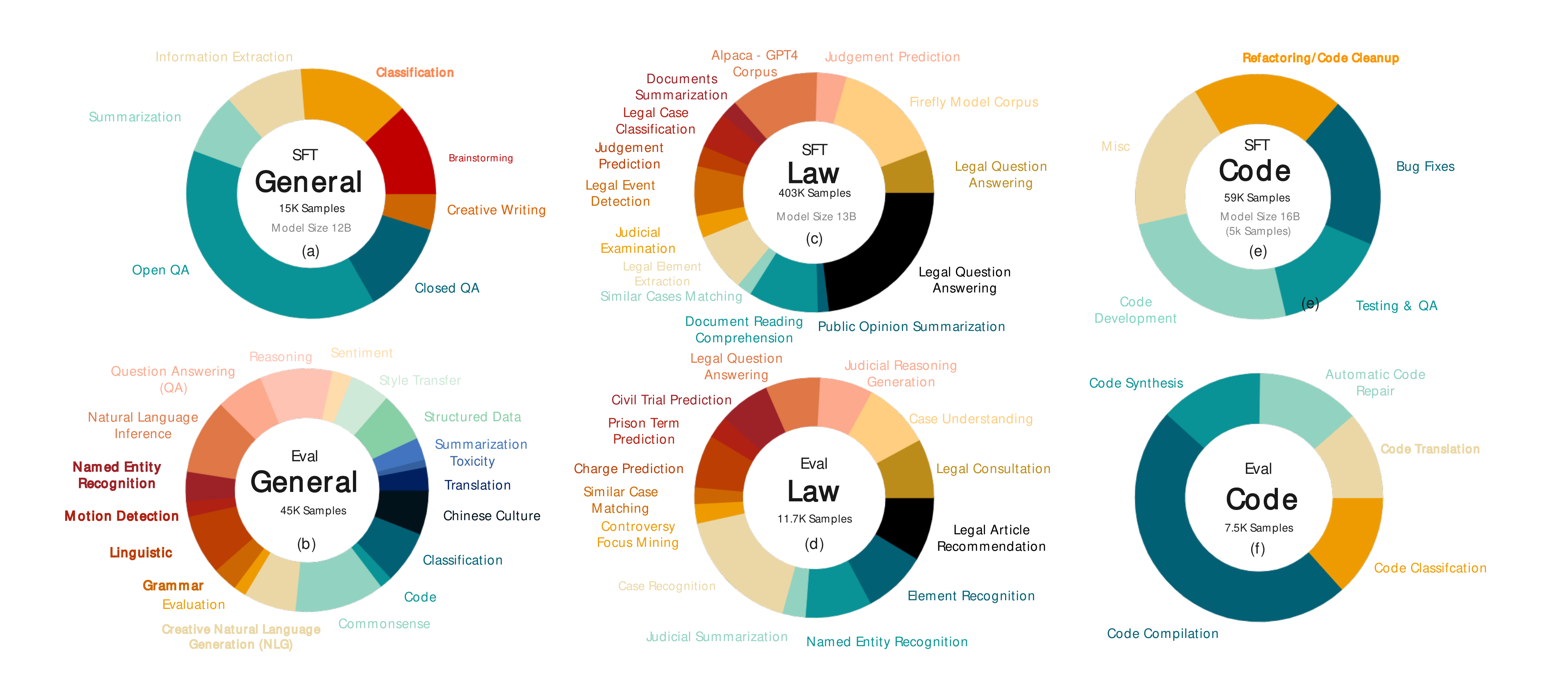}
    \caption{\textbf{Example \llm Data Distributions} - (a) General Domain (SFT)\cite{DatabricksBlog2023DollyV2}, (b) General Domain (Eval) \cite{li2024cifbenchchineseinstructionfollowingbenchmark}, (c)  Law (SFT)\cite{yue2023disc}, (d) Law (Eval)\cite{dai2024laiwchineselegallarge}, (e) Code (SFT) \cite{muennighoff2024octopackinstructiontuningcode}, (f) Code (Eval)\cite{khan2023xcodeevallargescalemultilingual}.}
    \label{fig:data-piechart}
\end{figure*}

\noindent(1) \underline{\textbf{I}nclusiveness:} \llms require data with broad and diverse coverage across multiple dimensions, including domains (e.g., general knowledge, specialized fields like finance, medicine, math~\cite{chen2025advancing}, and physics~\cite{li2024synthetic}), task types (e.g., question answering, summarization, code completion~\cite{wang2022self,mitra2024agentinstruct,shirgaonkar2024knowledge,abbas2024effective,SmallToLarge}), data sources (e.g., GitHub, Wikipedia~\cite{Pile,commoncrawl,raffel2020exploring,shen2023slimpajama}), languages~\cite{chen2024towards,shen2023slimpajama}, expression styles (e.g., academic, casual, formal~\cite{maini2024rephrasing,zhao2024decoratelm}), and data modalities (e.g., text~\cite{Pile,commoncrawl}, images~\cite{fuest2024diffusionmodels,diffusemix}, videos~\cite{CogVideoX,Hunyuanvideo,Wan}, tables~\cite{raffel2020exploring}).  


\noindent(2) \underline{\textbf{a}bundance:} LLMs require data with appropriate volume and balanced composition to prevent overfitting on homogeneous data. Specifically, abundance of data involves: $(i)$ constructing well-balanced datasets during pre-training~\cite{feng2024maximize,pan2024scalebio,xie2023doremi,liu2024regmix}, $(ii)$ adjusting data ratios to align with target applications during fine-tuning~\cite{ma2024task,fan2023doge}, and $(iii)$ continually enhancing domain-specific capabilities while maintaining acceptable general performance degradation in continual pre-training~\cite{que2024d,gu2024cmr}. Notably, the strength of LLMs lies not only in large-scale data~\cite{maini2024rephrasing,zhou2024jiuzhang3,commoncrawl,raffel2020exploring,Pile,shen2023slimpajama}, but also in constructing purposefully balanced datasets, which can further accelerate training and reduce computational cost.



\noindent(3) \underline{\textbf{a}rticulation:} LLMs require data that exhibit strong articulation, including three key aspects: $(i)$ the data should be well-formatted (e.g., proper punctuation and capitalization~\cite{chen2024data}), clean (free from duplicates, typos, and irrelevant content such as spam or gibberish~\cite{chen2024data}), and self-contained, featuring clear, fluent, and unambiguous language~\cite{maini2024rephrasing,zhao2024decoratelm}, $(ii)$ the data should be instructive~\cite{huang2024mustard,huang2024key,chen2025advancing}, i.e., offering sufficient context, guidance, and intermediate explanations that help the model connect questions to relevant background knowledge and understand the reasoning process. $(iii)$ the data should involve {step-by-step reasoning}\cite{li2025llms,ye2025limo,shen2025satori,hou2025advancing,zhu2024padprogramaideddistillationteach}, such that enhancing the \llms' reasoning capabilities by decomposing complex tasks into smaller, interpretable steps. 



\noindent(4) \underline{\textbf{S}anitization:} \llms require data to be \textit{sanitized}, meaning it is rigorously controlled and filtered to remove harmful elements while maintaining inclusiveness and neutrality. This involves four critical dimensions: $(i)$ \textit{Privacy compliance}, which requires the exclusion of personally identifiable information (e.g., ID numbers, phone numbers), inferred social relationships, and geolocation-related metadata~\cite{zeng2024mitigating,liu2023deid,lukas2023analyzing}; $(ii)$ \textit{Toxicity-free content}, ensuring the complete removal of hate speech, incitement to violence, and psychologically harmful language, as well as eliminating any discriminatory or aggressive semantic constructs~\cite{navigli2023biases}; $(iii)$ \textit{Ethical consistency}, which prohibits the presence of extremist ideologies, instructions for illegal activities, and stereotype-reinforcing narratives that may cause social harm~\cite{shen2024seal,slyman2024fairdedup,navigli2023biases}; and $(iv)$ \textit{Risk mitigation}, filtering out unverified medical claims, politically misleading information, and culturally insensitive expressions to prevent misinformation and value misalignment. Sanitized data must maintain a neutral tone and adopt an inclusive contextual framework, serving as a critical foundation for building safe LLMs~\cite{shen2024seal,slyman2024fairdedup}.


\subsection{Data Characters across \llm Stages}
\label{subsec:data-characters}

Next we specifically discuss the data characteristics across different \llm stages, together with the distinct techniques for data processing, storage, and serving (Table~\ref{tab:overview}). 


\hi{Data for Pretraining.} In the pre-training stage, \llms rely on TB-scale, diverse datasets to acquire broad language and even cross-modality understanding capabilities, while reducing the risk of overfitting. These datasets are typically sourced from a wide range of domains and formats, including web crawls (e.g., HTML pages and WARC files~\cite{commoncrawl}), open-source code repositories (e.g., raw source code files with metadata~\cite{github}), books (e.g., plain text or EPUB formats~\cite{zhu2015bookscorpus}), academic papers (e.g., LaTeX source or PDF-converted text~\cite{arxiv}), and interleaved image-text corpora (e.g., aligned captioned images in JSON or WebDataset format~\cite{laurenccon2024obelics}).

\hi{Data for Continual Pre-training.} Continual pre-training (or continued pre-training) typically involves datasets containing millions to billions of tokens, which are often over 100 times smaller than those used in the initial pre-training stage. The primary objective is to fill knowledge gaps and adapt the model to specific domains. Representative domain-specific datasets are like: (1) Finance: BBT-FinCorpus~\cite{lu2023bbt}, a large-scale and diverse financial datasets comprising approximately 300 GB of text; and (2) Healthcare: Medical-pt~\cite{MedicalGPT}, a Chinese-English medical dataset containing 360,000 entries curated from medical encyclopedias.

\hi{Data for Supervised Fine-Tuning (SFT).} Unlike pre-training, SFT relies on data presented in the form of instruction-response pairs, where the response includes not only the correct answer but also guidelines on tone, style, and reasoning steps to ensure user-friendly output. 

The SFT stage typically involves much smaller datasets compared to pre-training.  These datasets often consist of thousands to millions of labeled examples, with each example carefully crafted to guide the model in learning a specific, narrower set of tasks. For instance, in Figure~\ref{fig:characteristics},  (1) the summarization task constructs prompts using {\it problem descriptions and summarization objects}; (2) closed QA using {\it questions and corresponding knowledge texts}; (3) open QA tasks using {\it only questions} without knowledge text; and (4) captioning tasks using {\it task descriptions and images}. These prompts are paired with unique responses for model finetuning.

The composition of SFT datasets varies based on the application scenarios: 

\noindent \bfit{(1) General Instruction Following:} For \llms as general-purpose chatbots, SFT data include instructions for various daily tasks. Databricks-dolly-15K~\cite{DatabricksBlog2023DollyV2} is a corpus containing over 15,000 records. It encompasses seven types of tasks, including creative writing, closed QA, open QA, summarization, information extraction, classification, brainstorming. {This dataset is designed to enhance \llm to better adapt to specialized outputs that align with human-style requirements across diverse tasks. For example, in text summarization, it provides concise summary statements; whereas in text organization tasks, it structures outputs in table-of-contents format.} 

\noindent \bfit{(2) Specific Domain Usage:} For models specialized in fields such as law, finance, or medicine, the SFT data focuses on tasks pertinent to these fields. For example, DISC-Law-SFT~\cite{yue2023disc} is a legal SFT dataset containing 295k data entries from various legal scenarios, such as legal information extraction (32k), legal judgment prediction (16k), legal event detection (27k), and legal question-answering (93k). Similarly, Medical-SFT~\cite{MedicalGPT} is a medical SFT dataset (totaling 2,060k pieces), composed of  medical inquikry data (790k), online medical encyclopedia QA data (360k), English medical inquiry data (110k), medical knowledge graph QA data (79k). {For tasks such as legal question-answering and legal judgment prediction, the data is structured as triplets, comprising the prompt, response, and supporting reference information (e.g., legal provisions, case-based evidence, or regulatory documents). For the remaining tasks, they all take the form of instruction pairs composed of prompt and response.}

\hi{Data for Reinforcement Learning (RL).} {RL is generally divided into two types: one is RLHF (Reinforcement Learning with Human Feedback), and the other is Reasoning-oriented Reinforcement Learning (RoRL).} 

\noindent\bfit{(1) RLHF:} RLHF data is typically smaller than SFT data (e.g., thousands to dozens of millions of data samples), which involve more complex data annotations. Specifically, annotators compare multiple candidate responses to the same instruction and rank them according to human preference (e.g., levels from most helpful to least helpful). Collecting these preference pairs or rankings is more time-consuming than constructing instruction-response pairs in SFT. 

In the general domain, UltraFeedback~\cite{cui2024ultrafeedback} consists of 64,000 samples. For each sample, different models are used to generate 4 responses for each prompt (totaling 256,000 responses). GPT-4 is then employed to generate feedback for these four responses, which is used to help \llms to generate outputs that are in line with human standards and appropriateness. 

In specific domains such as healthcare, Medical-RLHF~\cite{MedicalGPT} has 4,000 random questions from a Chinese medical dialogue dataset. Each question is paired with a well-organized answer (i.e., the human doctor's reply) and a weaker answer from Llama-based model fine-tuned over synthesized QA samples. {These labeled data are used to train a reward model. During the training of the \llm, the reward model provides feedback based on the \llm's answers, guiding the training process towards generating high-quality responses.}

\noindent\bfit{(2) RoRL:} Compared to the complex annotated data in RLHF, RoRL allows the model to discover the best reasoning approach on its own through the correctness of the reward model. Specifically, it focuses on tasks requiring long-term reasoning, such as mathematical, coding, and logical designing experiments~\cite{guo2025deepseek}. {Under the premise of providing feedback on whether the answer is correct or not,   algorithm such as the Group Relative Policy Optimization (GRPO)~\cite{guo2025deepseek} and long-CoT RL~\cite{kimiteam2025kimik15scalingreinforcement} are adopted to train the model to independently discover the optimal problem-solving steps and converge.}

\hi{Data for Retrieval-Augmented Generation (RAG).} The RAG stage differs from above training stages, which involves large-scale dataset (reference corpus) for \llms to retrieve from during inference. {In this stage, data must be strictly reviewed to ensure authenticity and validity, while dynamic data requires real-time updates.} The domain of RAG datasets varies depending on the specific application scenarios. For instance, (1) in the medicine-specific \llm application (Medical-Graph-RAG), MIMIC-IV is used as the RAG dataset~\cite{wu2024medical}. This dataset contains data from over 65,000 ICU patients and more than 200,000 patients treated in emergency departments; (2) in the legal field, the RAG knowledge base used by DISC-LawLLM~\cite{yue2023disc} contains more than 800 national and local laws, regulations, and rules, as well as 24,000 legal-related exam questions. Besides, RAG data can include users' historical conversation records or personal information, in order to build a user-personalized \llm~\cite{ragchat,ragchat2,ragchat3}.

\hi{Data for \llm Evaluation.} Suitable evaluation datasets are essential for evaluating the performance of \llms. They provide representative data samples that reflect different aspects of an LLM's capabilities. 

In the general domain, the MMMU benchmark is used to assess the performance of LLMs across major multi-modal tasks in six key disciplines, covering 30 subjects and 183 subfields. It is built from 11,500 carefully curated questions and effectively tests models' perception, knowledge, and reasoning abilities~\cite{yue2024mmmu}.

In specific domains, typical evaluation datasets include those in coding, healthcare and law domains: (1) OpenAI's HumanEval dataset includes 164 programming problems, complete with function signatures, docstrings, bodies, and multiple unit tests. These problems are handcrafted to ensure they are not part of the training sets used for code generation models~\cite{humaneval}; (2) MedQA~\cite{medqa} contains a large number of medical exam questions from various regions, totaling 61,097 questions; (3) LexEval~\cite{li2024lexeval} constructs 23 evaluation tasks based on a legal cognitive classification framework, covering different aspects of legal knowledge, with at least 100 evaluation samples for each task.

\hi{Data for \llm Agents.} Beyond vanilla LLMs, agents strive for more advanced capabilities such as planning, tool orchestration and {multi-turn dialogue capability}~\cite{liu2024largelanguagemodelbasedagents}. These capabilities impose higher requirements on the training data for LLMs. First, many studies~\cite{wang2025stecastepleveltrajectorycalibration} aim to enhance planning abilities through interaction trajectory data, which refers to a sequence of records generated during the interaction between the agent and the environment, typically represented as $(\text{instruction } i, \text{ action } a_1, \text{ observation } o_1, \ldots, \text{ action } a_n)$. {UltraInteract~\cite{yuan2024advancing} takes the instruction as the root node, and uses both the correct actions and their corresponding incorrect actions as nodes to construct a preference trajectory tree, enabling the agent to learn the human preference of different actions}. Second, other studies focus on enhancing the agent's tool usage capabilities using tool usage data. For instance, AutoTools~\cite{shi2025tool} fine-tunes models on tool data that is labeled with special tags, such as $\texttt{<python>} code \texttt{</python>}$, thereby grounding language in concrete tool invocations. {Third, to enhance the agent's multi-turn dialogue capability, UltraChat~\cite{ding2023enhancing} employs an additional LLM to simulate user instructions and conversational content, thereby collecting multi-turn dialogue data.}

\begin{figure*}
    \centering 
    \hspace{-1em}
    \includegraphics[width=1.02\textwidth]{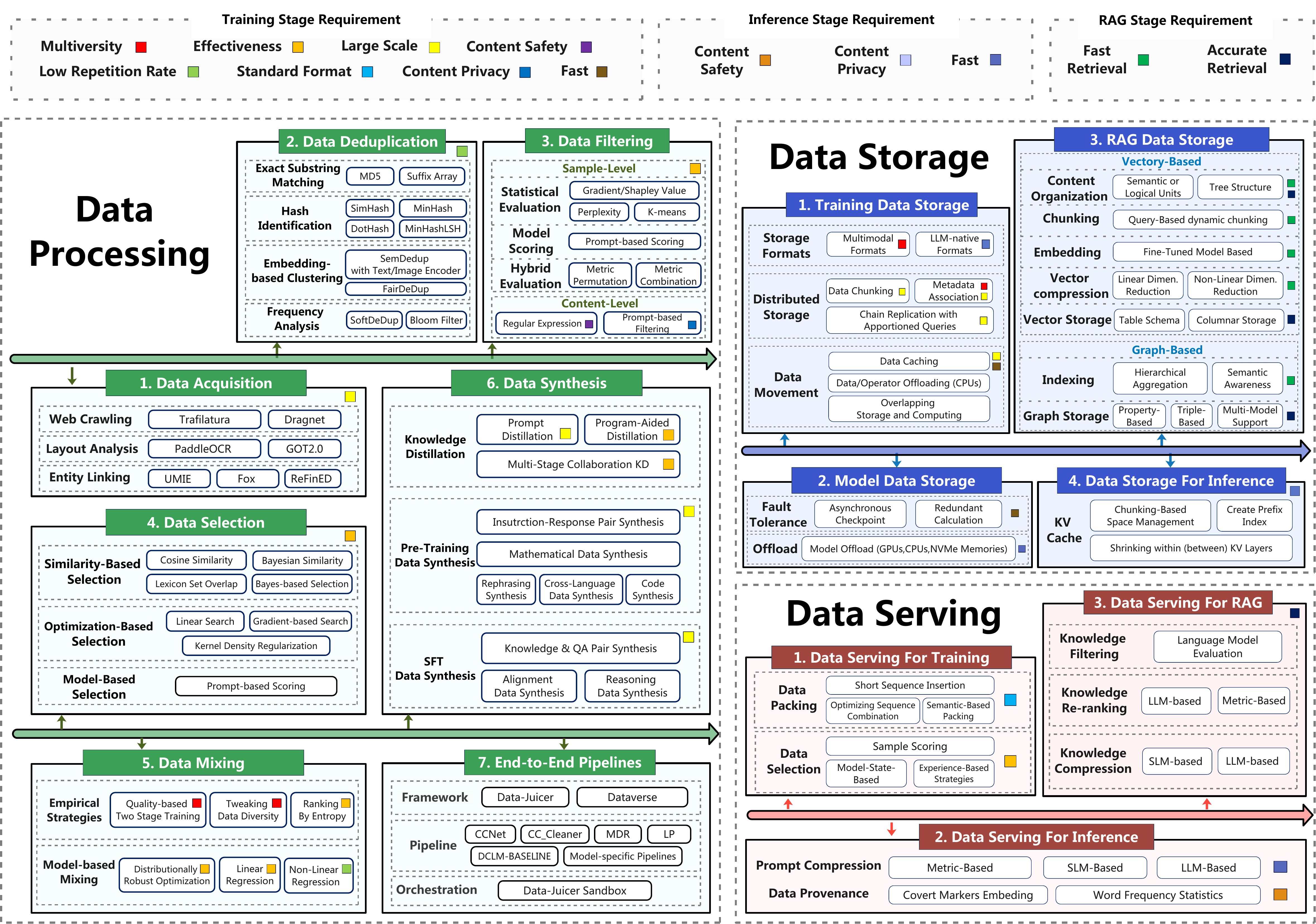}
    \vspace{-.5cm}
    \caption{Overview of \datallm Techniques.}
    \label{fig:motivation}
    \vspace{-1.em}
\end{figure*}

\subsection{Data Processing for \llm}
\label{sec:processing}

\begin{table}[!t]
\caption{Data Acquisition for \llms.}
\label{table:acquisition}
\resizebox{\linewidth}{!}{
\begin{tabular}{c c c c}
\hline
{\centering Method} & {\centering Objective} & {\centering Solution} & {\centering Tools} \\ \hline
\multirow{8}{*}{\makecell[c]{Website\\ Crawling}} & \multirow{3}{*}{\begin{tabular}[c]{@{}c@{}}HTML Textual \\ Content Extraction \end{tabular}} & Rule-based & Trafilatura~\cite{trafilatura} \\
 &  & Rule-based & BET~\cite{bet} \\
 &  & ML-based & Dragnet~\cite{dragnet} \\
 &  &  &  \\
 & \multirow{4}{*}{\begin{tabular}[c]{@{}c@{}}Automate Browser \\ Interactions \end{tabular}} & HTML parsing & Beautiful Soup~\cite{beautifulsoup} \\
 &  & Control web driver & Selenium~\cite{selenium}  \\
 &  & Wrap high-level API & Playwright~\cite{playwright}  \\
 &  & DevTools protocol & Puppeteer~\cite{puppeteer}  \\ \hline
\multirow{4}{*}{\begin{tabular}[c]{@{}c@{}}Layout-based\end{tabular}} & \multirow{4}{*}{\begin{tabular}[c]{@{}c@{}}Content Extraction \\ from Handwritten \\ or Non-text Data\end{tabular}} & Model pipeline & PaddleOCR  \\
 &  & Model pipeline & MinerU~\cite{wang2024mineru} \\
 &  & Multimodal \llm & GOT2.0~\cite{wei2024got} \\
 &  & Multimodal \llm & Fox~\cite{liu2024fox} \\ \hline
\multirow{4}{*}{\begin{tabular}[c]{@{}c@{}}Entity\\ recognition\\ \& linking\end{tabular}} & New Sample Derivation & Bi-Transformer & ReFinED~\cite{ayoola2022refined} \\
 & \multirow{2}{*}{Translation Consistency} & \multirow{2}{*}{\begin{tabular}[c]{@{}c@{}}Seq2seq Framework\\ using References \end{tabular}} & \multirow{2}{*}{AACTRANS~\cite{kolluru2022alignment}}  \\
 &  &  &  \\
 & Text-Image Integration & Multimodal \llm & UMIE~\cite{sun2024umie} \\
\hline
\end{tabular}
}
\end{table}

\subsubsection{Data Acquisition}
\label{subsubsec:acquisition}



Unlike classic machine learning, which primarily relies on collecting labeled data within a specific domain for supervised training (e.g., data for sentiment analysis and sentence similarity estimation), data acquisition for \llms typically (1) relies on large-scale web scraping to collect extensive data across diverse domains for unsupervised pretraining and (2) employs techniques such as layout analysis and entity linking to extract additional data from the collected content.






\begin{tcolorbox}[colback=gray!1,colframe=gray,lowerbox=visible]
  \textbf{Principles}
  \tcblower
  {Unlike classic ML data acquisition, \llms rely heavily on large-scale web scraping to ensure broad coverage and robust generalization. The main challenge is extracting high-quality textual content, often aided by layout-based and entity-linking methods. Managing time and resource efficiency at scale remains vital.}
\end{tcolorbox}



\hi{Data Sources.} The data is gathered from two primary sources:

\noindent\bfit{(1) Public Data}, often freely available under open licenses, include resources such as webpages~\cite{commoncrawl}, books~\cite{zhu2015bookscorpus}, and publicly accessible code repositories~\cite{Kocetkov2022TheStack}. 

\noindent$\bullet$ \underline{\textit{Webpage sources}} provide extensive pre-processed website content, such as 1.56T english text from crawled websites in C4~\cite{C4-T5}, 6.6B multilingual pages in mC4~\cite{xue2021mt5}, 6.3 trillion tokens of multilingual pages in CulturaX~\cite{nguyen2024culturax}.



\noindent$\bullet$ \underline{\textit{Digitized books}} supply structured, high-quality text, such as over 75,000 eBooks in Project Gutenberg~\cite{projectgutenberg}, over two million free ebooks in Open Library~\cite{openlibrary}, and film-aigned book descriptions in  BookCorpus~\cite{zhu2015bookscorpus}).

\noindent$\bullet$ \underline{\textit{Code repositories}} (e.g., GitHub~\cite{github}, GitLab~\cite{gitlab}, Bitbucket~\cite{bitbucket}) offer abundant programming data that can facilitate code search and analysis tasks, such as CodeSearchNet~\cite{husain2019codesearchnet} with 2M (comment, code) pairs.

\noindent\bfit{(2) Private Data} involve proprietary or confidential information not publicly available, such as internal company documents, customer support logs, application event logs, subscriber-only content (e.g., premium news articles, licensed scientific databases).  Collecting this data requires careful attention to ethical and legal constraints (e.g., GDPR, CCPA) and mandates removing sensitive details (e.g., employing anonymization or pseudonymization) and using secure pipelines (e.g., CI/CD systems) with encryption and role-based access controls. For instance, proprietary codebases and user-generated content (chat logs, Q\&A sessions) must be gathered under secure processes to maintain confidentiality.

\hi{Data Acquisition Methods.} As shown in Table \ref{table:acquisition}, there are three main techniques for data acquisition, including website crawling, layout analysis, and entity recognition and linking.

\noindent\bfit{(1) Website Crawling.} Most data are obtained through website crawling, which aims to extract textual content from crawled HTML files or multimodal image-text pairs using various extraction tools and browser automation assistants.  

Generally, we first parse the raw HTML to separate meaningful textual content from boilerplate elements. Second, since typical extraneous components (e.g., headers, footers, advertisements, sidebars) often contribute little to the data value (e.g., for \llm training), we execute scripts (using CSS selectors or XPath queries) to identify and extract critical elements like article text, headlines, dates, and author bylines. Third, once the relevant text has been scraped, we store it in structured format such as JSON, CSV, database (see data storage in Section~\ref{sec:storage}) for further processing. Specifically, for image elements encountered in HTML files, the image source URL is recorded, and the content of the \texttt{alt} attribute within the \texttt{<img>} tag is extracted and utilized as the corresponding image's textual caption.

\noindent$\bullet$ \underline{\emph{Rule-based Crawling.}} Most existing tools use heuristic rule-based matching algorithm. Trafilatura~\cite{trafilatura} is a heuristic algorithm based on hand-crafted rules (e.g., match HTML DOM nodes with the class equal to ``navbar'' to filter the navigation bar). BET~\cite{bet} employs the cumulative HTML tag distribution to find the largest region of fewest tags per text and extracts the corresponding text as the main content. 

\noindent$\bullet$ \underline{\emph{ML-based Crawling.}} Since many website regions cannot be easily classified by rules, some works~\cite{bevendorff2023contentextraction,trafilatura} design a HTML tag classifier to judge whether a DOM node contains textual content, where they adopt $L^2$ regularized logistic regression that inputs text density features and word frequencies in ''id`` and ''class`` attributes and outputs the probability that a given node contains textual useful content.

\noindent$\bullet$ \underline{\emph{Auxiliary Tools.}} Moreover, some auxiliary tools integrate user-friendly APIs for operating and interacting with HTML DOM trees. Beautiful Soup~\cite{beautifulsoup} is widely used to parse the raw HTML in Python. Selenium~\cite{selenium} automates browser actions and handles dynamic pages by controlling a web driver that communicates with the browser. Playwright~\cite{playwright} provides a high-level API to automate browser tasks while Puppeteer~\cite{puppeteer} communicates directly with the browser using the DevTools Protocol, allowing for headless browser interactions (e.g., in JavaScript-heavy websites).

\noindent\bfit{(2) Layout Analysis.} Layout analysis focuses on extracting textual content from handwritten or non-textual data (e.g., from the crawled ones), which can contain valuable information and require advanced layout analysis techniques for effective extraction. Existing methods include pipeline-based and end-to-end approaches.

\noindent$\bullet$ \underline{\emph{Layout Analysis Pipelines.}}  Intuitively, many works adopt OCR technology (e.g., Tesseract~\cite{Tesseract}) to convert raw data (e.g., scanned books) into machine-readable formats~\cite{PaddleOCR,wang2024mineru} in a pipeline manner, which consist of multiple small models. 
PaddleOCR~\cite{PaddleOCR} passes an image through a Layout Analysis model, which divides the image into different regions such as text, tables, and formulas for separate processing. The table area is sent to the Form Recognition module for structured recognition, and the text areas and formulas are input to the OCR engine for text recognition. Finally, the Layout Restoration module reconstructs all the regions in textual format using heuristic rules based on the relative location information of different extracted regions. 

Similarly, MinerU~\cite{wang2024mineru} works in a pipeline manner. It fine-tunes LayoutLMv3~\cite{huang2022layoutlmv3} for layout detection and YOLOv8~\cite{wang2024yolo} for formula detection to improve the system's generalization (handling a wider range of document types). The detected data are kept in markdown or JSON format. 


\noindent$\bullet$ \underline{\emph{End-to-End Models.}} End-to-End layout analysis refers to adopt multi-modal \llms to conduct end-to-end text acquisition. 
For instance, GOT2.0~\cite{wei2024got} is a acquisition model composed of $(i)$ a high-compression encoder that transforms the image to tokens, $(ii)$ a long-context decoder that outputs the corresponding OCR results, and $(iii)$ a linear layer acting as the connector to map the channel dimension between the vision encoder and the language decoder. Another example is Fox~\cite{liu2024fox}, which employs the natural content-aware CLIP-ViT~\cite{radford2021clip} and the artificial content-aware Vary~\cite{wei2023vary} as two vision encoders, enabling the model to perform fine-grained interactions and multi-page document understanding.
The end-to-end architecture reduces maintenance costs and enhances versatility, enabling the recognition of more complex elements (e.g., charts, sheet music) and supporting improved readability formats for formulas and tables (e.g., \LaTeX, Markdown). However, due to the use of \llms with larger parameter size (e.g, \textless 20M for PaddleOCR vs. 580M for GOT2.0 and 1.8B for Fox), the inference efficiency of these methods still needs improvement.

\noindent\bfit{(3) Entity Recognition $\&$ Linking.} Additionally, we can derive more valuable \llm samples by identifying and linking entities from the above extracted data. 
WEBIE~\cite{webie} introduces a large-scale, entity-linked information extraction dataset with 1.6M sentences from Common Crawl. It links entities using ReFinED~\cite{ayoola2022refined}, and applies distant supervision (DS) to extract 4.8M triples, where each triple consists of a subject, a relationship, and an object. 

Furthermore, to ensure the consistency of derived and origin samples (e.g., translation across English and other languages), Alignment-Augmented Consistent Translation (AACTRANS) model~\cite{kolluru2022alignment} uses a Seq2Seq framework that incorporates reference text in the target language to guide translations, ensuring consistency across related pieces of text. During training, aligned text pairs are augmented with reference-based word alignments to bias the model toward consistent translations. At inference, a common reference translation of the original sentence is used to align and translate related extractions using the AACTRANS model. 

However, AACTRANS fails to leverage shared knowledge across tasks, limiting the alignment performance. 
Instead, UMIE~\cite{sun2024umie} integrates text and visual inputs and produces structured outputs to learn linking knowledge from multiple tasks. The UMIE model is composed of four modules: (1) a text encoder for task instruction comprehension, (2) a visual encoder for image understanding, (3) a gated attention mechanism for cross-modal integration, and (4) a text decoder for structured output generation. Following different task instructors, UMIE is capable of performing various MIE tasks and generating corresponding structured outputs, thereby facilitating knowledge sharing.

{\it Notably, recent \llms could automatically learn the relationships among samples from randomly provided data, rendering the explicit entity linking an optional procedure in the data acquisition process~\cite{ding2024chatelentitylinking}.}

\subsubsection{Data Deduplication}
\label{subsec:deduplicate}

\begin{table}[]
\caption{Data Deduplication for \llms.}
\label{table:deduplication}
\resizebox{\columnwidth}{!}{%
\begin{tabular}{cccc}
\hline
\textbf{Method} & \textbf{Objective} & \textbf{Modality} & \textbf{Work} \\ \hline
\begin{tabular}[c]{@{}c@{}}Exact\\ substring\\ matching\end{tabular} & \begin{tabular}[c]{@{}c@{}}Deduplicate\\ samples with\\ identical substrings\end{tabular} & Text & \begin{tabular}[c]{@{}c@{}}MD5 \cite{dong2024baichuanseed}\\ Suffix Array \cite{nystrom2022deduplicating}\end{tabular} \\ \hline
\begin{tabular}[c]{@{}c@{}}Hashing\\ identification\end{tabular} & \begin{tabular}[c]{@{}c@{}}Deduplicate\\ samples with\\ similar substrings\end{tabular} & Text & \begin{tabular}[c]{@{}c@{}}SimHash \cite{SimHash}\\ MinHash \cite{MinHash, dong2024baichuanseed, nystrom2022deduplicating}\\ MinHashLSH \cite{shen2023slimpajama,silcock2022noise}\\ MinHash +\\ Bloom Filter \cite{khan2024lshbloom}\\ DotHash \cite{DotHash}\end{tabular} \\ \hline
\begin{tabular}[c]{@{}c@{}}Frequency\\ analysis\end{tabular} & \begin{tabular}[c]{@{}c@{}}Down-weighing\\ samples with\\ higher commonness\end{tabular} & Text & SoftDeDup \cite{he2024softdedup} \\ \hline
\begin{tabular}[c]{@{}c@{}}Embedding-\\ based\\ clustering\end{tabular} & \begin{tabular}[c]{@{}c@{}}Deduplicate\\ samples with\\ identical topics but\\ different formats\end{tabular} & \begin{tabular}[c]{@{}c@{}}Text +\\ Image\end{tabular} & \begin{tabular}[c]{@{}c@{}}SemDeDup \cite{abbas2023semdedup}\\ SemDeDup +\\ SSL Prototypes\cite{tirumala2023d4}\\ FairDeDup \cite{slyman2024fairdedup}\end{tabular} \\ \hline
\end{tabular}%
}
\end{table}


The collected raw data often contains significant redundancy, which can negatively impact \llm performance either by reducing its generalization ability to new or rarely-seen tasks \cite{nystrom2022deduplicating} or by memorizing and overfitting to the repeated subsets \cite{hernandez2022scaling,xie2023analysis}. Various deduplication methods have been proposed to detect and mitigate duplication, either by (1) completely removing duplicate samples \cite{dong2024baichuanseed, nystrom2022deduplicating, shen2023slimpajama, silcock2022noise, khan2024lshbloom, abbas2023semdedup, tirumala2023d4, slyman2024fairdedup} or by (2) down-weighing duplicate samples for data resampling \cite{he2024softdedup}. We classify these methods into four main categories.

\hi{Exact Substring Matching.} Exact substring matching methods identify and remove exactly identical samples across datasets, which can happen if (1) a sample references another sample (e.g., a report related to another), or (2) two individual datasets accidentally include the same sample (e.g., a webpage of a popular website). It is commonly used as a preliminary step to remove duplications. Relevant methods leverage techniques like hashing~\cite{dong2024baichuanseed} and suffix array~\cite{nystrom2022deduplicating} at the sample or sentence level.

\begin{tcolorbox}[colback=gray!1,colframe=gray,lowerbox=visible]
  \textbf{Principles}
  \tcblower
  Compared to structured classic ML data, LLM data is unstructured and requires careful identification and removal of duplicate or near-duplicate content from training datasets to improve efficiency, prevent overfitting, and mitigate bias using statistical metrics like perplexity or model evaluation. Challenges include (1) how to encode semantic texts into representations that could be precisely and efficiently compared and (2)  the scalability of the deduplication methods.
\end{tcolorbox}


\noindent $\bullet$ \underline{\emph{Sample-Level.}} \cite{dong2024baichuanseed} conducts sample-level deduplication by calculating the MD5 hashing value of each sample and deduplicate samples with identical MD5 values. 

\noindent $\bullet$ \underline{\emph{Sentence-Level.}} \cite{nystrom2022deduplicating} performs sentence-level deduplication by using Suffix Array, which combines all the samples into one sentence, computes the sentence Suffix Array, and deduplicates samples with common prefixes in the Suffix Array. Suffix Array \cite{manber1993suffix} is a data structure that stores the starting indices of string suffixes in lexicographical order. For instance, given the string ``patata'', its suffixes in lexicographical order are \emph{[``a'' (index 5), ``ata'' (index 3), ``atata'' (index 1), ``patata'' (index 0), ``ta'' (index 4), ``tata'' (index 2)]}, so its suffix array is \emph{(5, 3, 1, 0, 4, 2)}. As identically duplicate samples have the same prefix, they will become adjacent in the suffix array, making it easier to find the duplicates across the samples. In practice, they construct a suffix array on the sequence with a threshold of 50 tokens (empirically determined for significantly reducing the false positives), and find the duplicate samples with common prefixes in linear time.

\hi{Approximate Hashing-based Deduplication.} Hashing-based methods hash each sample into a fixed-length vector and deduplicate samples with significant vector overlap. Compared with the exact matching-based approach, it can identify near-duplicate samples with only a few words of difference (e.g., advertisements generated using the same template). Unlike normal hashing algorithms like MD5, hashes generated in this approach do not change significantly with even a bit of modification, making it possible to detect near-duplicate samples. There are various hashing algorithms, including SimHash \cite{SimHash}, MinHash \cite{MinHash}, DotHash \cite{DotHash}, and their variants \cite{shen2023slimpajama, silcock2022noise}.

\noindent $\bullet$ \underline{\emph{MinHash~\cite{MinHash}}} hashes samples into vectors using a series of hashing functions, where only the minimum value is retained for each function, and estimates similarity for each pair of vectors through Jaccard Index $Jaccard(X, Y) = \frac{X \cap Y}{X \cup Y}$, where X and Y represent sets of elements (For example, if X = {a, b, c, d} and Y = {b, c, d, e, f}, the Jaccard Index over X and Y would be $\frac{1}{2}$). \cite{MinHashVSSimHash} demonstrates that MinHash generally outperforms SimHash. In practice, \cite{dong2024baichuanseed} employed MinHash to the code data on both the sample and the repository levels for diversity and integrity, and \cite{nystrom2022deduplicating} employed MinHash on the sample level.

Moreover, MinHash has various variants for acceleration. MinHashLSH~\cite{shen2023slimpajama,silcock2022noise} improves MinHash by involving locality-sensitive hashing (LSH), which divides a vector into multiple bands and only compares the samples with partially identical vector bands instead of the whole vector, mitigating the computational overhead in sample comparison. LSHBloom~\cite{khan2024lshbloom} further improves MinHashLSH by using Bloom Filter, which hashes each band into a single integer value and inserting it into each corresponding Bloom Filter, and the sample will be flagged as a duplicate if any band's hashed value collides with an entry in the Bloom filter, accelerating duplicate samples searching while reducing memory usage with negligible false positive rate (e.g., 1e-5 in experiments). 

However, MinHash-based methods require building massive vector sets. When the number of samples and their lengths grow large, constructing vector sets becomes exceedingly expensive in terms of both time and space. Moreover, as the feature vector computation for each sample depends on this shared vocabulary, it is difficult to fully parallelize the process.

\noindent $\bullet$ \underline{\emph{SimHash~\cite{SimHash}.}} To address MinHash's issues, SimHash~\cite{SimHash} generates a sample’s feature vector \emph{solely from the words it contains}, converts each sample into a fixed-dimensional binary vector for similarity comparison. Specifically, it first hashes each token in the sample (e.g., by BPE tokenizer~\cite{bpetokenizer}) into a fixed-dimension vector of $\{0, 1\}^d$ (e.g., $[1, 0, 0, 1]$ and $[1, 1, 0, 0]$ ) weighted by the pre-defined weight $w$ (e.g., $w_1$ and $w_2$), where the weight is positive for $1$ and negative for $0$ (e.g., $[w_1, -w_1, -w_1, w_1]$, $[w_2, w_2, -w_2, -w_2]$). Then it added up these weighted vectors to a new vector of the same dimension $d$ (e.g., $[w_1 + w_2, -w_1 + w_2, -w_1 - w_2, w_1 - w_2]$). Finally, the values of the new vector are mapped to another vector of $\{0, 1\}^d$, where the positive values are mapped to 1 and 0 otherwise. The final vector is the fingerprint of each sample, and the similarity of the two samples is estimated by calculating the Hamming distance between their vectors. 

Compared with MinHash, SimHash stores and compares only one hash signature for each sample, greatly reducing the storage and computing overhead. However, keeping only one signature makes it harder to distinguish between two samples, especially those with low Hamming distances, requiring careful curation of data features.

\noindent $\bullet$ \underline{\emph{DotHash~\cite{DotHash}.}} Moreover, to further improve the deduplication accuracy and efficiency, DotHash~\cite{DotHash} assumes that uniformly sampled vectors in high-dimensional space are quasi-orthogonal. It encodes each sample into a combination of sample elements represented as fixed-length basis vectors, and the dot product of these vectors is an unbiased estimate of their intersection. For example, given two samples with their element basis vectors $a=\sum_{a\in A}\psi(a)\quad$ and $\quad b=\sum_{b\in B}\psi(b)$, the intersection is calculated by $\mathbb{E}[a\cdot b]=|A\cap B|$. 

However, \cite{SimiSketch} found that DotHash performs badly if the length of the basis vector is lower than the number of basis vectors, where quasi-orthogonal no longer holds. 

\hi{Approximate Frequency-based Down-Weighting.} To prevent the loss of potentially valuable information by retaining only one sample and removing the rest, SoftDeDup~\cite{he2024softdedup} deduplicates by reweighting samples, where samples with higher commonness are assigned lower sampling weights. Specifically, SoftDeDup computes the frequency of each n-gram across all the samples and calculates the commonness of each sample by multiplying the frequencies of all the n-grams that appear in the document. Samples with higher commonness are more likely to be duplicates and thus be down-weighted. 

\hi{Embedding-Based Clustering.} Except for samples with the same or similar substrings, some samples with similar semantics but different formats (i.e, expressed differently) may also negatively affect \llm training performance. For instance, for the following two sentences: $(i)$ {\it ``Unleash your potential with our lightweight, high-performance sports shoes – designed for comfort, speed, and style''}; $(ii)$ {\it ``Step into greatness with durable, breathable sports shoes perfect for running, training, and everyday adventures''}.  Both of the sentences are sports shoe advertisements but expressed differently, and such duplicates could degenerate model performance by making data imbalanced and introducing bias to the model. To address this issue, another approach leverages language models' embeddings (representing similar items as vectors close to each other in the vector space) for deduplication.

SemDeDup~\cite{abbas2023semdedup} identifies semantic duplicates by clustering embeddings and deduplicating those with high cosine similarities. It first encodes each sample into an embedding by leveraging the OPT \cite{OPT} text encoder and the CLIP \cite{CLIP, OpenCLIP} image encoder, and clusters the embeddings with K-means, so one can save time by finding duplicates within the cluster rather than the whole vector space. Then, within each cluster, it searches for semantic duplicates with cosine similarity above the pre-defined threshold. Finally, within each group of duplicates, it retains only the sample closest to the cluster centroid. As a multi-modal method, it can be applied to both text and image data, making it possible to deduplicate image data. In practice, \cite{abbas2024effective} leverages SemDeDup to deduplicate the image-text pair dataset LAION-400M \cite{LAION-400M}.

Like MinHash, SemDeDup also has many variants for performance improvement. \cite{tirumala2023d4} combines SemDeDup with the Self-Supervised Learning (SSL) Prototypes metric, which clusters the samples and retains the samples in each cluster based on their distance to their corresponding cluster centroid, where the samples closer to the centroid are more likely to be removed. FairDeDup~\cite{slyman2024fairdedup} modifies the logic of SemDeDup to improve the representation of underrepresented sensitive groups by prioritizing the retention of samples that align with sensitive concepts defined through user-provided prototypes, such as demographic subgroups. Within each cluster, instead of selecting the farthest sample from the centroid, it selects the sample that maximizes similarity to the least-represented group in the cluster to prevent samples with sensitive concepts from being pruned. 


\hi{Non-Text Data Deduplication.}  As \llms are increasingly applied to multimodal tasks (e.g., image-text retrieval, visual question answering), non-text data types such as images are becoming integral to \llm training datasets, necessitating dedicated deduplication techniques. Similar to texts, images can also be encoded into embeddings through neural networks designed for image-like data such as CNN, after which embedding-based deduplication methods can be applied. 
\textit{SemDedup}~\cite{abbas2023semdedup} adopts a semantic-based method by computing cosine similarity between image embeddings; two images are considered duplicates if their similarity exceeds a predefined threshold, which is tuned to balance detection precision and recall. In contrast, \textit{MINT-1T} employs a hash-based approach, using SHA256 checksums to identify and remove exact duplicates efficiently. Meanwhile, the \textit{DataComp} pipeline~\cite{DatacompMLLM} leverages the CNN-based near-duplicate detector~\cite{yokoo2021contrastive} to eliminate subtle duplicates and prevent evaluation set leakage. Models trained on these deduplicated image sets exhibit improved performance over baselines such as CLIP~\cite{CLIP} for higher precision and recall. 

\begin{table}[]
\caption{Data Filtering Methods for LLMs.}
\label{table:filtering}
\resizebox{\columnwidth}{!}{%
\begin{tabular}{cccc}
\hline
\textbf{Category} & \textbf{Objective} & \textbf{Methods} &  \\ \hline
\multirow{5}{*}{\begin{tabular}[c]{@{}c@{}}Sample-\\ level\\ Filtering\end{tabular}} & \multirow{5}{*}{\begin{tabular}[c]{@{}c@{}}Remove\\ low-quality\\ samples\end{tabular}} & Perplexity Measuring \cite{thrush2024improving,ankner2024perplexed,mekala2024smaller,li2023quantity,Superfiltering} &  \\
 &  & Influence Assessment \cite{lin2024data,he2024shed} &  \\
 &  & Clustering \cite{abbas2024effective,SmallToLarge} &  \\ \cline{3-4} 
 &  & Model Scoring \cite{wettig2024qurating,liu2023makes,shen2024seal} &  \\
 &  & Mixed Methods \cite{marion2023less,cao2023instruction,du2023mods} &  \\ \hline
\begin{tabular}[c]{@{}c@{}}Content-\\ level\\ Filtering\end{tabular} & \begin{tabular}[c]{@{}c@{}}Remove\\ partial-noising\\ samples\end{tabular} & \begin{tabular}[c]{@{}c@{}}Privacy Anonymization \cite{lukas2023analyzing,liu2023deid}\\ Image \& Video Filtering \cite{CogVideoX, Hunyuanvideo, Wan}\end{tabular} &  \\ \cline{1-3}
\end{tabular}%
}
\end{table}

\begin{figure}[!t]
    \centering
    \includegraphics[width=1\linewidth]{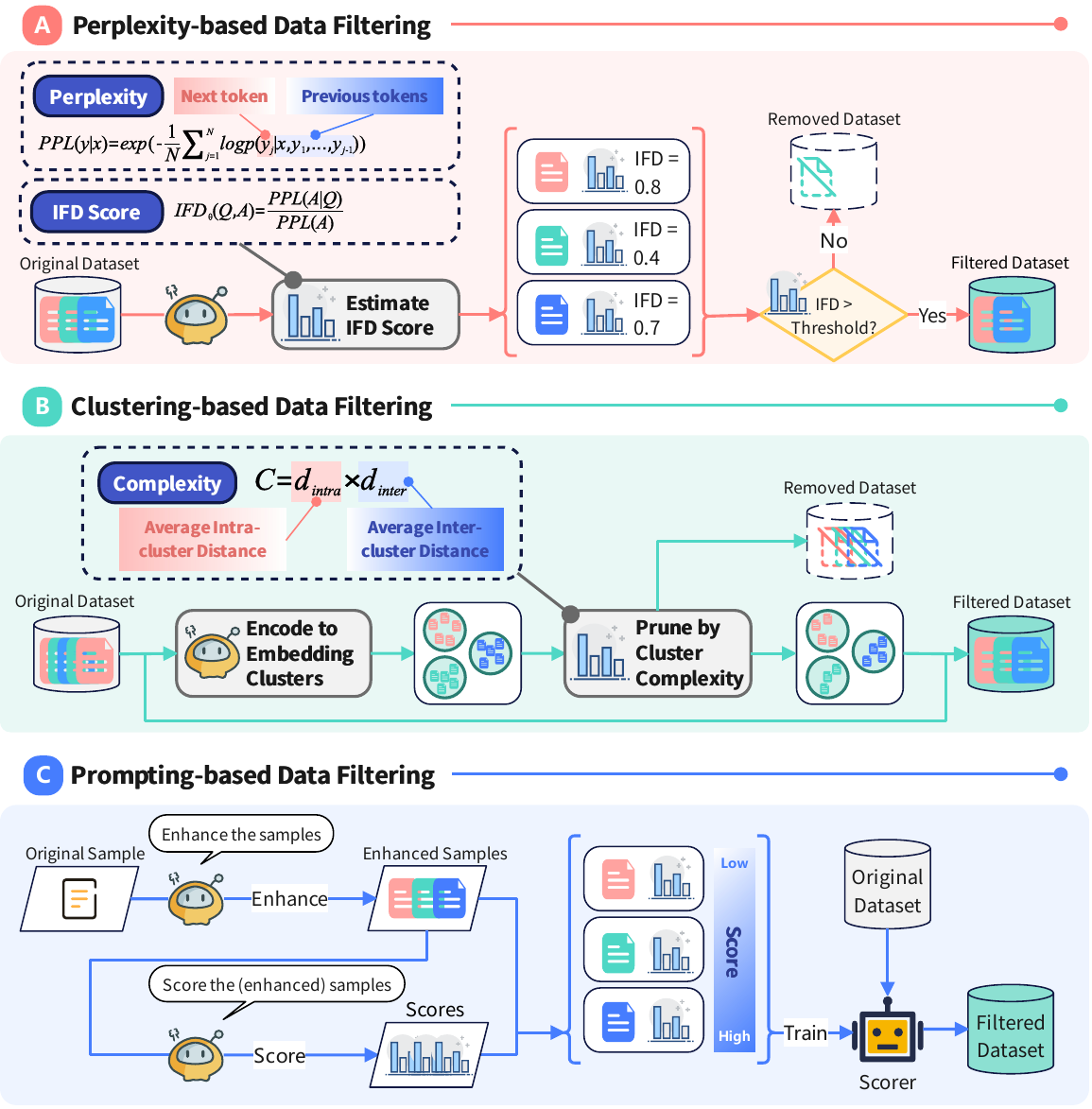}
    \caption{Example Data Filtering Workflows~\cite{Superfiltering, abbas2024effective,liu2023makes}.}
    \label{fig:data-filtering}
\end{figure}

\subsubsection{Data Filtering}
\label{subsubsec:filtering}

Data filtering removes low-quality or sensitive samples from the dataset to reduce computational overhead and protect privacy, while the model trained on the subset exhibits similar or even better performance than the one trained on the original dataset. To achieve this, one has to $(i)$ remove samples with low quality (\emph{Sample-level filtering}) or partial noisy information (\emph{Content-level filtering}), and $(ii)$ keep the selected samples diverse enough to cover various domains.

\hi{Sample-level Filtering} refers to evaluating samples using metrics or models and removing the samples that fail to meet the threshold (e.g., quality and diversity).
There are multiple metrics in this category:

\begin{tcolorbox}[colback=gray!1,colframe=gray,lowerbox=visible]
  \textbf{Principles}
  \tcblower
  Compared to classic ML data filtering, LLM data filtering emphasizes turning unstructured text into measurable metrics, with the main challenge being the effectiveness of evaluation methods, the standards of low-quality samples, and the computational complexity of these methods across massive datasets.
\end{tcolorbox}

\noindent\bfit{(1) Statistical Evaluation} uses various statistical methods to evaluate samples by directly applying statistical metrics to the samples (e.g., clustering results) or indirectly capturing characteristics from the models trained on the dataset (e.g., loss or perplexity from a surrogate model). Applicable statistical metrics include perplexity (and its variants), influence on model parameters, and clustering.



\noindent$\bullet$ \underline{Perplexity Measuring.} Perplexity measures the difficulty of a model generating the responses, represented as aggregated probabilities of the $j$-th response token given the question tokens and previous $j-1$ response tokens $\text{PPL}(y|x) = \exp\left(-\frac{1}{N}\sum_{j=1}^{N}\log p(y_{j}|x, y_1,...,y_{j-1})\right)$. The higher the perplexity value is, the harder the model generates the response. It is commonly used in selecting high-quality subsets in pre-training and fine-tuning phases. Based on the original perplexity, there have been several studies for improving the metric, including computing perplexities using a smaller-sized model for training a larger-sized model to reduce computational overhead, or employing advanced techniques such as Learning Percentage (LP) and Instruction-Following Difficulty (IFD) to identify and select challenging samples.

Specifically, \cite{thrush2024improving} uses an existing model to compute perplexity scores for multiple domains and selects pre-training samples from the domains with high correlation between the downstream benchmark error and the perplexity scores on the domain samples. The correlation is measured through a rank-based correlation coefficient $\gamma_j = \sum \text{sign}(y_k - y_l)(\text{rank}_j(x_{k,j}) - \text{rank}_j(x_{l,j}))$, where the rank difference reflects the model performance difference on the same sample, helpful in estimating $\theta^*$. They then rank the domains based on $\gamma_j$ and select samples from the top-ranked domains. To scale the process, a fastText classifier~\cite{FastText} is trained to distinguish selected documents, enabling page-level data selection.

To enhance efficiency, \cite{ankner2024perplexed} leverages a smaller-sized surrogate model to select high-quality pre-training subsets via perplexity score for training larger-sized models, greatly reducing the computational overhead in model training while still achieving the same performance as with the full dataset. They first train a surrogate model, a smaller-sized MosaicML \cite{MosaicML} model with 125 million parameters, on a random subset of the pre-training dataset to compute the perplexity scores for the remaining samples. Based on the perplexity scores, they find the optimal subset through a combination of selection criteria: $(i)$ the part of samples to keep (e.g., samples with low/medium/high perplexity scores), and $(ii)$ the fraction of samples to keep (e.g., 25\%, 50\%, 75\%). The subset is evaluated by training a larger-sized MosaicML model on it and analyzing the model's performance on downstream benchmarks. While the result shows that the smaller-sized model can effectively and efficiently filter data for the larger-sized model, they also find that the effectiveness highly depends on the dataset. For example, keeping the high perplexity samples exhibits better performance on the Pile dataset \cite{Pile}, while keeping the medium perplexity samples exhibits better performance on the Dolma dataset \cite{Dolma}.

Furthermore, there are some variants of perplexity-based evaluation. First, \cite{mekala2024smaller} proposes a perplexity-based metric, Learning Percentage (LP), to select samples that are more challenging for models to learn. Learning Percentage $\mathcal{LP}(i) = \frac{\mathcal{P}_{i-1} - \mathcal{P}_{i}}{\mathcal{P}_{0} - \mathcal{P}_{n}}$ measures the perplexity drop ratio of a sample between the specific epoch $i$ and the whole training procedure. The key idea is that models tend to learn easier samples first and harder samples later, so one can find harder samples that are not thoroughly learned during early epochs. The authors use $\mathcal{LP}(1)$ (the learning percentage after the first epoch) to rank the training samples from the hardest to the easiest and split them into three equal-sized parts. It shows that the smaller-sized variant of the model can effectively select samples for the larger-sized variant, and models of all sizes trained on the harder part outperform the ones trained on all the samples. 

Also based on perplexity, \cite{li2023quantity} proposes the Instruction-Following Difficulty (IFD) metric to select samples that are more difficult for models to follow. IFD ($\text{IFD}_{\theta}(Q, A) = \frac{PPL(A|Q)}{PPL(A)}$) measures the influence of the questions (instructions and inputs combined) on generating corresponding responses by comparing the perplexity of the response with or without the question strings $PPL(A|Q)$ and $PPL(A)$. A higher IFD score suggests higher model following difficulty. The authors first build a pre-experienced subset by clustering and resampling the samples from the WizardLM \cite{xu2023wizardlm} and Alpaca-GPT4 \cite{peng2023instruction} datasets, on which they train the model for one epoch to obtain initial knowledge. The model is then used to calculate the IFD score on all the samples, and the ones with high IFD scores are prioritized.

Superfiltering \cite{Superfiltering} further enhances \cite{li2023quantity} by employing the surrogate model from \cite{ankner2024perplexed}. Instead of training a smaller-sized model, the authors directly use GPT-2 \cite{radford2019language} as the surrogate model to calculate IFD scores on the same datasets. Compared to their previous work \cite{li2023quantity}, the adoption of surrogate model simplifies the procedure and accelerates the filtering process.

\noindent$\bullet$ \underline{Influence Assessment.} Another data filtering approach is to assess the influence of a sample on \llm model performance or learning process by measuring how the metrics change \emph{when the sample is upweighted or removed}. The samples with substantial impact on the model parameters are regarded as influential and thus are selected.

DEALRec \cite{lin2024data} identifies influential and challenging fine-tuning samples through two metrics: $(i)$ \textit{Influence Score} for assessing the influence of a specific sample on the model performance. It starts by measuring the influence on parameter change, where a surrogate model is trained on the full dataset to estimate how the model parameters would change when certain sample is removed or upweighted, expressed by $\hat{\theta}_{-s}-\hat{\theta}\approx\frac{1}{n}H_{\hat{\theta}}^{-1}\nabla_\theta\mathcal{L}(s,\hat{\theta})$, where $H_{\hat{\theta}}$ is the Hessian matrix and $\nabla_{\theta} \mathcal{L}(s, \hat{\theta})$ is the loss gradient of sample $s$. The formula is then evolved to measure the influence on empirical risk change, expressed by $I_{\text{remove, loss}}(s, \mathcal{D}) = \frac{1}{n}\sum_i\frac{1}{n}\nabla_\theta\mathcal{L}(s_i,\hat{\theta})^\mathrm{T}H_{\hat{\theta}}^{-1}\nabla_\theta\mathcal{L}(s,\hat{\theta})$; $(ii)$ \textit{Effort Score} for assessing the difficulty for the surrogate model to learn a specific sample for generalization to new samples, defined as $\delta_s = \| \nabla_{\phi} \mathcal{L}^{\text{LLM}}(s) \|_2,$ where $\Phi$ is the model parameter. A higher effort score suggests greater difficulty. The final score combines the above two scores, written as $I_s = \text{Influence Score} + \lambda \cdot \text{Effort Score}$.

Besides, SHED~\cite{he2024shed} utilizes the Shapley value \cite{Shapley}, which estimates the contribution of a member to the group, to calculate the influence of a sample on the model performance and select representative samples with high influence. The method first clusters the samples and selects the ones closest to each cluster centroid as the representative samples to reduce computational overhead. It then calculates the Shapley value for each representative sample $i$ by iteratively removing $n$ samples from the dataset until all the samples have been removed and calculating the contribution of the removed $n$ samples in each iteration $a$ to the model performance compared with the previous iteration, written as: $c_{(an+1..(a+1)n)\in D_p}=v(D_p\setminus\{1..an\})-v(D_p\setminus\{1..(a+1)n\})$. The process will be repeated for $k$ times for higher accuracy, after which the Shapley value for each representative sample $i$ is defined as $S_i \approx \frac{1}{k} \sum_{k} \frac{c_i(k)}{n}$. Finally, the subsets can be selected either by selecting from the top-rank samples or weighted sampling the samples through $\Pr(i) = \frac{e^{fS_i}}{\sum_{i} e^{fS_i}}$, where $f$ controls the trade-off between quality and diversity.

\noindent$\bullet$ \underline{Clustering.} A common approach to select high-quality and diverse subsets is to encode the samples into embeddings in the latest space and cluster them using cosine similarity, where similar samples are usually clustered into the same group. Selecting within the clusters reduces redundancy, while selecting across the clusters increases diversity.

Density-Based Pruning (DBP) \cite{abbas2024effective} selects high-quality and diverse subsets by clustering samples into clusters and resampling the samples based on the cluster complexity. They encode the samples into embeddings using a pre-trained vision model DINOV2-L/14 \cite{Dinov2} and cluster them using K-means. For each cluster, they calculate the average intra-cluster cosine-distance to the internal centroid $d_{intra}$ and inter-cluster cosine distance to the other centroids $d_{inter}$, and the cluster complexity as a product of the two distances $C = d_{intra} \times d_{inter}$. The cluster complexity is later converted to probability using softmax to resample the samples across clusters, where clusters with higher complexity have higher weights.

Rather than the sample embedding itself, SmallToLarge~\cite{SmallToLarge} selects a diverse subset by clustering the samples based on their loss trajectories. It first trains a smaller-sized surrogate \llm model on the whole dataset to obtain the loss trajectories of each training sample, defined as $\mathcal{L}_{i}(\bm{\phi}^{(t)}) = -\log p_{\bm{\phi}^{(t)}}(\mathbf{y}_{i}|\mathbf{x}_{i})$, where $\bm{\phi}^{(t)}$ is the model parameters at time $t$. These samples are then clustered based on loss trajectories and randomly resampled to form a diverse subset.

\noindent\bfit{(2) Model Scoring} uses \llms for evaluating sample quality. The quality criteria can either be specified $(i)$ explicitly via \llm prompt engineering or $(ii)$ implicitly learned from human-labeled data.

QuRating \cite{wettig2024qurating} selects high-quality pre-training samples by prompting \llm to compare pairs of samples along the four quality criteria (writing style, fact \& trivia amount, educational value, and the expertise required to understand), training a rater on the scalar quality ratings, and filtering samples using the rater. Initially, GPT-3.5-turbo is prompted on each pair of samples to judge which one is better on each quality criterion, where the binary confidence $p_{B \succ A} \in [0, 1]$ that the sample B is preferred over the sample A is recorded. The pairwise binary confidence is then translated into sample quality ratings $p_{B \succ A} = \sigma(s_B - s_A)$ through the Bradley-Terry model. A QuRater model is later trained on these quality ratings to predict quality ratings for new samples on each criterion. The new samples are resampled with the probability $p(d_i) \propto \exp\left(\frac{s_i}{\tau}\right)$, where $\tau$ adjusts the trade-off between quality and diversity.

Rather than prompting the models to compare samples, Data-Efficient Instruction Tuning for Alignment (DEITA) \cite{liu2023makes} prompts \llm models to evolve and score the samples for building sample scorers. The authors first prompt ChatGPT to evolve the samples along instruction complexity and response quality, and again prompt ChatGPT to score these evolved samples. They then train scorers on the evolved samples with their corresponding scores to enable their scoring abilities. Finally, they use these scorers to score new samples and multiply the scores to form the final score, where the new samples are resampled based on the final scores for diversity. 

Model scoring methods also help mitigate bias and toxicity. \llm often exhibit harmful biases due to the massive and unchecked datasets they are trained on, which can have various biases, ranging from gender and racial stereotypes to cultural and socioeconomic prejudices~\cite{navigli2023biases}.
Safety-enhanced Aligned LLM Fine-tuning (SEAL) \cite{shen2024seal} selects high-quality and safe fine-tuning samples through a safety-aligned selector. The selector is trained based on a safety-aligned model, Merlinite-7b \cite{Merlinite-7b}, using bi-level optimization, which minimizes the safety loss on the safe dataset while minimizing the fine-tuning loss on the filtered dataset during training to ensure the selector always prioritizes safe and high-quality samples during selection. After the selection, the top-p\% samples will be selected.

\noindent\bfit{(3) Hybrid Methods.} Instead of relying on a single method, some methods mix various kinds of data filtering methods and evaluate each permutation of these methods or parameters to find the best combination of methods or parameters that further boosts model performance.

\cite{marion2023less} selects high-quality pre-training data based on three metrics: $(i)$ Perplexity, $(ii)$ EL2N $\chi(x_i, y_i) = \mathbb{E} \| f(x_i) - y_i \|_2$ for measuring the prediction probability discrepancy between the reference model and the ground truth, and $(iii)$ Memorization factor $score(M, N) = \frac{1}{N} \sum^N_i 1(z_{M+i} = \hat{z}_{M+i})$ for measuring the fraction of N tokens correctly generated after prompting the model with the first M tokens~\cite{biderman2024emergent}. 
For each metric, they retain samples based on two criteria: $(i)$ the fraction of samples to keep (10\%, 30\%, 50\%, and 70\%) and $(ii)$ the part of samples to keep, e.g., the bottom (for Perplexity and L2-Norm Error) and top (for Memorization). They train \llm for each case and select the best-performing one, and the result shows that {\it Perplexity} effectively removes the ``easiest'' samples, improving model performance and outperforming other metrics.

Instead of comparing metrics and choosing the best of them, InstructionMining \cite{cao2023instruction} combines various metrics (e.g., including input/output length, reward score, perplexity, etc.) into one linear function with each metric as indicator, written as $log L_{loss} \propto L_{0}+\beta_{0}+\beta_{1}I_{1}+\beta_{2}I_{2}+\cdots+\beta_{n}I_{n}+\epsilon$. The $\beta$ parameters are estimated using least squares. In practice, it evaluates fine-tuning samples on a fine-tuned model LLaMA-2-7B \cite{Llama-2} and selects samples by finding the optimal set of samples to keep using the hyperparameter optimizer BlendSearch \cite{BlendSearch}.

MoDS~\cite{du2023mods} considers diversity into selection and iteratively selects high-quality, diverse, and necessary subsets and adds the samples the \llm model performs poorly on during fine-tuning using a reward model and the K-Center greedy algorithm \cite{K-Center}. The method is conducted mainly in three steps: $(i)$ Use a reward model to score the quality of each (instruction, input, output) triplet in the dataset, where the low-quality ones are filtered out, forming a high-quality dataset. $(ii)$ Use the K-Center greedy algorithm \cite{K-Center} to select the samples in the high-quality dataset that are farthest apart from each other in the BERT \cite{bert} embedding space, forming a diverse seed dataset. $(iii)$ Fine-tune a pre-trained \llm model on the seed dataset to enable its instruction-following ability and generate responses for the high-quality dataset. The generated responses are evaluated using the same reward model, and those with low quality scores, which means the model is weak at generating such responses, will be collected. The collected samples with their original responses will be selected again using the K-Center greedy algorithm and then added to the seed dataset, forming the final dataset.



\hi{Content-level Filtering.} To avoid removing too many critical samples from the dataset and weakening the model performance, some works only filter out noise or sensitive content within the samples. For noise removal, common methodologies include removing or replacing specific characters (e.g., remove invisible or invalid characters, unescape HTML characters and detect punctuation misuse), removing unnecessary texts (e.g., the texts that appear as decorating elements on the web pages such as ``print'', ``likes'' and ``loading'' ), and cleaning harmful information (e.g., spam, gambling, pornographic content and site links)~\cite{yang2023baichuan}. 



For privacy anonymization, \llms can memorize private and sensitive information (e.g, user identity details or clinical health data) from datasets during pre-training and fine-tuning, which can be leaked through specially crafted prompts, thereby posing significant privacy risks. \cite{lukas2023analyzing} demonstrates that it is possible to extract, reconstruct, and infer personally identifiable information (PII) from \llm models by identifying the most frequent PII appearing in model responses or by prompting models with partial information about a specific individual. From a data management perspective, these privacy threats can be mitigated by identifying and filtering out potential sensitive information in the datasets.

DeID-GPT~\cite{liu2023deid} utilizes existing \llms to identify and remove PII from unstructured medical text without changing its meaning. In their case, the \llms are prompted to de-identify information from clinical notes in accordance with HIPAA privacy regulations. An example prompt is: \textit{``Please de-identify the following clinical notes by replacing any terms that could be a name, an address, a date, or an ID with the term `[redacted]'.''}

Instead of using general \llms, \cite{lukas2023analyzing} uses Named Entity Recognition (NER) models such as spaCy~\cite{spaCy} and Flair \cite{akbik2019flair} to tag PII in the samples and removes or replaces them with hashed tags, entity tags like ``[NAME]'' or ``[LOCATION]'', or a simple tag like ``[MASK]''. The last tag was adopted to maximize privacy, as the other ones are still vulnerable to membership inference by linking the samples.

The rise of multi-modal LLMs, particularly large video generation models, drives the need for robust video data filtering. CogVideoX \cite{CogVideoX} employs a pipeline focusing on coherent motion, removing videos with poor dynamics. It defines negative labels for artificial edits, low motion connectivity, visual flaws, and excessive text. A manually annotated subset trains six Video-LLaMA\cite{Video-LLaMA}-based filters, while optical flow and aesthetic scores ensure motion coherence and visual appeal, refining the dataset to approximately 35M high-quality 6-second clips.  

HunyuanVideo \cite{Hunyuanvideo} uses a multi-step pipeline: splitting videos into clips, encoding embeddings, deduplication, and resampling. Filters include motion (OpenCV-based optical flow), OCR (text removal), clarity (visual blur detection), aesthetic (Dover\cite{Dover}-based scoring), and source (YOLOX\cite{Yolox}-like watermark/border removal). This process generates five progressive training sets with increasing thresholds.  

Wan \cite{Wan} applies pre- and post-processing pipelines. Pre-processing filters unsuitable data using OCR, aesthetic evaluation (LAION-5B~\cite{LAION-400M}), NSFW scoring, watermark detection, and resolution thresholds, removing approximately 50\% of low-quality data. Samples are clustered for diversity, manually scored, and an expert model selects high-quality, naturally distributed data. Videos are classified into six tiers, prioritizing smooth motion. Post-processing refines images by selecting top 20\% via an expert model and manually curating gaps. For videos, top candidates are filtered by visual quality and motion complexity, ensuring balance and diversity across 12 themes.

\subsubsection{Data Selection}

\begin{table}[]
\caption{Comparison of Different Data Selection Methods.}
\label{tab:data_selection_comparison}
\resizebox{\columnwidth}{!}{%
\begin{tabular}{ccc}
\hline
\textbf{Method} & \textbf{Stage} & \textbf{Evaluation Metric} \\ \hline
Similarity & \begin{tabular}[c]{@{}c@{}}Pre-training,\\ Fine-tuning\end{tabular} & \begin{tabular}[c]{@{}c@{}}Cosine Similarity \cite{xie2024efficient}\\ Bag-of-Words Similarity \cite{xie2023data}\\ Lexicon Set Overlap \cite{qin2024enabling}\\ Bayes-based Selection \cite{brandfonbrener2024color}\end{tabular} \\ \hline
Optimization & Fine-tuning & \begin{tabular}[c]{@{}c@{}}Linear Search \cite{Dsdm}\\ Gradient-Influence Search \cite{LESS}\\ Kernel-Density Regularization \cite{liu2024tsds}\end{tabular} \\ \hline
Model & Pre-training & Logits-based LM-Score \cite{zhang2024autonomous} \\ \hline
\end{tabular}%
}
\end{table}

Different from previous reviews~\cite{dataselectionsurvey,wang2024survey}, we define data selection as the process of choosing subsets of already well-cleaned data samples in order to adapt \llms to specific domains (e.g., medical or legal \llms).



\begin{tcolorbox}[colback=gray!1,colframe=gray,lowerbox=visible]
  \textbf{Principles}
  \tcblower
  Unlike traditional ML data selection, LLM data selection focuses on aligning the topics of the text samples, requiring encoding semantic topics into measurable distributions. However, managing computational efficiency and ensuring robust generalization across diverse tasks remain critical unresolved issues. 
\end{tcolorbox}

\hi{Similarity-based Data Selection.} One class of methods aims to select subsets similar to the specified target data. 

\noindent$\bullet$ \underline{\emph{Cosine Similarity:}} Domain-Adaptive Continual Pre-training (DACP)~\cite{xie2024efficient} adapts a general-purpose LLM to a target task by selecting domain-specific unlabeled data based on similarity (cosine similarity), novelty (perplexity), and diversity (entropy). For the similarity part, it identifies data most similar to the task-specific labeled data by encoding both into embeddings (using \cite{spaCy}) and choosing domain samples that align with the task's embedding distribution.



\noindent$\bullet$ \underline{\emph{Bag-of-Words Similarity:}} DSIR~\cite{xie2023data} selects a subset of unlabeled pre-training data matching the target distribution by computing feature distributions ($\hat{p}_{\text{feat}}$, $\hat{q}_{\text{feat}}$) for raw and target data represented as bag-of-words, estimating importance weights $w_{i}=\frac{\hat{p}_{\text{feat}}(z_{i})}{\hat{q}_{\text{feat}}(z_{i})}$, and resampling raw data with probability $\frac{w_i}{\sum^N_{i=1}w_i}$.



\noindent$\bullet$ \underline{\emph{Lexicon Set Overlap:}} \cite{qin2024enabling} selects the subset with the most shared lexicons using the Domain Specific Score (DSS), which quantifies the relevance of a dialogue set \(T\) to specific domains by measuring the overlap between \(T\) and domain lexicons \(L = \{l_1, l_2, \dots, l_m\}\), calculated as \(\text{DSS}(T, L) = \frac{1}{m} \sum_{i=1}^{m} \frac{|T \cap l_i|}{n}\), where \(n\) is the number of tokens in \(T\).



\noindent$\bullet$ \underline{\emph{Bayes-based Selection:}} CoLoR-filter \cite{brandfonbrener2024color} formulates pre-training subset selection as a Bayesian optimization problem, which selects a subset $S$ by maximizing downstream likelihood $\text{Pr}(D_{\text{down}}|S)$. It uses two auxiliary models:  A ``prior'' model ($\theta_{\text{prior}}$) trained on a large general dataset $D_{\text{down}}$ and a ``conditional'' model ($\theta_{\text{prior}}$) fine-tuned on the union of the large general dataset and a small downstream dataset $D_{\text{prior+down}}$. The selection criterion for a data point $x_i$ is the conditional loss reduction (CoLoR): $\text{CoLoR}(x_i) = -\log \Pr(x_i|\theta_{\text{prior+down}}) - (-\log \Pr(x_i|\theta_{\text{prior}}))$. The key idea is to score samples based on the likelihood difference between these two models and select the ones that exhibit higher likelihood under the conditional model and larger conditional loss reduction. 

\hi{Optimization-based Data Selection.} Optimization-based data selection methods select subsets towards reducing model loss and improving model performance on the target tasks.

\noindent$\bullet$ \underline{\emph{Linear Search.}} Model-Aware Dataset Selection with Datamodels (DsDm)~\cite{Dsdm} selects the optimal subset of training data that minimizes the model's loss on target tasks by employing linear datamodel \cite{Datamodels}, a parameterized function that maps a subset of training data to the model outputs for the specified target, to estimate how the inclusion of each training sample would affect the model's loss on the target, reducing computational overhead. In practice, a linear datamodel $\tau_{\theta_x}(1_S) = \theta_x^\top 1_S$ with parameters $\theta_x$ and a characteristic vector $1_S$ (a binary vector indicating which samples are in $S$) is adopted to map the subset $S$ to the model loss on a sample $x$ through $L_x(S) = \mathbb{E}[\ell(x; A(S))]$. For each target, the characteristic vector $1_S$ is adjusted to reflect the subset, and the parameters $\theta_x$ are estimated using a regression loss function like mean squared error over the training subset. After training, the datamodel selects the subset $S$ of the size $k$ that minimizes the loss $\hat{L}_{D_{\text{targ}}}(S) = \frac{1}{n} \sum_{i=1}^n \tau_{\theta_{x_i}}(1_S)$ for the target task.

\noindent$\bullet$ \underline{\emph{Gradient-Influence Search.}} Low-rank Gradient Similarity Search (LESS) \cite{LESS} identifies the most impactful subset of data for fine-tuning \llms by analyzing gradient similarities. It first fine-tunes the model on a random subset (e.g., 5\% of data) for a few epochs using LoRA to reduce trainable parameters and accelerate gradient computation, and saves the checkpoints after each epoch. Next, LESS computes Adam LoRA gradients for each training sample, projects them into lower-dimensional gradient features via random projection, and stores them in a gradient datastore. For downstream tasks, it calculates gradient features of few-shot validation samples and estimates the influence of each training sample $\bm{z}$ on a validation sample $\bm{z}'$ using cosine similarity: $\text{Inf}_{\text{Adam}}(\bm{z}, \bm{z}') \triangleq \sum_{i=1}^{N} \bar{\eta}_{i} \cos(\nabla \ell(\bm{z}'; \bm{\theta}_{i}), \Gamma(\bm{z}, \bm{\theta}_{i}))$, where $\Gamma(\bm{z}, \bm{\theta})$ is the Adam update. The training samples with the highest influence scores are selected for fine-tuning.

\noindent\(\bullet\) \underline{\emph{Kernel-Density Regularization.}} Task-Specific Data Selection (TSDS) \cite{liu2024tsds} identifies high-quality pre-training or fine-tuning data for particular tasks by balancing two objectives: $(i)$ distribution alignment with the target task data and $(ii)$ diversity to avoid near-duplicates, accomplished via kernel density estimation (KDE) regularization. Concretely, one begins with a small set of target task samples \(Q = \{q_i\}_{i=1}^M\) and a large candidate pool \(D = \{x_j\}_{j=1}^N\), both of which are embedded into a shared metric space (e.g., using gradient-based or semantic embeddings). The optimization for distribution alignment is conducted by solving for probability mass $\gamma_{ij}$ (transported from \(q_i\) to \(x_j\)): $\min_{\gamma \in \mathbb{R}_{\geq 0}^{M \times N}} \frac{\alpha}{C} \sum_{i=1}^M \sum_{j=1}^N \gamma_{ij} d_{ij} + (1-\alpha) G_{\text{KDE}}(\gamma) \quad \text{s.t.} \quad \sum_{j=1}^N \gamma_{ij} = \frac{1}{M}, \forall i \in [M]$, where $d_{ij}$ is the distance between $q_i$ and $x_j$ in the metric space, and $G_{\text{KDE}}(\gamma)$ is the regularization term that adds diversity and penalizes over-density using KDE: $G_{\text{KDE}}(\gamma) = M \max_{i,j} \rho_j \left| \gamma_{ij} - \frac{1/\rho_j}{M \sum_{j'} 1/\rho_{j'}} \right|$, where $\rho_j = \sum_{x' \in D} (1 - f(x_j, x')^2/h^2$ is the density estimate for candidate $x_j$ (higher for near-duplicates). Afterwards, it samples $x_j$ with probability $p_j = \sum_i \gamma_{ij}^*$.

\hi{Model-based Data Selection.} 
These methods aim to determine subsets guided by prompting the LLM itself. 

Autonomous Data Selection (AutoDS)~\cite{zhang2024autonomous} prompts the \llm to assess and select mathematical and educational samples from a larger dataset. For each sample, the \llm is asked two questions: $(i)$ Is it mathematically relevant, and $(ii)$ It it educationally valuable. The \llm responds to each question with ``Yes'' or ``No'', and the logit of each response is extracted to compute the LM-Score: $\text{LM-Score}(\cdot) = \frac{\exp(\text{logit}(\text{`YES'}))}{\exp(\text{logit}(\text{`YES'})) + \exp(\text{logit}(\text{`NO'}))}$, and the composite score: $\text{LM-Score}(Q_1, Q_2) = \text{LM-Score}(Q_1) \cdot \text{LM-Score}(Q_2)$. The composite score ranks and selects high-quality math samples. 






\subsubsection{Data Mixing}

\begin{table*}[ht]
\centering
\caption{Comparison of Data Mixing Methods for \llms.}
\label{tab:Mixing}
\resizebox{\linewidth}{!}{%
\begin{tabular}{cccc}
\hline
\textbf{Taxonomy} &
  \textbf{Stage} &
  \textbf{Methods} &
  \textbf{Traits} \\ \hline
\multirow{3}{*}{\begin{tabular}[c]{@{}c@{}}Before Training\\ (Human Experience)\end{tabular}} &
  \multirow{3}{*}{Pre-training} &
  Multi-Source Data Adjusting &
  \multirow{2}{*}{Intuitive and easy to implement, suitable for rapid experimentation.} \\
 &
   &
  ~\cite{feng2024maximize}, \cite{shen2023slimpajama} &
   \\
 &
   &
  Entropy-Based Mixing~\cite{ge2025bimixbivariatedatamixing} &
  Low computation cost with quality quantification by entropy. \\ \hline
\multirow{15}{*}{\begin{tabular}[c]{@{}c@{}}Before Training\\ (Model-Based Optimization)\end{tabular}} &
  \multirow{2}{*}{Pre-training} &
  \multirow{2}{*}{Linear Regression Model~\cite{liu2024regmix}} &
  Only 10\% of DoReMi's~\cite{xie2023doremi} computational resources are required. \\
 &
   &
   &
  Simultaneously train hundreds of small models to accelerate optimization. \\
 &
   &
   &
   \\
 &
  \multirow{2}{*}{Pre-training} &
  \multirow{2}{*}{Bivariate Data Mixing Law~\cite{ge2025bimixbivariatedatamixing}} &
  Avoid iterative training of proxy models (low computational costs). \\
 &
   &
   &
  Show relation between loss and training steps. \\
 &
   &
   &
   \\
 &
  Continual Pre-training &
  Chinchilla Scaling Law~\cite{que2024d} &
  Support knowledge transferring to new domains ($\downarrow$ over 95\% training costs). \\
 &
   &
   &
   \\
 &
  Pre-training &
  Exponential Functions~\cite{ye2024data} &
  Support datasets without explicit domain division. \\
 &
   &
   &
   \\
 &
  \multirow{3}{*}{Continual Pre-training} &
  \multirow{3}{*}{Power-law Function~\cite{gu2024cmr}} &
  Compared to single-objective optimization like \cite{que2024d} \\
 &
   &
   &
  \cite{gu2024cmr} ensures that domain performance improvement \\
 &
   &
   &
  does not compromise general capabilities. \\
 &
   &
   &
   \\
 &
  Pre-training &
  Classification Model~\cite{liang2024data} &
  Reverse engineering for finding the suitable data recipe of \llms. \\ \hline
\multirow{5}{*}{\begin{tabular}[c]{@{}c@{}}During Training\\ (Bilevel Optimization)\end{tabular}} &
  \multirow{2}{*}{Pre-training} &
  Calculate domain contribution by &
  \multirow{2}{*}{Requires a proxy model, performances well in OOD datasets.} \\
 &
   &
  gradient inner products\cite{fan2023doge} &
   \\
 &
   &
   &
   \\
 &
  \multirow{2}{*}{Fine-tuning} &
  Dynamically adjust weights by &
  Multiple applications like multilingual training, \\
 &
   &
  gradient alignment values~\cite{pan2024scalebio} &
  instruction following, large-scale data reweighting \\ \hline
\multirow{3}{*}{\begin{tabular}[c]{@{}c@{}}During Training\\ (Distributionally Robust Optimization)\end{tabular}} &
  Pre-training &
  Group DRO~\cite{xie2023doremi} &
  For pre-training, smooth adjusting to prevent abrupt weight changes \\
 &
   &
   &
   \\
 &
  Fine-tuning &
  Task-level DRO~\cite{ma2024task} &
  For fine tuning, quick response to task difficulty changes \\ \hline
\end{tabular}%
}
\end{table*}

Since \llms rely on massive and diverse datasets, the composition of these datasets significantly impacts model performance~\cite{na2024scalable}. For instance, as shown in Figure~\ref{fig:data-piechart}, we can see \llms require different ratios of domain data to achieve capabilities such as {medical diagnosis, coding, and solving math problems}. To this end, data mixing refers to the strategy of (1) combining datasets from different domains, sources or structures in specific proportions to train \llms or (2) making \llms give different proportions of attention on different domains (e.g., by changing the sampling probabilities) in the training session. Effective data mixing ensures that the model captures broad generalization capabilities while balancing performance across tasks and domains~\cite{feng2024mixture}. 
Existing data mixing methods can be classified into two main categories:

\begin{tcolorbox}[colback=gray!1,colframe=gray,lowerbox=visible]
  \textbf{Principles}
  \tcblower

{Unlike traditional ML models like BERT (trained on smaller, domain-specific data with homogeneous distributions), \llms require massive multilingual or multi-domain corpora, raising the critical challenge of optimizing dataset mixing ratios for performance. Current methods use heuristic experimentation or formulate ratio-performance relationships (e.g., validation loss), but cost-effective determination of optimal ratios, beyond heuristics, remains unresolved due to high cost demands for functional approximations.}
\end{tcolorbox}


\hi{Before-Training Mixing (Human Experience).} This method provides empirical data mixing strategies such as setting different ratios of datasets based on various factors (e.g., complexity and diversity of the datasets) that likely improve \llms' abilities. 

First, to study the effect of data mixture, there are works that experiment heuristically on different data ratios for pre-training of \llms. {\cite{feng2024maximize} suspects training sequence from simple to complex data would improve \llms' performance, thus} introduces a two-stage data mixing strategy for \llm pre-training: (1) It first blends web-crawled data with minimal high-quality content (1.9\% math, 15\% code), testing ratios (\(<\)35\% high-quality) and selecting optimal mixtures via evaluations on CommonsenseQA~\cite{talmor2018commonsenseqa} and HumanEval~\cite{humaneval}.  (2) It then filters low-quality data, boosting math (24\%→29\%), code (20\%→29\%), and instructional alignment data. Ratios are similarly optimized through empirical validation.
The method iteratively refines proportions using down-sampled Megatron-8B~\cite{shoeybi2019megatron} for efficiency, then scales findings to a 25B model, balancing diversity-quality tradeoffs with reduced experimental overhead.  Similarly, Slimpajama~\cite{shen2023slimpajama} explores the impact of data source diversity and weight distribution on model performance by adjusting the proportions of data from multiple sources, such as Commoncrawl~\cite{commoncrawl}, C4~\cite{raffel2020exploring}, Github~\cite{github} .

{Second, we can utilize metrics to judge different datasets and mix them.} To calculate the best result rather than just try different combinations, Bimix ~\cite{ge2025bimixbivariatedatamixing} adopts entropy metrics (e.g., Shannon entropy~\cite{shannon1948mathematical}, conditional entropy~\cite{shannon1948mathematical}) as the quality scores which are then normalized to compute the proportions of each domain (e.g., conditional entropy, written as as $H_i\left(X_i^{(t+1)} \mid X_i^{(t)}\right) = -\sum_{x \in X_i^{(t)}} \sum_{x' \in X_i^{(t+1)}} P(x, x') \log P(x' \mid x)
$, where $X_i^{(t+1)}$ $X_i^{(t)}$ are sets of tokens at positions $t+1$ and $t$ separately, $x$ and $x'$ are tokens belonging to them, $P(x, x')$ is the joint probability, $P(x' \mid x)$ is the conditional probability.

\hi{Before-Training Mixing (Model-Based Optimization).} This category of methods design linear or non-linear models that depict $(i)$ the relation between the distribution of each domain, $(ii)$ validation loss, and $(iii)$ some other variables like training steps, based on which they find the optimal settings through various model-based techniques. 

\noindent\bfit{(1) Linear Regression Model:} Some methods utilize pairs like data mixtures and corresponding model performance to fit a linear regressing model, such that finding the best data mixture ratios.

Typically, REGMIX~\cite{liu2024regmix} defines the domains by source (like ArXiv, FreeLaw, etc.), which uses Dirichlet distribution (which
controls the distribution of probabilities across multiple categories with a parameter) to generate all kinds of data distribution of several domains to train a small-scale proxy model to collect performance data, which is then used to fit a linear regression model (LightGBM~\cite{ke2017lightgbm}) to predict the optimal data mixing distribution. Then REGMIX uses both the best distribution and the average of top-100 distributions to verify on variations of TinyLlama~\cite{zhang2024tinyllama} with additional layers with versions of 1B and 7B.
    
\noindent\bfit{(2) Non-linear Regression Model:} There are also many methods that design non-linear regression models for data mixing {by considering more complex training characters.}


\noindent$\bullet$ \emph{Bivariate Data Mixing Law.} Based on observations of validation loss changes due to variables like domain proportion {(where the data come from different sources like Pile-CC)} and training steps, BiMix~\cite{ge2025bimixbivariatedatamixing} proposes Bivariate Data Mixing Law that depicts the relation among domain's proportion, training steps and validation loss, which can be written as $L_{i}\left(r_{i}, s\right)=\frac{A_{i}}{r_{i}^{\alpha_{i}}}\left(\frac{B_{i}}{s^{\beta_{i}}}+C_{i}\right)$, where $A_{i}$,$B_{i}$,$C_{i}$ are domain-dependent scaling coefficients, $\alpha_{i}$ and $\beta_{i}$ are power-law exponents that control the influence of domain proportion and training steps respectively, $s$ represents the training step count. It utilizes the law to fit the actual data curves by fixing the domain's proportion or training steps and varies the other one to get validation loss by training a small model (decoder-only transformers based on the DoReMi~\cite{xie2023doremi} architecture with 280M parameters). After depicting the relation, we model the task as an optimization problem (resolvable by Lagrange multipliers) and then verify on larger \llm (decoder-only transformers based on the DoReMi~\cite{xie2023doremi} architecture with 1B parameters).


\noindent$\bullet$ \emph{Chinchilla Scaling Law.} D-CPT~\cite{que2024d} establishes a mathematical relationship {which could be used to find the best mixture of general and domain-specific data} between validation loss, model size, data size, and domain data mixing ratios based on Chinchilla Scaling Law~\cite{10.5555/3600270.3602446} to optimize domain-specific continual pre-training as $L(N, D, r)=E+\frac{A}{N^{\alpha}}+\frac{B \cdot r^{\eta}}{D^{\beta}}+\frac{C}{(r+\epsilon)^{\gamma}}$ ($N$ is model parameter count, $D$ is training data volume (number of tokens), $r$ is domain corpus ratio, $E,A,B,C,\alpha,\beta,\gamma,\eta,\epsilon$ are fitting parameters), with a variation which introduces K which describes the difficulty to learn the domain's knowledge as $L(N, D, r)=E+\frac{A}{N^{\alpha}}+\frac{B \cdot r^{\eta}}{D^{\beta}}+\frac{C}{(r+\epsilon)^{\gamma}}+\frac{F}{K^{\mu }}$ ($F$ is a fitting parameter). It fits formula parameters through small-scale experiments to predict performance under different training configurations and find the suitable ratio to minimize the domain validation loss while ensuring the generalization loss does not exceed the specified threshold.

\noindent$\bullet$ \emph{Exponential Functions.} Data Mixing Law~\cite{ye2024data} establishes an exponential relationship between validation loss and data mixing ratios of several domains {(e.g., public datasets like Pile-CC, Books3)}, $L({r}) = c + k \exp\left( \sum_i t_i r_i \right)$, where \( L({r}) \) is the validation loss, \({r}\) represents the mixing ratios of different domains, and \(c\), \(k\), and \(t_i\) are learnable parameters. That is, it experiments on a small model with the exponential relationships to predict the best data domain mixing ratios on \llm performance with scaling laws, which combines training step scaling laws ($L(S) = c + kS^\alpha$, where \( S \) is the number of training steps, and \( \alpha \) is a fitting parameter.), which is used to infer the validation loss at target training steps from results at smaller steps, and model size scaling laws ($L(N) = c + kN^\beta$, where \( N \) is the number of model parameters, and \( \beta \) is a fitting parameter), which is used to infer the validation loss for large model sizes from smaller model sizes.  

\noindent$\bullet$ \emph{Classification Model.} \cite{liang2024data} aims to find the data proportion of closed-source model by data proportion detection, which first generating large-scale data from the \llm, then using a classification model to categorize the generated data and compute perplexity, deriving the proportions of pre-training data based on the Data Mixing Law~\cite{ye2024data} (which is a mathematical formula describing the relationship between the proportion of pre-training data and the model's loss in different domains.).

\noindent$\bullet$ \emph{Power-law Function.} CMR~\cite{gu2024cmr} aims to optimize the continual pre-training by finding the best ratio of generic dataset and domain-specific dataset. Based on the research before and the data observed on different sizes of models with different ratios of data, the relationships between loss and mixture ratio, and training volume fit in power-law forms, which are described as $L(R)=\alpha \cdot R^{s} +\beta$ and $L(T)=\alpha_{1}  \cdot T^{s_{1}} +\beta_{1}$, where $\alpha$, $\beta$, $s$, $\alpha_{1}$, $\beta_{1}$ and $s_{1}$ are fitting parameters. Based the relationships, they propose a metric \emph{Critical Mixture Ratio}, which is the maximum data mixing ratio that balances between (1) {significantly reducing domain loss while (2) keeping the increase in general loss within a pre-defined tolerance range}.
Based on the two aspects, the ratio is defined as $R^{*}=\max \{R \mid R \in F\}$, where $R$ is the ratio of generic dataset and domain-specific dataset, $F$ is feasible mixture ratios which comprises all mixing proportions that satisfy the constraints of the general loss function.

\hi{During-Training Mixing (Bilevel Optimization).} This method adopts a closed-loop optimization technique that {ensures model parameters are well optimized~\cite{colson2007overview}.} Generally, Bilevel optimization
involves two nested optimization problems: 
(1) the inner-level problem ensures model parameters are optimized under given weights (e.g., minimizing weighted training loss), while (2) the outer-level updates weights through backpropagation of validation loss, forming a closed-loop optimization.




Typically, ScaleBiO~\cite{pan2024scalebio} reconstructs the data sampling weight optimization problem into a bilevel optimization problem, where outer-level problem is adjusting data weights to minimize validation loss; and the inner-level problem is adjusting model parameters to minimize weighted training loss {and it could be applied to tasks like multilingual training (mixture of languages) and instruction following (mixture of quality).} 
ScaleBio first experiments on small models. Then it extends to larger models like LLaMA-3. ScaleBiO initialize the weights equably for all data sources. In each iteration, it randomly selects a subset of data sources to update their weights: for the selected data sources, it adjusts the weights by optimizing the gradient of the validation loss, prioritizing the increase of weights for data that contribute significantly to model performance, while decreasing the weights for data that have less impact on performance. After updating the weights, retrain the model parameters and repeat the process until convergence. 
    
To enhance the efficiency of BiO-based data mixing, DoGE~\cite{fan2023doge} defines $(i)$ inner-level problem as that under the condition of fixed data mixing ratios, optimize the proxy model parameters to minimize the weighted sum of domain losses; and $(ii)$ outer-level problem as \emph{adjusting the data mixing ratios such that the model parameters obtained through inner-level problem optimization achieve optimal performance on the target loss}. The method is executed on a small-scale proxy by following steps: Initially, it sets the domain weights as a uniform distribution. In each iteration, it dynamically adjusts the weight of each domain based on the gradient alignment value (calculated as the inner product of the gradient of current data domain and the sum of gradients from all data domains), which measures the contribution of the data from the current domain to the gradient direction of all other domains' data. Using the updated weights, it resamples the data and updates the model parameters. Repeat the process for multiple iterations until the weights stabilize, then  apply to actual \llm pre-training.

\hi{During-Training Mixing (Distributionally Robust Optimization).} To search for a robust data mixing strategy (which can be sub-optimal but with low uncertainty), some methods adopt Distributionally Robust Optimization (DRO) for data mixing. DRO achieves robustness against distributional uncertainty by optimizing for the worst-case scenario within a set of distributions (referred to as the uncertainty set or ambiguity set). 

\noindent$\bullet$ For \llm pre-training, DoReMi~\cite{xie2023doremi} defines the worst case as domains where the proxy model underperforms compared to the reference model, which initially sets the domain weights as a uniform distribution and each domains contains several sample sets, and uses it to train Transformer decoder-only LM with 280M parameters and computes loss in each example set, which provides a reference point to measure the improvement potential (the loss difference) of the proxy model in each domain. Next, DoReMi trains a small-scale proxy model (also Transformer decoder-only LM with 280M parameters) by adjusting the domain data weights through DRO, which dynamically adjusts the domain weights and tilt the weights toward domains with larger losses (compared to the reference model). Finally, validate performance of weighted domain data on large models (Transformer decoder-only LM with 8B parameters). 

\noindent$\bullet$ For \llm fine-tuning, tDRO~\cite{ma2024task} defines the worst case the same as DoReMi, which computes the relative loss for each domain with a proxy model (e.g. Qwen1.5-0.5B~\cite{bai2023qwen}); and they compare the training loss of domain data with the reference model (e.g., Qwen1.5-0.5B), and evaluate each domain's potential for model improvement, and update the domain weights accordingly, giving more attention to high-loss domains. Finally, the updated weights are normalized to form a new sampling distribution and repeat the process to get final data distribution.

\subsubsection{Data Distillation and Synthesis}
\label{subsubsec:synthesis}


Synthetic data, which mimics real-world scenarios, is particularly valuable for resolving problems such as $(i)$ data scarcity (e.g., augmenting data for a small dataset)~\cite{xu2023wizardlm}, $(ii)$ privacy concerns (e.g., replacing sensitive data with synthesis data)~\cite{xie2024differentially}, $(iii)$ the need for diverse and high-quality datasets (e.g., generating examples for underrepresented cases)~\cite{liu2024augmenting}, $(iv)$ lack of reasoning data (e.g., for code, chain of thought), $(v)$ human alignment (e.g., label better \llm's response by human beings or \llms).

\begin{tcolorbox}[colback=gray!1,colframe=gray,lowerbox=visible]
  \textbf{Principles}
  \tcblower
  {Traditional ML methods use rule-based templates, basic augmentation (lexical substitution, back-translation), or statistical models to create limited synthetic data, addressing data scarcity/class imbalance. While \llm-driven synthesis employs \llms to produce diverse, high-quality data, tackling data scarcity, privacy concerns, and diverse training needs. Key paradigms include: (i) sample-driven generation, (ii) domain-aligned synthesis, and (iii) reasoning-centric formatting. Challenges involve ensuring rigorous reasoning chain synthesis and optimizing cost-quality balance in data production.}
\end{tcolorbox}

Despite the advantages, synthetic data can  negatively impact \llm training, such as when characteristics like toxicity are inherited from the source model or even amplified~\cite{shimabucoro2407llm}. Thus, it is vital to design data synthesis methods for \llms~\cite{zhu2024synthesize}. As shown in Figure~\ref{fig:motivation}, we discuss methods dealing these problem through the diverse \llm stages, including pre-Training, SFT, Reinforcement Learning and RAG.






\hi{Knowledge Distillation.} {Due to \llms' massive parameter scale and high resource demands which make practical deployment challenging, so we utilize} knowledge distillation ({such as designing paradigms to prompt \llm to generate high-quality data}) to training a student \llm with less parameters to mimic the target model's generation ability. 

\noindent$\bullet$ \emph{Task-Specific Prompt Distillation.} To significantly reduce inference costs and latency while maintaining performance, 
{\cite{shirgaonkar2024knowledge} employs task-specific prompts: (1)
Chain-of-Density (CoD): Iteratively adds entities to summarize for enhanced density.
(2) Chain-of-Thought (CoT): Guides reasoning tasks (e.g., math) through stepwise logic.
Using GSM8K~\cite{cobbe2021gsm8k} data and Llama-3.1-405B-Instruct, synthetic data is generated for fine-tuning smaller models (Llama-3.1-8B/70B-Instruct) paired with simplified prompts, balancing efficiency and task specialization.}

\noindent$\bullet$ \emph{Code Verification and Error Correction Distillation.} Existing knowledge distillation methods (e.g., Chain-of-Thought Fine-tuning) rely on synthetic data generated by \llms, but such data often contains incorrect intermediate reasoning steps which can mislead small models during learning, hindering the improvement of their reasoning capabilities.

{Pad~\cite{zhu2024padprogramaideddistillationteach} proposes Program-aided Distillation (PaD) to address error-prone synthetic data in knowledge distillation with
(i) Programmatic Reasoning: LLMs generate executable code (e.g., math problems as Python calculations) instead of natural language CoT, with Python compilers auto-filtering logic errors.
(ii) Error-Injection Training: Models learn error correction by fixing synthetically injected AST-based errors (e.g., NameError).
(iii) Semantic Validation: Decoding selects steps via semantic alignment scoring (e.g., cosine similarity) to prevent error propagation.
PaD replaces flawed CoT steps with verifiable program logic, enhancing small models' reasoning robustness through code-based distillation and self-correction mechanisms.}

\noindent$\bullet$ \emph{Multi-stage Collaboration Distillation Between Student models.}
In domains with high annotation costs (e.g., biomedical parsing) or complex task structures (e.g., syntactic/semantic parsing), labeled data is extremely scarce, making traditional supervised fine-tuning ineffective.
{MCKD~\cite{zhao2023multistage} introduces Multi-stage Collaborative KD (MCKD) for low-resource generation as 3 steps.
(i) Initialization: GPT-3.5 generates pseudo-labels for unlabeled data.
(ii) Collaborative Distillation:
Splits data into two subsets for cross-labeling via paired T5-Base models, reducing noise overfitting.
Iteratively refines labels over 3 iterations.
(iii) Final Training: Trains a single model on refined labels.
Achieves near-supervised performance with 50 labeled examples (vs. 500 required traditionally) through multi-stage noise reduction and collaborative pseudo-label optimization.}







\hi{Pre-training Data Augmentation.} The pre-training stage of \llm requires a vast amount of data and it can be costly to synthesize such data with powerful models like GPT-4. Therefore, there are techniques like distillation~\cite{zhou2024jiuzhang3}, or simply mixing synthetic data into the whole corpus.



\noindent$\bullet$ \emph{Distilled \llm for Mathematical Data Synthesis.} 
{JiuZhang3.0~\cite{zhou2024jiuzhang3} proposes an \llm-based synthesis method for high-quality math problems:
(i) Model Distillation, fine-tunes DeepSeekMath-7B on GPT-4-generated QA pairs (with curated prompts and math texts) to mimic GPT-4’s generation.
(ii) Uses gradient similarity to prioritize task-relevant data.
(iii) Refines the model with filtered data to produce aligned outputs.
The final math synthetic corpus are generated by the refined model based on the multi-source corpus (e.g., Wikipedia) and prompt sets.}



{


\noindent$\bullet$ \emph{Fintuned LLM for Instruction-Response Pair Synthesis.} In order to study the effect of supervised pre-training, 
{Instruction PT~\cite{cheng2024instruction} introduces an Instruction Synthesizer (Mistral-7B finetuned on 40+ task categories) to augment raw text with few-shot multi-task instructions (e.g., "Summarize school activities" → QA/reasoning pairs). Unlike GPT-style pre-training, it integrates structured task execution (QA, classification) alongside language modeling. This hybrid approach boosts data efficiency (500M model $\approx$ 1B baseline) and multi-task adaptability from pre-training.}


\noindent$\bullet$ \emph{\llm Prompting for Mathematical Data Synthesis.}
Current math-specialized \llms rely on SFT with problem-solving data (e.g., step-by-step solutions). However, since CPT improvements in math are far less significant than SFT gains. 

To study the impact of problem-solving data in continual pre-training, \cite{chen2025advancing} proposes enhancing models' mathematical reasoning capabilities by augmenting problem-solving data (e.g., step-by-step solutions for common math problems) during pre-training, rather than relying solely on traditional math corpora (e.g., theorem texts). 
First, a student model (Llama2~\cite{Llama-2}) is utilized to generate answers from the collected math problems. Then, it uses a teacher model (Llama2~\cite{Llama-2} with more parameters) detects errors in a student model's solutions and generates corrective steps guided by prompts. This teaches the target \llm self-checking and error-correction skills. Experiments indicate continual pre-training excels at learning complex reasoning (e.g., multi-step equation solving) than SFT, where MathGPT-8B using only 100B well-generated math-related tokens can exhibit capabilities comparable to Qwen2-Math-72B~\cite{yang2024qwen2technicalreport}. 



\noindent$\bullet$ \emph{LLM Prompting for Rephrasing Synthesis.} {To introduce more diversity to the data, some methods rephrase the data to different styles of texts like Q\&A or concise definition.} 
{WRAP~\cite{maini2024rephrasing} leverages instruction-tuned models (e.g., Mistral-7B) to rephrase web text (C4) into four formats:
(i) simple vocabulary and sentence structures that are understandable to young children.
(ii) Standardized encyclopedia-style expression.
(iii) Complex terminology and concise academic sentence structures.
(iv) multi-turn dialogue.
Mixing rephrased and original data trains \llms to adapt to diverse formats (e.g., zero-shot QA), achieving 3× faster training and 50\% lower perplexity on Pile benchmark~\cite{Pile} via hybrid real-synthetic data synergy.}







\noindent$\bullet$ \emph{LLM Prompting for Cross-language Synthesis.} \llms like Llama-3 exhibit deficiencies in cross-language tasks and multidisciplinary scientific reasoning, while continual pre-training often triggers catastrophic forgetting (e.g., performance degradation in original capabilities like English tasks). \cite{chen2024towards} proposes to synthesize data so as to enhance Llama-3's Chinese proficiency and scientific reasoning capabilities while mitigating catastrophic forgetting. 
They utilize Mistral-7B~\cite{mistral} to generate multidisciplinary scientific question-answer pairs (e.g., Q\&A on ``explaining the electrostatic repulsion principle of ion double layers in electrolyte solutions'') from seed data collected and classified into multiple disciplines by TinyBERT~\cite{jiao2019tinybert} and BERT-Tiny-Chinese~\cite{BERTTinyChinese} from Dolma’s CC~\cite{Dolma} and C4~\cite{dodge2021documenting}. And generate coding problems with LeetCode algorithm tasks as seeds by Magicoder-S-DS-6.7B~\cite{wei2023magicoder} .These are mixed with Chinese, English, and synthetic data in a 1:7:2 ratio, significantly boosting scientific reasoning.

Additionally, through substitution experiments (validating data strategies using TinyLlama-1.1B~\cite{zhang2024tinyllama} as a proxy model), they find that (1) a 20\% synthetic data ratio with an error rate below 30\% yields optimal results; and (2) a curriculum progressing from simple to complex topics outperforms random training.


\noindent$\bullet$ \emph{Code Interpreter + \llm Prompting for Code Synthesis.} Current code generation models rely heavily on large teacher models (e.g., GPT-4) to generate synthetic training data, leading to poor scalability, high costs. And most datasets focus on direct code completion or text-to-code translation, but lack Input-Output (I/O) case-based reasoning tasks (e.g., inferring code from example mappings like ``hello'' $\rightarrow$ ``olleh'').
This gap results in weak generalization for inductive programming challenges.

{To bridge this gap, Case2Code~\cite{shao2025case2code} generates training data through four steps: (i) Extract executable Python functions (with input/output parameters) from open-source repositories; (ii) Use lightweight LLMs (e.g., InternLM2-7B) to analyze function logic and generate diverse input samples; (iii) Execute functions to obtain real outputs and filter invalid results; (iv) Convert I/O pairs into natural language prompts with diversified templates for improved generalization. This method leverages "code interpreter + lightweight LLM" to cost-effectively produce 1.3M training samples, eliminating reliance on expensive teacher models.}


\noindent$\bullet$ \emph{LLM-based Clustering for Synthetic Data Evaluation.} In order to study the impact of diversity of large-scale synthetic data,
{~\cite{chen2024diversity} introduces an LLM-based clustering method to quantify synthetic data diversity and analyze its impact on model performance. (i) Builds hierarchical topic trees from web-crawled data via GPT-4 (e.g., Quantum Computing → Qubit Types → Superposition); (ii) Generates diverse datasets by varying topics, prompts (styles, target audiences, etc.) and LLMs (GPT-4o, Llama-3, etc.). Experiments across different diversity combinations show synthetic data diversity positively correlates with model performance on benchmarks like HellaSwag~\cite{zellers2019hellaswag} and ARC-Challenge~\cite{ferre2021first}.}

\noindent$\bullet$ \emph{LLM Prompting for Multimodal Image-Text Synthesis.} Current approaches for synthesizing multimodal pre-training data typically employ two main approaches: (1) the generation of images conditioned on textual input using text-to-image models, and (2) the augmentation of uncaptioned or simple-captioned source images via multimodal models.
In the domain of text-to-image synthesis, current methods use diffusion models~\cite{fuest2024diffusionmodels} for image generation.
Examples include DiffuseMix~\cite{diffusemix}, which enhances datasets by augmenting image samples through the blending of original and diffusion-generated images, and EDA~\cite{eda_diff}, which applies diffusion models to produce variations of real images that retain semantic consistency while augmenting the dataset.
Concerning image captioning, several studies focus on improving the quality of image-text pairs. LaCLIP~\cite{fan2023improvingclip} uses ChatGPT to rewrite existing captions, thereby introducing greater diversity in linguistic expression while maintain the core semantic content. A limitation of this method is the potential for visual semantic loss due to the language model's lack of direct access to the image. To mitigate this, VeCLIP~\cite{lai2024veclip} incorporates a multimodal LLM (LLaVA) to provide a detailed visual description of the image contents (e.g., color and shape attributes, objects, and relations among objects). This description is then fused with the original caption by a LLM to yield a more comprehensive final caption.
To simultaneously synthesize both image and text samples, CtrlSynth~\cite{CtrlSynth} proposes a system comprising three modules: the Florence-large~\cite{xiao2023florence2} vision tagging model to extract basic visual elements of an image (e.g., color and shape attributes, objects, and relations among objects), the Qwen2-7B-Instruct~\cite{yang2024qwen2technicalreport} language model to generate synthetic text which meets the requirements in the instruction, and the stable-diffusion-x1-base-1.0~\cite{podell2023sdxlimprovinglatentdiffusion} text-to-image model to generate novel and diverse image samples based on text prompts.

\hi{SFT Data Augmentation.} The SFT stage of \llm training mainly focus on improvement of specific domains (math, medicine, etc.), aligning \llm's knowledge to instructions, enhancing reasoning ability, etc. Current methods take \llms as the main method to generate data with some designed frameworks. Many works \cite{huang2024key, liu2024augmenting, mitra2024agentinstruct} take existed datasets as seeds to synthesize mimic datasets. 

\noindent$\bullet$ \emph{LLM-based Knowledge and Q\&A Pairs Synthesis.} 
{To enrich or enhance the diversity of data for better model performance, there are various prompt frameworks such as building topic taxonomy~\cite{li2024synthetic} and iterative synthesis~\cite{huang2024key}.}

For example, to cover various domains of human knowledge, 
{GLAN~\cite{li2024synthetic} introduces a knowledge-classification framework for synthetic text generation by GPT-4. 
(i) Organize knowledge domains (natural sciences/humanities) into disciplines (math/programming) by; (ii) Develop course outlines with units (e.g., "Intro to Calculus") and core concepts (e.g., "Limits"); (iii) Use GPT-4 to create diverse questions by combining concepts, then generate answers with faster GPT-3.5. This structured approach ensures systematic coverage of knowledge areas while balancing generation quality and efficiency.}

Though this could enhance understanding of \llm about many domains, but to get better enhancement still needs to focus on one aspect, like math, 
{KPDDS~\cite{huang2024key} identifies mathematical problem themes (e.g., algebra, geometry) and core skills (e.g., factoring) using GPT-4, then constructs a matrix mapping theme co-occurrence probabilities to guide logical problem generation. GPT-4 synthesizes new questions based on these themes and solutions, which are evaluated for quality (clarity, coherence) and refined via GPT-4 voting. The method further diversifies questions through variations and applies iterative voting to optimize output. This structured approach ensures contextually coherent, avoiding random combinations.}

Instead of combining elements {like KPDDS (e.g., combining algebra and geometry to synthesize problems)}, 
{MMIQC~\cite{liu2024augmenting} enhances mathematical reasoning by iteratively generating complex, diverse problems from existing ones for fine-tuning. Using a seed dataset, GPT-4 creates problems via added constraints, variables, or extended reasoning. A filtering mechanism ensures logical consistency, problem-solution alignment, and correctness, with validated data expanding the dataset iteratively.}


\noindent$\bullet$ \emph{LLM-based Alignment Data Augmentation.} Domain knowledge is one thing, and lead \llm's knowledge align with instruction is another thing that could be done to get better performance {through techniques like few-shot prompting. }

{AgentInstruct~\cite{mitra2024agentinstruct} uses LLMs to create scalable, diverse Q$\&$A data. GPT-4 converts raw input (text/code) into structured formats (argument passages, API lists) to enable diverse instruction creation. Multiple GPT-4 agents generate varied task instructions and answers following a detailed taxonomy (e.g., reading comprehension, coding tasks). GPT-4 and Claude-3 then refine tasks by adding complexity (e.g., integrating dense context or escalating difficulty), ensuring high-quality, adaptable outputs.}

Similarly, SELF-INSTRUCT~\cite{wang2022self} aligns \llm's knowledge to prompts by generating task instructions and examples: Starting with a small set of manually written seed tasks, a \llm (e.g., GPT-3) is prompted to generate new task instructions covering various task types, such as classification, question-answering, and generation. Next, different strategies are employed to generate inputs and outputs based on the task type. For instance, for classification tasks, possible class labels (e.g., "positive" and "negative") are generated first, followed by inputs corresponding to each label. For open-ended tasks, a question description is generated first, followed by an answer. The generated data undergoes multiple rounds of filtering, including removing duplicates or invalid data and ensuring input-output alignment.

\begin{table}[]
\centering
\caption{Data Synthesis for \llm.}
\label{tab:synthesis}
\resizebox{\linewidth}{!}{%
\begin{tabular}{ccc}
\hline
\textbf{Stage}      & \textbf{Category}               & \textbf{Methods}                                                 \\ \hline
\multirow{3}{*}{Distillation} &
  \multirow{2}{*}{Reasoning Augmentation} &
  Cot~\cite{shirgaonkar2024knowledge} \\
                    &                                 & Prompt with Tools~\cite{zhu2024padprogramaideddistillationteach} \\
                    & Data Augmentation               & Prompt with Multi-Agent~\cite{zhao2023multistage}                \\ \hline
\multirow{2}{*}{Pre-Training} &
  \multirow{2}{*}{Data Augmentation} &
  Distillation + Fine Tuning + Prompt~\cite{zhou2024jiuzhang3} \\
 &
   &
  Prompt~\cite{cheng2024instruction}, \cite{chen2025advancing}, \cite{maini2024rephrasing}, \cite{chen2024towards},\cite{shao2025case2code}, \cite{chen2024diversity} \\ \hline
\multirow{5}{*}{SFT} &
  Data Augmentation &
  Prompt~\cite{li2024synthetic}, \cite{huang2024key}, \cite{liu2024augmenting}, \cite{mitra2024agentinstruct} \\ \cline{2-3} 
 &
  \multirow{4}{*}{Reasoning Augmentation} &
  Prompt~\cite{huang2024mustard}, \cite{hou2025advancing}, \cite{shen2025satori} \\
                    &                                 & Human Label~\cite{lightman2023let}                               \\
                    &                                 & Automated Label~\cite{wang2023math}                              \\
                    &                                 & High Quality Reasoning Data~\cite{ye2025limo}, \cite{li2025llms} \\ \hline
\multirow{3}{*}{RL} & Prompts Optimization            & Prompt~\cite{wang2022self}                                       \\
                    & \multirow{2}{*}{Human Feedback} & RLHF~\cite{bai2022training}                                      \\
                    &                                 & RLHF By LLM~\cite{zheng2023judging}                              \\ \hline
RAG                 & Privacy Protection              & Prompt~\cite{zeng2024mitigating}                                 \\ \hline
\end{tabular}%
}
\end{table}

\hi{SFT Reasoning Data Augmentation.} Synthesize reasoning data (e.g., code, chain of thought) {through techniques like Chain-of-thought(CoT), or utilizing verification tools for more rigorous reasoning.}

\noindent$\bullet$ \emph{Prompting \llm To Math Reasoning With Verify Tool.} Also for math, MUSTARD~\cite{huang2024mustard} utilizes mathematical proof tools to get reasoning enhancement. First, fundamental concepts from the field of mathematics are selected as seeds, and GPT-4 generates corresponding problems through two types of solutions: (1) One is a natural language explanation of the reasoning process, and (2) the other is a formal language solution that can be verified (e.g., code compatible with mathematical proof tools). Next, formal solutions are verified using mathematical proof tools to ensure the correctness of the reasoning and answers. For content that fails verification, the model adjusts based on feedback and re-verifies until a correct result is generated. 

\noindent$\bullet$ \emph{CoT Data Synthesis By LLM Exploring.} Works mentioned above highly rely GPT-4 for its advanced ability for math to generate problems and solutions to fine-tune for higher reasoning ability. While more recent research try to enhance \llms' reasoning ability by technique like Chain-of-Thought (CoT, which let \llms use tokens to output their reasoning steps) and synthesis or label finer reasoning data for training.

By generating CoT data that covers a wide range of reasoning paths through a trial-and-error self-verification loop, \cite{hou2025advancing} breaks the traditional limitation of relying solely on correct reasoning paths. Specifically, multiple \llms (e.g., Qwen-7B, Llama-3-8B) are utilized to generate diverse solutions for the same mathematical problem (20-50 responses per problem) to encourage models to explore incorrect paths (e.g., wrong formulas, logical leaps) while retaining complete error analysis. Then a verifier \llm (e.g., GPT-4) performs critical analysis on each response:
(a) For incorrect paths, annotate the error steps and generate correction suggestions (e.g., ``Step 3 misapplies the cosine theorem, which should be replaced with the Pythagorean theorem'').
(b) For correct paths, extract key reasoning steps to form a concise CoT.
Merge corrected incorrect attempts with correct paths to construct multi-branch CoT.

{Similarly, Satori~\cite{shen2025satori} introduces Chain-of-Action-Thought (COAT), a reasoning framework with meta-action tokens (Continue / Reflect / Explore) enabling dynamic pauses, logic verification, and strategy shifts with a two-stage pipeline: (i) Multiple LLM agents generate COAT-formatted reasoning chains to fine-tune a base model for COAT-formatted syntax mastery. (ii) Partial rollbacks ($\le$5 steps) from historical reasoning (correct/incorrect paths) append $<$reflect$>$ to trigger revised reasoning with reinforcement learning (RL) combined with rewards for answer correctness, error correction, and penalties for failures. The RL-enhanced model is distilled into base models (e.g., Llama8B) for iterative refinement.}


These works propose framework by letting \llm reason by themselves, and we also have works that label reasoning data for fine tuning to get reasoning ability.

\noindent$\bullet$ \emph{Reasoning Data Labeling.} \cite{lightman2023let} compares the effects of outcome supervision (provides feedback based solely on the correctness of the final answer) and process supervision (provides feedback for each step in the reasoning process) on mathematical reasoning tasks by comparing manually labeling the reasoning steps generated by GPT-4 with outcome supervision. The results showed that process supervision model achieved significantly higher problem-solving accuracy (78.2\%) compared to outcome supervision model (72.4\%)

But this would cost too much manual effort, so MATH-SHEPHERD ~\cite{wang2023math} proposes a method to automatically generate process-annotated data for training Process Reward Models (PRM, which evaluate the quality of each reasoning step). First, complete the remaining reasoning and answers multiple times for the initially generated reasoning steps with \llm, then each step is scored based on two metrics:
(1) Hard Estimation (whether the correct answer is generated, with values of 0 or 1).
(2) Soft Estimation (the proportion of correct answers generated through this step).
These scores assess the step's ability to derive the correct answer.

\noindent$\bullet$ \emph{High Quality and Well Format Data Are The Keys To Better Reasoning.} Moreover, LIMO~\cite{ye2025limo} and \cite{li2025llms} state that high quality and well-formatted reasoning data are keys to high performance. \cite{ye2025limo} emphasizes stimulating complex reasoning capabilities in \llms through a small number of high-quality training examples with questions and reasoning chains. 
Powerful models (such as R1, DeepSeek-R1-Distill-Qwen32B) are used for evaluation and synthesis, retaining problems that remain challenging. 
Each problem is accompanied by detailed solutions and reasoning chains (from official solutions, expert solutions, and \llms-generated Cot, etc.) and filtered by rules-based and LLM-assisted methods.

\cite{li2025llms} finds that the overall structure of the reasoning steps is more important than the specific content. With problems from Numina-Math~\cite{numina_math_datasets} etc. and long CoT generated by DeepSeek-R1~\cite{guo2025deepseek} and QwQ-32B-Preview~\cite{team2024qwq} as data to fine-tune. With modification of the fine-tune data, reveals that training the model with incorrect answer samples results in an accuracy drop of only 3.2\% compared to training with correct samples. However, shuffling 67\% of the reasoning steps in the training samples leads to a 13.3\% drop in accuracy on AIME 2024 problems relative to training with correct samples.


\hi{Reinforcement Learning} The RL stage of \llms find the most human-preferential responses within the multiple responses generated by \llm of one instruction. Works like \cite{bai2022training, zheng2023judging} manually label the responses or let \llms do the job.

Label better \llm's response by human or \llms. To align the model's responses with human expectations, \cite{bai2022training} gathers helpful and harmless data through open-ended conversations. Then, a preference model is trained to score the responses in the data, providing a basis for reward optimization in reinforcement learning. The preference scores guide the optimization of the language model's responses. Next, the latest model generates new data, continuously updating the preference model to improve performance on high-quality data. To improve efficiency, \cite{zheng2023judging} proposes a new chatbot evaluation method using language models as "judges" to compare and score chatbot responses, with the goal of automating the evaluation process and reducing human involvement. It introduces two benchmarks: one focusing on multi-turn conversation performance and another collecting user preferences via crowdsourcing. The method also addresses potential biases, such as preferences for answer order or length, through strategies like swapping answers, using few-shot examples or Chain-of-Thought. The approach demonstrates that language models can achieve high consistency with human evaluators, providing a scalable and interpretable framework for efficient chatbot assessment.

\hi{Retrieval-Augmentation Generation.} The RAG stage mainly offers knowledge and documents from outside to avoid additional training cost. Main works in this stage of data synthesis focus on privacy issues.

Replace sensitive data with synthesis data. In order to mitigate the privacy issue, \cite{zeng2024mitigating} proposes a two-stage synthetic data generation and privacy-enhancing method for the RAG stage of \llm. 

In the first stage, key information is extracted from the original data (such as ``symptom description'' and ``treatment plan'' in medical dialogues), and \llm is used to generate synthetic data based on key information but does not contain sensitive details. 

In the second stage, \llms are applied to the synthetic data, and rewriting strategies are employed to eliminate potential privacy leaks (such as removing specific names or obfuscating descriptions). 

This process of evaluation and rewriting is repeated to ensure that the generated data retains its key utility while completely avoiding privacy concerns.

\subsubsection{End-to-End Data Processing Pipelines}
\label{subsubsec:pipelines}

With above data processing methods, we separately introduce existing frameworks that support common processing operations; practices of integrating some of these methods within pipelines in real-world \llm data preparation; together with some preliminary pipeline orchestration methods.

\begin{tcolorbox}[colback=gray!1,colframe=gray,lowerbox=visible]
  \textbf{Principles}
  \tcblower
  When designing data processing pipelines, several critical factors must be considered: (1) the trade-off between data quality and quantity; (2) dependencies across the processing operations (e.g., text extraction necessarily preceding operations like deduplication and filtering); (3) efficiency optimization (e.g., conducting computationally intensive steps like model-based filtering after lightweight processing steps like URL filtering).
\end{tcolorbox}


\subsubsection*{2.2.7.1 Typical data processing frameworks}

Data processing frameworks provide built-in libraries, operators, and intuitive interfaces that can benefit the design of data processing pipelines for different \llms. Here we showcase three typical data processing frameworks.


(1) Data-juicer \cite{chen2024data} is an open-source framework designed for customizable, high-quality, and efficient data processing. It offers a diverse range of pre-built data processing operators such as data formatting, mapping, filtering, and deduplication. Additionally, the framework features visualization and automatic evaluation, enabling users to receive immediate feedback on their data pipeline. To manage large-scale datasets effectively, Data-juicer is optimized for distributed computing, ensuring robust performance and scalability. 

(2) Dataverse~\cite{park2024dataverse} is an open-source framework designed to simplify custom ETL (Extract-Transform-Load) pipeline development through an easy-to-use block-based interface that enables users to easily customize by adding, removing, or rearranging blocks. The platform offers a diverse range of pre-built data processing operators, including deduplication, decontamination, bias mitigation, and toxicity reduction, while also supporting the integration of data from multiple sources. Similar to Data-juicer, Dataverse integrates with Apache Spark for distributed processing and supports AWS integration for cloud scalability.


(3) \cite{sun2024integrated} introduces a data processing framework that allows users to customize data processing pipelines using a comprehensive suite of operators categorized in two main modules: (1) The processing module consisting of data reformatting (read and import strctured data), cleaning (removed undesired data such as HTML tags and translate text), filtering, and deduplication (using MinHashLSH in Section~\ref{subsec:deduplicate}) operators; (2) The analyzing module featuring refined data probing and automatic evaluation. 


\subsubsection*{2.2.7.2 Typical data pipelines}

Data processing pipelines aim to orchestrate a subset of data processing operations (in a specific order) that transform raw data into high-quality \llm training data (mostly for the pre-training stage). Here we showcase three representative pipelines.

\begin{figure*}
    \centering \includegraphics[width=.95\textwidth]{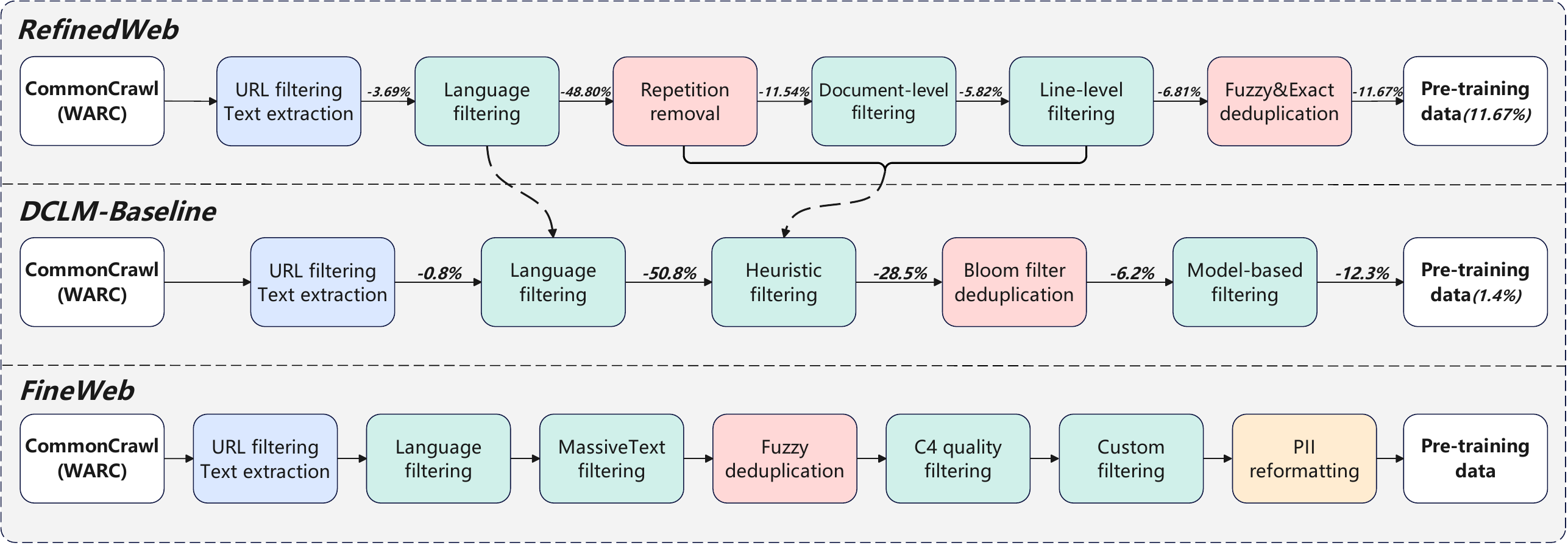}
    \vspace{-.25em}
    \caption{Typical data processing pipelines for \llms.}
    \label{fig:typical_pipeline}
\end{figure*}

\noindent$\bullet$ \underline{The MacroData Refinement (MDR) pipeline} is designed to construct the RefinedWeb Dataset, which has been used for pre-training Falcon LLMs~\cite{penedo2023refinedwebdatasetfalconllm}. MDR refines web-scale data from Common Crawl~\cite{commoncrawl} through three main operations. 

\noindent{\it (i) Data acquisition:} MDR first applies a lightweight URL filter to exclude irrelevant links before any computationally intensive steps. It then extracts text from WARC files using warcio and Trafilatura~~\cite{trafilatura}, followed by language identification (i.e., removing content with limited natural language) using fastText~~\cite{FastText} as implemented in CCNet~\cite{wenzek2019ccnetextractinghighquality}. 

\noindent{\it (ii) Data filtering:} To eliminate low-quality content, MDR employs both (1) document-level filtering~\cite{rae2021scaling} and (2) line-level filtering, which removes noisy content such as social media counters or navigation links. 


\noindent{\it (iii) Data deduplication:} Despite prior filtering, substantial content duplication remains, which can degrade model performance. MDR performs both {\it fuzzy deduplication} using MinHash and {\it exact deduplication} with suffix arrays to minimize redundancy. To address computational limits, the Common Crawl corpus is partitioned into 100 segments, with deduplication performed per segment. Additionally, to avoid cross-part redundancy, URL-level deduplication is applied by excluding URLs already retained in earlier segments.

Overall, MDR follows three core design principles: {\it (i) scale first}, by maximizing data volume from Common Crawl to support large model training; {\it (ii) strict deduplication}, as rigorous redundancy elimination is critical for training efficiency and generalization; and {\it (iii) heuristic filtering}, favoring rule-based filters over ML-based ones to reduce bias and maintain transparency.

\noindent$\bullet$ \underline{The DCLM-Baseline pipeline} also processes data from the Common Crawl dataset. Different from MDR, in addition to text extraction and language identification, it applies efficient heuristic filtering~\cite{penedo2023refinedwebdatasetfalconllm} to exclude irregular content (e.g., toxic words or webpages from illegal sources). Next, DCLM-Baseline adopts a Bloom filter for data deduplication, ensuring its scalability with large datasets. Finally, over the processed data with much smaller size, it conducts model-based quality filtering (most computationally intensive) to remove {low-quality content. Specifically, a fastText classifier trained on instruction-formatted data, including OH-2.5 (OpenHermes 2.5) and ELI5 (ExplainLikeImFive), is used to retain the top 10\% of documents.} 




\noindent$\bullet$ \underline{The FineWeb pipeline} (for preparing a 15T-token pretraining dataset) starts with text extraction from WARC files using Trafilatura~\cite{trafilatura}, which is more custom than directly using WET format data and language filtering with fastText. Different from the above pipelines, it conducts MassiveText filtering, i.e., heuristic quality filters and repetition filters on paragraph, line, and gram level~\cite{rae2021scaling}. Besides, it conducts fuzzy deduplication using individual MinHash deduplication for each CommonCrawl snapshot, {as this approach matches RefinedWeb's performance, whereas global deduplication yields little improvement over non-deduplicated data.} After deduplication, given the observation that the C4 dataset yields superior performance on some benchmarks despite its smaller size, a selection of C4~\cite{raffel2020exploring}'s {heuristic} filters is applied {to drop low-quality content such as unpunctuated lines and policy statements}. Finally, {to further enhance data quality,} additional custom heuristic filters are developed through a systematic process. Moreover, personal identifiable information (PII) such as email addresses is anonymized using regex patterns in the public release of the dataset.

Compared to MDR and DCLM-Baseline, the FineWeb pipeline is considerably more complex due to its integration of multiple layers of filtering, each inspired by empirical evaluations and comparisons with other datasets such as C4 and RefinedWeb. Its design reflects a trade-off that prioritizes performance over simplicity.

\subsubsection*{2.2.7.3 Orchestration of data pipelines}

The above data pipelines are mostly designed by experience. Instead, Data-Juicer Sandbox~\cite{chen2024data2} proposes a ``Probe-Analyze-Refine" workflow, which involves systematically exploring the impact of various data processing operations and their orders on model performance, combining effective operations into data recipes, and optimizing data utilization through duplication analysis and diversity analysis. The orchestrated pipelines are validated through applications on state-of-the-art models like Mini-Gemini (for image-to-text generation) and EasyAnimate (for text-to-video generation).

\subsection{Data Storage for \llm}
\label{sec:storage}

In this section, we introduce storage techniques for \llms, which we categorize accroding to the tasks they address, including (1) data formats, (2) data distribution, (3) data organization, (4) data movement, (5) data fault tolerance, and (6) KV cache.

\subsubsection{Data Formats}
Data formats are file formats for training data and models.
For \llms, appropriate file formats for data and models can enhance storage efficiency, accommodate multimodal data, be suitable for model training, ensure security, and influence compatibility across different frameworks.

\begin{tcolorbox}[colback=gray!1,colframe=gray,lowerbox=visible]
  \textbf{Principles}
  \tcblower
Compared to traditional machine learning, LLMs place greater demands on data being multi-modal and in a unified format. The main challenge is how to achieve high data reading efficiency in multi-modal scenarios. Current methods address this using techniques like sequential storage. 
\end{tcolorbox}

\hi{Training Data Format.} 
For training data, file formats are required to have good storage efficiency (e.g., TFRecord~\cite{tfrecord}), be adaptable to large amounts of data (e.g., MindRecord~\cite{mindspore}), and sometimes be suitable for model training (e.g., \texttt{tf.data.Dataset}~\cite{tfdata}).

\noindent\underline{(1) Pure-Text Formats.} Common formats such as CSV, JSON, TSV, and TXT are often used to store pure-text \llm data (though they are not limited to such content). However, for large-scale training datasets (at the PB scale), these formats incur significant storage overhead {due to the lack of compression (e.g., not supporting binary encoding)}, leading to storage waste and slow data loading during \llm training. 

To address these issues, TFRecord~\cite{tfrecord} is based on Protobuf (a highly efficient binary serialization protocol) and stores data in a row-based format. As a binary format, its size is significantly smaller than JSON or CSV. Besides, data can be written and read in a streaming manner, making it especially suitable for scenarios like training where data is consumed sample by sample.

\noindent\underline{(2) Multimodal Formats.} 
Pure-text formats are not well-suited for multimodal datasets containing images, videos, and text. To address this, file formats such as TFRecord~\cite{tfrecord} in TensorFlow and MindRecord~\cite{mindspore} in MindSpore have been developed to natively support efficient multimodal data storage.

\noindent$\bullet$ Unlike traditional formats (e.g., COCO JSON~\cite{cocodataset}, which store image metadata in separate JSON files), TFRecord~\cite{tfrecord} allows users to encapsulate images, labels, and metadata within a single \texttt{tf.train.Example}, eliminating the need for separate label files. Moreover, as multimodal datasets substantially increase data volume, TFRecord supports data sharding, enabling the creation of distributed files that can be assigned across multiple servers to facilitate parallel training.

\noindent$\bullet$ MindRecord organizes data into two types of files: $(i)$ the data file, which contains a file header, scalar data pages (e.g., image labels and filenames), and block data pages (e.g., image and text) to store training data; and $(ii)$ the index file, which maintains indexing information based on scalar data  to support efficient retrieval and dataset analysis.

\noindent\underline{(4) Tensor Data Formats.} Compared to the storage formats mentioned above, tensor formats represent data as multi-dimensional arrays. On GPUs or TPUs, such multi-dimensional structures can be partitioned and processed in parallel, making them highly suitable for large-scale computation. For example, \texttt{tf.data.Dataset}~\cite{tfdata} can organize various raw data types (e.g., images, text) into a unified tensor format, ready for direct use by models. However, tensor formats, {due to their dense multi-dimensional storage, incur large storage overhead and offer poor readability}, and are typically adopted only in model training.

\hi{Model Data Format.} Model storage formats need to pay attention to security (e.g., Safetensors~\cite{casey2025empiricalstudysafetensorsusage}) and are usually closely tied to their respective model training frameworks~\cite{torchptpth, tensorflow, huggingface}.

\noindent$\bullet$ Pickle (.pkl~\cite{pickle}) is a Python-specific format supported by almost all Python frameworks and can store any Python object, not limited to model parameters, making it convenient for saving model states and other custom information.  

\noindent$\bullet$ Safetensors~\cite{casey2025empiricalstudysafetensorsusage} was introduced by Huggingface to address the security concerns inherent in Python's Pickle-based serialization. While Pickle serializes both the data and behavior of Python objects—enabling arbitrary code execution during deserialization—safetensors avoids this risk by focusing exclusively on tensors and their associated metadata. This design ensures safe deserialization without the possibility of executing malicious code. Additionally, safetensors supports memory mapping (mmap), which significantly enhances the efficiency of model loading.

\noindent$\bullet$ PyTorch-specific formats (e.g., \texttt{.pt}, \texttt{.pth}~\cite{torchptpth}) are optimized for model storage. Typically, \texttt{.pth} files are used to save training checkpoints, including model parameters, optimizer states, and epoch information, while \texttt{.pt} files are used to store only the model parameters.

\noindent$\bullet$ TensorFlow offers two common saving formats~\cite{tensorflow}: (1) \texttt{SavedModel} format for saving the entire model, including computation graph, weights, optimizer; (2) \texttt{.ckpt} for storing model weights, optimizer states, and training metadata, and is used to save and restore {progress} during training. 

\noindent$\bullet$ ONNX~\cite{onnx} is a cross-framework deep learning model format that supports interoperability across frameworks like PyTorch, TensorFlow, and Caffe2. It offers cross-platform and cross-framework advantages, but does not store training state information.

\noindent$\bullet$ The Hugging Face Transformers library~\cite{huggingface} adopts a modular storage design, i.e., model weights are stored in binary \texttt{.bin} files, model configurations are stored in \texttt{.json} or \texttt{.txt} files.

\subsubsection{Data Distribution}
\label{DataDistribution}
With the development of \llms, the scale of LLM training datasets and the number of parameters of LLMs themselves are growing rapidly (e.g., 9.5 PB data form Common Crawl ~\cite{ilyankou2024ccgpxextractinghighqualityannotated}, DeepSeek-R1~\cite{guo2025deepseek} has 617B parameters). A single node cannot store such large-scale data, and the data needs to be distributed across multiple nodes. The key technologies involved mainly include (1) distributed storage systems and (2) heterogeneous storage systems.

\begin{tcolorbox}[colback=gray!1,colframe=gray,lowerbox=visible]
  \textbf{Principles}
  \tcblower
Compared to traditional machine learning, the data (e.g., training data and model data) used in LLMs including both is growing exponentially. The main challenge lies in how to efficiently store and manage such large-scale data. Current approaches address this through distributed and heterogeneous storage systems. 
\end{tcolorbox}

\hi{Distributed Storage Systems.} 
Distributed storage systems refer to storing a large-scale datasets across multiple nodes (e.g., JuiceFS~\cite{juicefs}, 3FS~\cite{3FS}). Traditional distributed file systems (such as HDFS~\cite{borthakur2008hdfs}) often come with high costs. Moreover, most distributed file systems still use the POSIX protocol when loading the training data for \llms, which bring about significant software overhead.

JuiceFS~\cite{juicefs}, a typical distributed file system based on object storage, uses object storage (e.g., S3~\cite{S3}) as the backend to store data. Compared to traditional distributed file systems (file or block storage), distributed file systems based on object storage enables simpler horizontal scaling. 
It does not need complex directory hierarchy (File Storage) and does not involve complex management logic (Block Storage), thereby significantly reducing storage costs (approximately 20\% of the cost of traditional file systems).

\begin{figure}[!t]
    \centering
    \includegraphics[width=1.\linewidth]{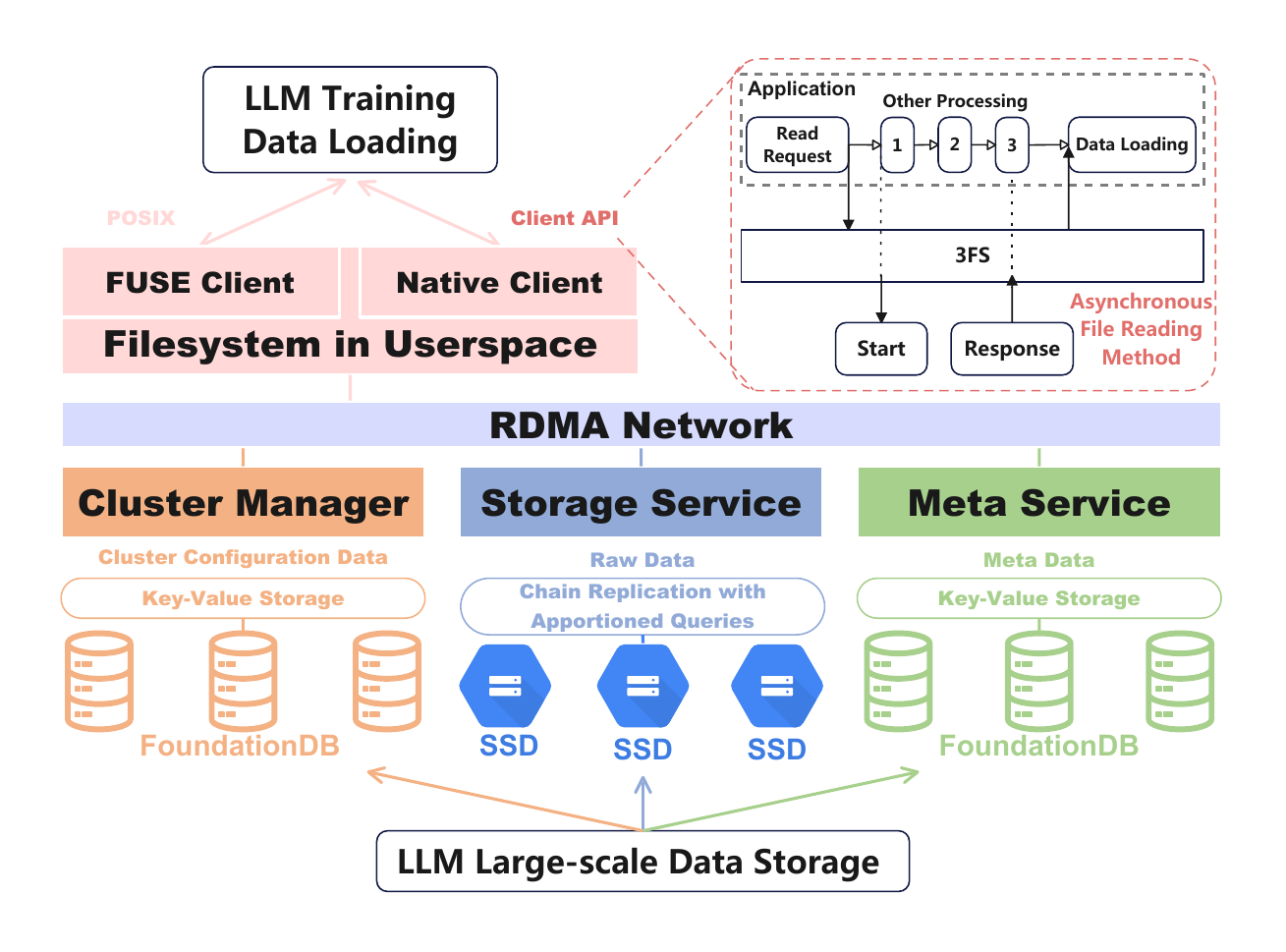}
    \caption{The storage architecture of 3FS~\cite{3FS}.}
    \label{fig:3FS}
\end{figure}

As shown in Figure~\ref{fig:3FS}, 3FS~\cite{3FS} employs a large number of SSDs for distributed data storage and uses the CRAQ algorithm to ensure data consistency. Specifically, a piece of data is saved as multiple same chunks, which together form a Chain. For read requests, they can be sent to any chunk in the Chain, and the chunk will return the data. For write requests, the writing operation is carried out sequentially on each chunk. When a certain chunk malfunctions, instead of using the incremental data generated during the abnormal period to overwrite the data as in traditional methods, it first moves the chunk to the end of the chain. Only when the chunk returns to normal will the entire content of other samples be copied to the abnormal chunk. These operations, while ensuring data consistency, will cause a certain delay in write operations. However, they have almost no impact on read operations, which are more important for \llm training.

Meanwhile, 3FS~\cite{3FS} discovers that in the context of LLM training, the File Cache significantly consumes system memory, thereby degrading overall I/O performance. To address this, 3FS adopts an asynchronous data loading approach, disables file caching and exclusively utilizes Direct I/O for data access, significantly reducing memory pressure. Moreover, it performs system-level alignment of buffer pointers, offsets, and lengths to satisfy Direct I/O requirements, thereby avoiding additional memory copies caused by user-side alignment operations.

\hi{Heterogeneous Storage Systems.}
Heterogeneous storage systems refers to deploying the model state across diverse storage media (e.g., GPUs, CPUs, NVMes Memory). When deploying the model, The Zero Redundancy Optimizer (ZeRO)~\cite{rajbhandari2020zero} deploys model states across multiple GPUs. However, simply distributing the model across multiple GPUs often significantly increases computational costs.

Some methods~\cite{rajbhandari2021zero,rhu2016vdnn,ZeROf-Offload,yang2024protrain}  alleviate GPU memory pressure by storing data in host memory or NVMe SSD.
vDNN~\cite{rhu2016vdnn} utilizes a per-layer memory management approach based on a sliding window that dynamically allocates memory at runtime based on the computational demands of the current layer. 
Its memory transfer mechanism includes both static and dynamic policies: the static policy offloads feature maps of all layers or only convolutional layers, while the dynamic policy determines which layers and convolutional algorithms to offload at runtime, balancing trainability and performance based on network characteristics. vDNN fully utilizes CPU memory by offloading intermediate feature maps that are not immediately needed and prefetching them prior to backpropagation. 
ZeRO-Infinity~\cite{rajbhandari2021zero} offloads model states to CPU (e.g. activations) and NVMe memory, effectively alleviating the GPU memory bottleneck. 
To further reduce memory pressure, it introduces a memory-centric tiling technique that lowers the working memory requirements for \llm training, enabling the execution of large operators without relying on model parallelism. 

However, both vDNN and ZeRO-Infinity only utilize CPU's memory without leveraging its computational capabilities. In contrast, ZeRO-Offload~\cite{ZeROf-Offload} retains the parameters and forward/backward computations on the GPU while offloading the remaining computations (such as optimizer calculations) to the CPU, thereby harnessing the CPU's computational power.

Unlike the aforementioned methods that often rely on manual parameter tuning (e.g., specifying offloading targets like CPU or NVMe), ProTrain~\cite{yang2024protrain} introduces a model- and hardware-aware automated framework. It incorporates a Memory-Aware Runtime Profiler for monitoring real-time memory and compute loads, partitions parameters into persistent (resident on GPU) and non-persistent (offloaded/loaded on demand) chunks based on their usage patterns, and reduces redundant data copying via pre-allocated chunk buffers.

\subsubsection{Data Organization}
\label{sec:RAGstorage}

Data organization refers to data operations (e.g., content organization in vector-based organization) during the storage stage that are designed to optimize retrieval accuracy and efficiency in RAG systems.
When \llm answers questions, issues like hallucination~\cite{ji2023survey} and lack of timeliness often arise.
To address these limitations, RAG~\cite{lewis2020retrieval} (e.g., vector-based retrieval and graph-based retrieval) have been introduced. They provide models with real-time, reliable context during inference. And both retrieval methods are based on the relevant data organization operations (e.g., vector-based organization and graph-based organization).

\begin{tcolorbox}[colback=gray!1,colframe=gray,lowerbox=visible]
  \textbf{Principles}
  \tcblower
Compared to traditional machine learning, LLMs require RAG knowledge to access real-time information. The main challenge is how to ensure both the efficiency and accuracy of retrieval. Current methods address this through vector-based and graph-based data organization techniques. However, existing RAG systems still fall short of meeting the high-quality retrieval demands at the enterprise level, where the document scale can reach millions of pages.
\end{tcolorbox}

\hi{Vector-Based Organization} Vector-based organization refers to converting data into vector form for efficient retrieval. It processes the original data through multiple stages (e.g., Content Organization, Chunking, Embedding, Compression and Storage).

\noindent\underline{(1)  Content Organization.}
For the source data, organizing the content can enhance its logical structure, thereby facilitating improved efficiency and accuracy in retrieval. Works like Dense x retrieval~\cite{chen2023dense}, APS~\cite{hosseini2024scalable} refine text into independent semantic units, which could be described as the minimal sentence that include all the necessary context information from the original text to express its meanings, and Thread~\cite{an2024thread} reorganizes documents into logical units, with each unit containing prerequisites, headers, body content, linkers (describing possible paths for next step), and metadata, enabling a logical and structured representation of the document's content, which significantly enhances the system’s logical coherence and processing efficiency especially in complex tasks (e.g.,  troubleshooting and dynamic operational workflows). 

Similarly, \cite{che2024hierarchical} organizes the content of scientific papers into a hierarchical tree structure, where the root node of the tree is the paper's title and child nodes are different sections, such as the introduction and methods. The relationship between parent and child nodes represents the global-local content relationships, such as the connection between the abstract and introduction. Then it traverses the paths from the root node to the leaf nodes to extract important contextual information.

\noindent\underline{(2)  Chunking.}
In vector-based retrieval, embedding long texts may reduce retrieval efficiency. Thus, an effective chunking strategy is required to divide the text into appropriately sized segments for encoding. The optimal chunk length needs to balance retaining fine-grained semantics and maintaining sufficient context, since a too long text might suffer from significant semantic compression during embedding, while too short a text would increase processing costs. 

Allowing overlap between consecutive chunks ensures that important information at the boundaries is not lost and the continuity of context is maintained. Different from traditional chunking, MoG~\cite{zhong2024mix} adopts a dynamic chunking strategy, which chunks data when building the knowledge base, where MoG dynamically determines the optimal granularity {(e.g., sentence-level, paragraph-level, or section-level)} of the knowledge source based on the input query through a trained router. The router, implemented as an MLP, assigns weights to different granularities to guide snippet selection. MoGG~\cite{zhong2024mix} extends MPG by converting reference documents into graphs and redefining granularity as hopping ranges, enabling effective retrieval of dispersed information for complex queries.

\noindent\underline{(3)  Embedding.}
In vector-based retrieval, the original input (text, images, audio, or other domains) is transformed into dense vector representations using models specifically adjusted for each data type. These representations encapsulate the underlying semantic meaning of the original content, and are then stored in a vector database for storage and retrieval. Various embedding models are used to correctly encode semantic information:

\noindent$\bullet$ \emph{BGE} uses a bilingual joint training framework that combines language-specific subword tokenization and specialized adaptation layers. This design aligns semantic representations across languages, improving cross-lingual retrieval accuracy~\cite{chen2024bge}.

\noindent$\bullet$ \emph{STELLA} features a cross-instance attention aggregation mechanism that explicitly captures inter-sentence dependencies during pretraining. Besides the general embedding model, STELLA offers an extra dialogue model in incomplete query situations where the user input has problems such as semantic omission and reference digestion. This reduces the embedding dimensions and inference latency, making it especially effective for large-scale tasks~\cite{stella}.

\noindent$\bullet$ \emph{GTE} introduces a dual-negative sampling strategy within its contrastive learning paradigm. Though introducing negative samples usually works in series of embedding models, this strategy incorporates more reverse contrastive terms within a fixed batch, strengthening the model's ability to distinguish subtle semantic differences.~\cite{li2023towards}.

\noindent\underline{(4)  Compression.}
Vector retrieval in \llms differs from regular vector retrieval in that semantically similar vectors are often high-dimensional, so dimensionality reduction techniques are needed to reduce storage pressure.

\noindent$\bullet$  \bfit{Linear Dimensionality Reduction.}
Locally-adaptive Vector Quantization (LVQ)~\cite{aguerrebere2023similaritysearchblinkeye} centralizes the data and scales each vector individually, calculating the quantization bounds adaptively in a localized manner, fully utilizing the quantization range to compress the vectors. This method is typically suitable for compressing vectors with around 100 dimensions, but it performs poorly when the vector dimension is very large, such as tens of thousands.

LeanVec~\cite{tepper2024leanvecsearchingvectorsfaster} combines linear dimensionality reduction with LVQ for vector compression. In ID(In-distribution) scenarios, LeanVec uses PCA, while in OOD(Out-of-distribution) scenarios, it introduces the LeanVec-OOD optimization method, which minimizes the square of the inner product between the query vector and the representation error to find the optimal projection subspace for both the dataset and the query set, thereby reducing the vector dimension. However, LeanVec is a simple linear dimensionality reduction method, and its performance may be affected in terms of accuracy when reducing the dimensionality drastically. 

LeanVec-Sphering~\cite{tepper2024gleanvecacceleratingvectorsearch} modifies the loss function, transforming the problem of finding the projection matrix into an optimization problem under the Mahalanobis distance, which allows for more effective discovery of the optimal projection matrix, thereby better preserving the similarity structure between vectors when processing high-dimensional vectors. 

\noindent$\bullet$  \bfit{Non-linear Dimensionality Reduction.}
GleanVec~\cite{tepper2024gleanvecacceleratingvectorsearch} uses spherical k-means clustering in the data partitioning stage to group vectors based on direction, capturing the data’s structural features. By associating cluster labels with vectors, it narrows the search range and reduces unnecessary calculations during inner product computation. In the local linear dimensionality reduction stage, GleanVec applies the LeanVec-Sphering method to reduce dimensionality within each cluster, preserving the inner-product relationship, which simplifies calculations while maintaining accuracy.

\noindent\underline{(5)  Storage.}
After the above steps, the data will be stored in vector form in a vector database. During \llm inference, the model vectorizes the input and uses similarity metrics such as cosine similarity or dot product to retrieve the most relevant data from the database.

{Faiss}~\cite{faiss}, when storing vectors, relies on the chosen index type. The Flat Index stores all vectors directly, such as IndexFlatCodes, which stores vectors in a flat array and supports sequential IDs. It is ideal for small datasets with high-precision requirements. The IVF Index clusters vectors with a coarse quantizer and stores them in inverted lists, supporting user ID operations and optionally using a DirectMap for efficient access. This reduces the search range and speeds up retrieval, making it suitable for large datasets. The PQ Index compresses vectors by splitting them into sub-vectors and quantizing them with a k-means quantizer (e.g., PQ6x10), trading accuracy for reduced storage space, making it suitable for high storage demands and lower precision needs.

In the Milvus~\cite{milvus}, vector storage differs based on the number of vectors per entity. For single-vector entities, vectors are stored continuously without row IDs. Since vectors are sorted by row ID and have the same length, a vector can be directly accessed using its row ID, reducing storage overhead and improving query access efficiency. For multi-vector entities, vectors are stored in a columnar format. For example, for entities A and B, each with two vectors, the storage format is (A.v1, B.v1, A.v2, B.v2). This columnar storage enables more efficient data processing by vector dimension, facilitating batch operations and improving processing performance. 

Weaviate~\cite{weaviate} utilizes a graph data model to manage data entities, storing vectors as node attributes linked to these entities. For example, in the case of text data, vectors generated by a text embedding model are associated with their corresponding text entity nodes, enabling efficient graph traversal and multi-hop queries based on vector similarity. Additionally, Weaviate can store vectors alongside structured attributes. For instance, the vectors of e-commerce products, along with structured attributes such as price and category, are stored in the corresponding entity nodes. This allows for hybrid queries that combine vector similarity and structured attribute conditions, enhancing query flexibility and practicality. 

LanceDB~\cite{lancedb} uses a columnar storage format called Lance to store data. Compared to traditional Parquet formats, Lance introduces the concept of a table schema. A single row in LanceDB can store images, text, audio, video, and any number of vectors corresponding to different parts of the original data, and it can be dynamically updated. This makes LanceDB particularly suitable for storing multi-modal data. Currently, LanceDB is used for handling various RAG tasks.

\hi{Graph-Based Organization.} Unlike vector-based organization, which helps LLM find knowledge related to a user's query through fuzzy searching, graph-based data explicitly represents entities and their relationships, enabling the identification of precise matching information in the database.
We will introduce graph-based organization from two aspects: indexing and storage.

\noindent\underline{(1)  Indexing.}
In the indexing phase, it is necessary to establish an efficient indexing architecture to address the issue that directly retrieving raw triples is inefficient for complex queries such as multi-hop reasoning or path search, because the inherent sparsity in the graph structure often leads to significant query latency.

GraphRAG~\cite{edge2024local} adopts community clustering and hierarchical summarization strategies. It uses the Leiden algorithm to detect tightly connected subgraphs, called communities, in the knowledge graph. Then, it generates hierarchical summaries for each community. Once a certain element in a triple is retrieved, the index collects relevant community summaries and sends them for inference. For example, it can condense hundreds of triples related to "quantum mechanics" into a semantic summary: "Quantum mechanics is the fundamental theory describing the behavior of matter and energy at microscopic scales". 

Furthermore, LightRAG~\cite{guo2024lightrag} integrates deduplication functionality to identify and merge identical entities and relations from different paragraphs. In real-time update scenarios, LightRAG introduces the Delta Index mechanism, which builds local indexes only for newly inserted edges and entities, using background merging threads without the need for community reconstruction, significantly reducing overhead related to community detection compared to GraphRAG.

MiniRAG~\cite{fan2025minirag} proposes a semantic-aware heterogeneous graph indexing mechanism, integrating text chunks and named entities into a unified structure, reducing the reliance on large language models for complex semantic understanding. The low semantic calculating  requirement while deploying grants MiniRAG a more excellent performance on resource-constrained devices compared to other methods.

\noindent\underline{(2)  Storage.} Graph data is usually stored in graph databases in three models: property graph models~\cite{propertygraph}, RDF (Resource Description Framework) models~\cite{AmazonNeptune}, and multi-model~\cite{arangodb}.

 Neo4j, JanusGraph, and TigerGraph use property graph models~\cite{propertygraph} to store graph-based data. A property graph model consists of "nodes" and "edges," where both can contain attributes (key-value pairs). This model uses query languages like Cypher and GSQL, designed for relationship modeling and querying, making them highly suitable for complex relationship queries during RAG in LLMs. 

Amazon Neptune~\cite{AmazonNeptune} supports both property graph models and RDF models for graph-based data storage. The RDF model, based on triples (subject, predicate, object), represents entities, attributes, and relationships in a way that enhances knowledge reasoning. By combining these two models, Neptune can meet diverse knowledge storage needs, such as rapid queries and deep reasoning. 

ArangoDB~\cite{arangodb} uses a multi-model approach to store graph-based data. It supports multiple data models (e.g., document, key-value pair, graph), allowing the selection of appropriate storage and query methods depending on the requirements. This allows ArangoDB to store graph data (relationship information), document data (context or factual information), and key-value pairs (configuration or metadata) in the same database, facilitating LLMs to extract relationships from knowledge graphs while also retrieving document-type data (e.g., specific context information).


\subsubsection{Data Movement}
\label{subsubsec:DataTransmission}
Data movement refers to the process of moving data from storage nodes to computing nodes. This process can achieve high data movement performance by caching data. {Meanwhile, offloading data and operators to multiple nodes for computation can improve the speed of data preprocessing.} 
Additionally, the highest overall performance can be achieved by overlapping data storage and computation operations to jointly schedule storage and computing resources. 

\begin{tcolorbox}[colback=gray!1,colframe=gray,lowerbox=visible]
  \textbf{Principles}
  \tcblower
Compared to traditional machine learning, LLMs involve massive data transfers from storage nodes to compute nodes. The main challenge is how to accelerate the data moving rate. Current methods address this through data caching, compute-storage overlap, and data/operator offloading. 
\end{tcolorbox}

\hi{Caching Data} in advance can increase the data moving rate. However, if a fixed cache policy is used, in order to meet the IO requirements of training, the configured storage capacity often far exceeds that required for storing the dataset~\cite{TectonicShift}. Therefore, a dynamically adjustable cache policy is needed. Some methods~\cite{kumar2020quiver, gu2022fluid, TectonicShift} dynamically adjust the cache mechanism by analyzing the characteristics and requirements  of \llm jobs in real time.

{Quiver}~\cite{kumar2020quiver}
optimizes cache sharing strategies based on the following IO characters during model training: (1) data shareability (due to significant overlap in data access within and across jobs), (2) substitutability (the I/O order does not affect job correctness, enabling small caches to improve performance by substituting data and reducing thrashing), and (3) predictability (using mini-batch processing times to estimate job sensitivity to I/O performance for informed cache allocation).

Fluid~\cite{gu2022fluid} dynamically  adjusts cache capacity according to I/O conditions, optimizing the online training speed for each individual \llm job. Specifically, Fluid uses a coordinator to monitor the processes of LLM jobs. It calculates the number of samples within a specific time window based on the batch sizes fed back by the jobs, and thus obtains the real-time training speed. Subsequently, based on the concept of the TCP congestion control algorithm~\cite{TCP}, it adopts a trial-and-error approach to dynamically adjust the cache capacity. When the training speed increases, the cache capacity is increased according to the preset scaling-up factor and scaling step. Conversely, when the training speed decreases, the cache capacity is decreased according to the preset scaling-down factor and scaling step.

Meta proposes Tectonic-Shift~\cite{TectonicShift}, a hybrid storage architecture that integrates flash memory with the traditional HDD-based distributed file system Tectonic. Tectonic-Shift organizes data segments into buckets for storage in flash memory and determines segment admission and reinsertion by comparing bucket priorities (computed from both historical and predicted future access patterns) against dynamically adjusted thresholds. It also optimizes the segment size (e.g., 256 KB) of CacheLib~\cite{cachelib} to improve flash memory utilization.

\hi{Data/Operator Offloading} refers to offloading data preprocessing operations such as shuffling, sampling, and augmentation, to multiple devices in order to improve processing speed. Currently, data preprocessing pipelines (e.g., tf.data) are typically performed on the CPU, whose efficiency is often lower than the training speed achieved by Machine Learning (ML) accelerators like GPUs and TPUs. So enhancing the efficiency of data preprocessing to match the high-speed processing capabilities of ML accelerators has become a challenge~\cite{graur2024pecan}. 

Some research~\cite{graur2022cachew,audibert2023tf} offload data preprocessing tasks to remote CPU servers. {Cachew}~\cite{graur2022cachew} divides the input dataset of each job into independent subsets for processing by remote CPU nodes. Additionally, users can specify locations for caching and reusing data in the input pipeline. The scheduler makes decisions during runtime based on specific metrics and algorithms through automatic scaling and caching strategies. The automatic scaling strategy adjusts the number of worker nodes according to client-reported metrics. The automatic caching strategy compares the processing times of different cache locations and selects the optimal caching scheme.
The tf.data service~\cite{audibert2023tf} addresses input data bottlenecks by horizontally scaling CPU nodes and leveraging a coordinated read mechanism to mitigate straggler issues caused by input size variability in distributed training. Specifically, it is comprised of four key components: a dispatcher, a pool of workers, clients, and an orchestrator. {The dispatcher manages dataset assignment to workers using various sharding strategies, for example, the OFF strategy performs no sharding, the DYNAMIC strategy applies disjoint first-come-first-served sharding, and several static sharding strategies are also supported. Workers are responsible for actual data processing. Clients issue data processing requests to the workers. Orchestrator deploys the aforementioned three components as containers within the same Borg~\cite{tirmazi2020borg} unit.}

Although the above method of offloading to remote CPU servers can alleviate data stalls, the cost of remote CPUs is high, and the resources of ML accelerator nodes are not fully utilized. {Pecan}~\cite{graur2024pecan} introduces two strategies, {\it AutoPlacement} and {\it AutoOrder}, to alleviate input data preprocessing bottlenecks and reduce training costs. The {\it AutoPlacement} strategy dynamically schedules data preprocessing workers across ML accelerator hosts and remote CPU servers. It first establishes a baseline batch processing time for model training, incrementally adds local workers, and then prunes redundant remote workers to determine the optimal combination of local and remote resources. The {\it AutoOrder} strategy analyzes the transformation operations within the input data pipeline, reordering them to place data-reducing transformations {(such as sampling, filtering, or image cropping)} earlier and data-expanding ones {(such as image padding and one-hot encoding)} later. While adhering to user-specified ordering constraints, this reorganization improves the preprocessing throughput of individual workers.

Different from the works that are only compatible with a single training framework as mentioned above (e.g., Cachew and tf.data service can only work with TensorFlow). Powered by native composable operators {(e.g., data loading, transformation, and filtering functions)}, Cedar~\cite{zhao2024cedar} can flexibly support different ML frameworks and libraries, enabling users to effortlessly build data pipelines. 

\hi{Overlapping of storage and computing} means that the data loading and computation processes in \llm training alternate. In LLM training, which proceeds in data batches, ideally the data loading unit can prepare the next batch while the computing unit processes the current one, reducing overall training time. However, if a data isn't cached locally, its need to load the data through remote I/O bandwidth. When this bandwidth is insufficient, computation pauses to wait for data loading, creating an IO bottleneck. Some researches optimize the pipeline at different training stages (e.g., the pre-training and SFT stage~\cite{zhao2023silod}, the RL stage~\cite{zhong2025optimizing}).

SiloD~\cite{zhao2023silod} leverages the characteristics of the pipelined execution of data loading and computation at the pre-training and SFT stage to build an enhanced performance evaluator. When data loading becomes the bottleneck, it uses a learned model (IOPerf) to quantify the cache and remote I/O demands of different training jobs,providing support for resource allocation in the pipelined of data loading and computation.

Compared with the pre-training and SFT stages, the RL stage requires an additional training of the reward model to evaluate the output of the original model. This leads to a greater amount of computational resources remaining idle (pipeline bubbles) during the RL stage.
RLHFuse~\cite{zhong2025optimizing} takes advantage of the independence between the original and reward models during the training stage to break the training task into sub-tasks of micro-batches. In the case of differences in the sizes and parallel strategies of the two models, it first transforms the problem to ensure that each stage of the two models uses the same number of GPU resources, and then uses the simulated annealing algorithm~\cite{kirkpatrick1983optimization} to generate a fused pipeline schedule.

\subsubsection{Data Fault Tolerance}
\label{datafault}
Data fault tolerance refers to the ability to quickly resume from the point of interruption during model training by storing checkpoints or performing redundant computations in the event of training interruptions.

\begin{tcolorbox}[colback=gray!1,colframe=gray,lowerbox=visible]
  \textbf{Principles}
  \tcblower
Compared to traditional machine learning, LLMs place greater emphasis on fault tolerance during training due to their large model sizes and the high cost of retraining. The main challenge is how to quickly resume normal training in the event of an interruption. Current methods address this by saving checkpoints or using redundant computation. 
\end{tcolorbox}

\hi{Checkpoints.} 
Some methods store the model state as checkpoints to handle training interruptions. However, restoring model states across multiple platforms or frameworks may encounter compatibility issues. At the same time, frequently saving model checkpoints can consume a large amount of storage space, especially during large-scale model training.

For compatibility issues, {PaddleNLP}~\cite{paddlenlp} has developed a unified model storage technology. It stores model weights, optimizer weights, and other data in a unified {\it safetensors} format, eliminating the need to differentiate distributed strategies during checkpoint storage. Specifically, when the distributed training strategy changes (e.g., switching between data parallelism and model parallelism) or the number of machines is adjusted, Unified Checkpoint enables training to resume using only a single complete checkpoint, without requiring separate checkpoints for each configuration.

\noindent\underline{(1)  Asynchronous Storage.} 
Apart from standardized checkpoint storage, for frequently saving model, some researches~\cite{CheckFreq, bytehdfs} aim to accelerate checkpoint saving through asynchronous storage without affecting the model's training speed. 

CheckFreq~\cite{CheckFreq}  employs a two-stage checkpointing technique designed to capture model state copies in memory for asynchronous storage while ensuring model parameter consistency through pipelining with subsequent iteration computations. Specifically, when idle GPU memory is available, it prioritizes snapshotting on the GPU to reduce costs; otherwise, it stores checkpoints in CPU memory and adjusts the checkpoint frequency accordingly.

In the training of \llms on the MegaScale system~\cite{bytehdfs}, HDFS is used to store the model state. When storing model states, there are problems of {\it balancing the checkpoint frequency and dealing with the HDFS bandwidth bottleneck} during model recovery in the training process. To address this, MegaScale adopts a two-phase storage approach: (1) GPU worker nodes quickly write the on-chip state to the host memory and continue training; (2) a background process asynchronously transfers the state to HDFS to reduce interference with training. When resuming training, a worker node in the specified data parallel group reads the shared state partition and broadcasts it to other nodes, reducing the HDFS load and alleviating bandwidth pressure.

\noindent\underline{(2)  Hierarchical Management} 
refers to storing model checkpoints across a multi-level storage system, storing the checkpoints that may be needed in the closer storage nodes, aiming to improve recovery speed.
{Gemini}~\cite{wang2023gemini} stores checkpoints in a hierarchical storage system composed of local CPU memory, remote CPU memory, and remote persistent storage. It introduces a near-optimal checkpoint placement strategy for CPU memory. By analyzing the relationship between the number of machines and checkpoint replicas, it flexibly adopts group placement or ring placement to maximize the likelihood of recovery from CPU memory in the event of failures. 
ByteCheckpoint~\cite{ByteCheckpoint} manages checkpoint files using an architecture combining SSD and HDD storage servers. New checkpoint files are stored as "hot" data on SSDs for quick access due to evaluation task downloads after creation. Once the evaluation is completed and there are no training anomalies, their access frequency drops, and they become "cold" data, being migrated to HDDs to free up SSD space and ensure the hot storage can efficiently store currently frequently accessed checkpoint files.

\hi{Redundant Computations}
Unlike checkpoint, some methods~\cite{bamboo,oobleck,recycle} are based on parallel computing and redundantly compute the state data of the model, enabling quick recovery of the training state from non-failed nodes in case of failures.

Inspired by the RAID disk redundancy technology~\cite{RAID}, Bamboo~\cite{bamboo} enables each computing node to perform computations not only on the neural network layers it is responsible for, but also on some layers of its neighboring nodes as redundant computations. When a node is preempted, its predecessor node has all the information required for training, allowing the training to continue without wasting previous computational results.

Unlike Bamboo's node-based redundant computation, Oobleck~\cite{oobleck} uses pipeline templates to define training pipeline execution, specifying node allocation, stage numbers, and model layer-GPU mappings. During training, at least \(f + 1\) logically-equivalent yet physically-heterogeneous pipelines are instantiated from these templates, considering the fault tolerance threshold \(f\) and batch size. When a pipeline node fails, Oobleck leverages other pipelines' model state redundancy and reinstantiates the pipeline to resume training.

Unlike Bamboo and Oobleck, which use pre-set redundant computations in standby, ReCycle~\cite{recycle} leverages the computational redundancy inherent in parallel training to reassign the tasks of failed nodes to nodes with the same processing in other data-parallel groups. This unique approach enables quick resumption of training without the need for spare servers.

\subsubsection{KV Cache}
\label{KVCache}
\llms use auto-regressive generation, where each token depends on prior ones. KV Cache avoids redundant computation by reusing stored key-value pairs, improving efficiency. However, its memory grows with sequence length, making efficient cache management crucial.

\begin{tcolorbox}[colback=gray!1,colframe=gray,lowerbox=visible]
  \textbf{Principles}
  \tcblower
Compared to traditional machine learning, LLMs require KV cache to accelerate inference. The main challenge lies in efficiently managing the cache as the KV size grows rapidly. Current methods address this by indexing KV, shrinking KV, and managing KV placement or cache space. 
\end{tcolorbox}

\hi{Cache Space Management} refers to separating the logical structure of the KV cache from its physical storage implementation, which facilitates memory allocation and improves memory utilization.
vLLM~\cite{vllm} and vTensor~\cite{vtensor} divide the KV cache into fixed-size blocks and store them in a non-contiguous manner. vLLM manages these blocks through a mapping mechanism, while vTensor stores the fixed-size KV cache blocks non-contiguously in physical memory. This decouples the logical and physical KV blocks, utilizing a block table to manage dynamic memory allocation by tracking the mapping relationships and fill states.

\hi{KV Placement}
refers to using a perception strategy to store frequently used KV in faster storage media (such as GPU memory), while storing less frequently used KV in slower storage media (such as SSD), or releasing them directly.
RAGCache~\cite{jin2024ragcache} provides a prefix-aware PGDSF replacement policy that prioritizes cache nodes based on access frequency, size, and recomputation cost. And stores frequently accessed data in fast GPU memory and less frequent data in slower host memory, maximizing cache efficiency. 
CachedAttention~\cite{gao2024cost} leverages the inference job scheduler to observe the jobs waiting for execution. To improve cache efficiency, the KV cache of a pending job is prefetched into the host memory from disk before execution. Meanwhile, KV caches that are no longer required are evicted, based on the jobs waiting to be executed.

\hi{KV Shrinking}
KV Cache Shrinking refers to trimming or reducing the KV Cache in order to lower memory usage and improve inference efficiency.
CacheGen~\cite{liu2024cachegen} uses a customized tensor encoder to encode the KV cache into a more efficient bitstream, thereby reducing bandwidth usage. It also compresses the KV cache using techniques such as block-based encoding, hierarchical quantization, and arithmetic encoding, while dynamically adjusting the compression level and transmission method based on network conditions to ensure low latency and high generation quality. 

Unlike CacheGen, which only considers intra-layer redundancy, MiniCache~\cite{liu2024minicache} is based on the similarity of KV cache states in adjacent layers. It decomposes the state vectors into magnitude and direction components, calculates the direction vectors using SLERP~\cite{SLERP}, and merges the KV caches of adjacent layers to form a merged cache that contains information such as direction vectors, magnitudes, and angles.

Compared with the traditional method of storing the complete KV data, HCache~\cite{gao2024faststaterestorationllm} only stores the hidden states (the size of the hidden states is only half that of the KV cache, and recomputing the KV cache from the hidden states can reduce the computational load). When restoring the state, a bubble-free restoration scheduler is used to concurrently execute the transmission of hidden states and the recomputation from hidden states, maximizing the overall resource utilization.

\hi{KV Indexing}
refers to the process of constructing an indexing architecture for the KV Cache to accelerate the query process of the KV Cache.
{ChunkAttention}~\cite{ye2024chunkattention} organizes the KV cache into a prefix tree using a prefix-aware KV cache (PAKV), sharing key-value tensors of common prefixes to accelerate the corresponding KV query process.
~\cite{zheng2024batchllm} proposes Prefix Sharing Maximization (PSM): By dynamically reordering data columns and rows, it maximizes prefix sharing among requests to improve cache hit rates. Column Reordering sorts columns based on value frequency and size, prioritizing those with more shared prefixes. Row Sorting groups requests with identical prefixes together, further enhancing cache reuse.

\subsection{Data Serving for \llm}
\label{sec:serving}

Data service encompasses data preprocessing operations carried out after data is transferred from storage to computing nodes and before its actual utilization by the LLM, aiming to facilitate more effective data consumption by the LLM. These data preprocessing operations include: data shuffling, data compression, data packing, and data provenance.

\subsubsection{Data Shuffling}
\label{sec:DataSelection}

Data shuffling in data serving means that different data needs to be selected and provided to \llms at various stages (e.g., in different epochs for pretraining). For example, corresponding training data needs to be supplied according to the training requirements during the training stage; during the RAG stage, corresponding knowledge needs to be supplied based on the degree of relevance to the questions.

\begin{tcolorbox}[colback=gray!1,colframe=gray,lowerbox=visible]
  \textbf{Principles}
  \tcblower
Compared to traditional machine learning, LLM applications are divided into multiple stages, each requiring different types of data to be fed into the model. The main challenge is how to select data that meets the specific requirements of LLMs. In the training stage, current methods provide training data by scoring based on data samples or model states, or by using empirical training strategies. In the RAG stage, data is selected through metrics, rules, or models to supply relevant knowledge to the LLM.
\end{tcolorbox}

\hi{Data Shuffling for Training.}
As \llms continuously trained over new tasks, it may begin to lose its ability to retain early task knowledge, a phenomenon known as catastrophic forgetting~\cite{mccloskey1989catastrophic, mcclelland1995there}. To address this, some data supply methods are employed to manage datasets during the training process and provide high-quality data. Meanwhile, some methods, instead of altering the dataset, propose reasonable learning strategies.

\noindent\underline{(1)  Data Pruning.} Data pruning refers that during the training process, partial shuffling is carried out on the training dataset, and high-quality data is retained, so that the model is trained on the data that has not been fully learned and is of high quality.

\noindent\bfit{Sample Scoring.} Some methods~\cite{fayyaz2022bert,attendu2023nlu} prune datasets by scoring samples, selecting high-scoring samples for subsequent training. \cite{fayyaz2022bert} applies the EL2N metric to identify important examples in a dataset, written as $\chi(x_i, y_i) = \mathbb{E} \| f(x_i) - y_i \|_2$, where $f(x_i)$ is the model's prediction and $y_i$ is the true sample. Based on the computed EL2N values, it periodically prunes irrelevant data during training. \cite{attendu2023nlu} extends the EL2N metric to evaluate sample importance, written as $\hat{\chi}_{ema}(x, y) \leftarrow \alpha \cdot \hat{\chi}_{nlu}(x, y) + (1 - \alpha) \cdot \hat{\chi}_{ema}(x, y)$, where $\alpha$ is a smoothing parameter. Based on extended EL2N values, it periodically selects data subsets for training.

\noindent\bfit{Model State Scoring.} Unlike the aforementioned approach of scoring samples and prune the dataset, some methods~\cite{tan2024data,albalak2023efficient,wu2024mixture,luo2024velocitune} prune the distribution of dataset by scoring the model's state (such as training loss and learning status). 

Moving-one-Sample-out (MoSo) \cite{tan2024data} identifies and selects the most informative \llm pre-training samples by assessing the influence of a specific sample on the training loss. The MoSo score measures how the training loss over the dataset $\mathcal{S}$ excluding $z$ (i.e., $S \setminus z$) would change when the sample $z$ is removed.
This approximation measures the agreement between $z$ and $S \setminus z$, where the sample is considered important and receives a higher score if the gradient of $z$ is consistently aligned with the average gradient. 

Similarly, Velocitune~\cite{luo2024velocitune} is a dynamic domain weight adjustment method based on learning velocity, which is defined as $V_t[i] = \frac{\ell_t[i] - \ell_{\text{target}[i]}}{\ell_{\text{init}[i]} - \ell_{\text{target}[i]}}$, where $V_t[i]$ denotes the learning velocity for domain $i$ at step $t$, $\ell_t[i]$ is the current loss for domain $i$, $\ell_{\text{target}[i]}$ is the target loss for domain $i$, predicted by the scaling law~\cite{kaplan2020scaling}, $\ell_{\text{init}[i]}$ is the initial loss for domain $i$, calculated before training starts. The method calculates the learning velocity of each domain and dynamically adjusts the sampling weights, giving more attention to domains with slower learning progress, thereby achieving a balanced learning effect. 

Some methods~\cite{albalak2023efficient,wu2024mixture} combine reinforcement learning based on scoring the model to adjust the dataset.
ODM~\cite{albalak2023efficient} is based on the multi-armed bandit algorithm. It regards each data domain as an arm and uses classical reinforcement learning methods. By taking the training loss as the reward function, it optimizes the data mixing ratio online to adapt to training dynamics. That is, it dynamically adjusts the sampling weights of each data domain and preferentially selects data with high information gain and large losses. 

MOS~\cite{wu2024mixture} proposes a {\it scoring network} that dynamically adjusts the sampling probabilities of different datasets based on the model's current learning state, combined with reinforcement learning, to alter the distribution of training data. This adjustment is guided by three reward functions: \noindent\textit{$(i)$ Transferability} for measuring the similarity (e.g, cosine distance) between datasets as the reward. \noindent\textit{$(ii)$ Learning difficulty} for measuring the perplexity changes. \noindent\textit{$(iii)$ Learning trajectory} for smoothing the reward values using Exponential Moving Average (EMA) to more stably optimize the sampling distribution.

\noindent\underline{(2)  Training Strategy.} 
In addition to directly prune the dataset during training, appropriate learning strategies can also alleviate catastrophic forgetting. ~\cite{dong2023abilities} found that {different abilities vary with data volume, with mixed data improving abilities at low resources and causing conflicts at high resources.} Thus, DMT~\cite{kim2024strategic} is proposed, which first fine-tunes on a specific dataset and then fine-tunes on mixed data to effectively balance general and specialized abilities and mitigate conflicts and forgetting. It proposes a strategy where training data are sorted based on criteria like input length, attention weights and training loss, allowing the model to gradually learn from simple tasks to more complex ones.

\hi{Data Selection for RAG.}
In the RAG stage, it is necessary to retrieve the stored knowledge (see details in \ref{sec:RAGstorage}) and provided the retrieved results to the \llm. During this process, it needs to ensure the effectiveness of the retrieved results in order to obtain better answers from the \llm~\cite{Ma_2023}. Currently, the retrieval quality is mainly guaranteed through RAG knowledge filtering and RAG knowledge re-ranking.

\noindent\underline{(1)  RAG Knowledge Filtering.}
RAG knowledge filtering refers to filtering out documents with poor relevance after retrieval. Some methods~\cite{Ma_2023,DBLP:journals/corr/abs-2306-16092,chang2024mainragmultiagentfilteringretrievalaugmented} use a model as a judge to filter documents. \cite{Ma_2023} uses small language models (SLMs) as filters, performing preliminary predictions and evaluating difficulty. For easy samples, the SLM's predictions are used as the final decision; for difficult samples, the top N most likely labels are selected from the SLM's predictions for subsequent re-ranking. In Chatlaw~\cite{DBLP:journals/corr/abs-2306-16092}, after retrieving relevant information, the \llm evaluates the retrieved content. Only content that is deemed highly relevant after evaluation is used to generate the final response, effectively reducing interference from irrelevant or incorrect information. MAIN-RAG~\cite{chang2024mainragmultiagentfilteringretrievalaugmented} collaboratively filters and scores retrieved documents by leveraging multiple LLM agents to enhance relevance and reduce noise. The framework adopts a dynamic filtering mechanism that uses score distributions to adjust relevance thresholds, ensuring high recall of relevant documents while minimizing computational overhead.

\noindent\underline{(2)  RAG Knowledge Re-ranking.} 
After filtering, multiple documents may remain, requiring re-ranking of the retrieval results to place the most relevant ones at the top for more accurate model output. Research on~\cite{eibich2024aragog} shows that using a large model for re-ranking performs better than methods like Maximum Marginal Relevance (MMR) and Cohere re-ranking. For large model re-ranking, general-purpose large language models (e.g., GPT) can be used directly, or specialized zero-shot re-ranking models such as Cohere rerank ~\cite{cohere} or RankVicuna~\cite{pradeep2023rankvicuna} can be employed. The latest ASRank~\cite{abdallah2025asrank} leverages pre-trained \llm to compute the matching probability between document answers and answer cues, scoring and re-ranking the retrieved documents.

\subsubsection{Data Compression}
\label{sec:DataCompression}
Data compression refers to compressing the input data for the model.
Previous studies have shown that prompts are crucial for triggering \llm domain-specific knowledge, and prompts are typically designed based on specific tasks (including chain-of-thought, context learning, and historical dialogues). As the complexity of chain-of-thought, context learning, and RAG increase, longer prompts are required~\cite{jiang2023llmlingua}. However, overly long prompts may lead to higher response latency, increased costs, and even exceeding the maximum token limit. Existing methods mainly compress the model inputs in two aspects. Some methods~\cite{xu2023recomp,cheng2024xrag,shi2024compressing,jung2024familiarity,rau2024context} compress the retrieved results in the RAG stage and then put them into the prompt, while other methods compress the entire prompt~\cite{jiang2023llmlingua,jiang2023longllmlingua,pan2024llmlingua,mu2024learning,chevalier2023adapting}.

\begin{tcolorbox}[colback=gray!1,colframe=gray,lowerbox=visible]
  \textbf{Principles}
  \tcblower
Compared to traditional machine learning, LLMs often require longer inputs, and in some cases, the input must be compressed to fit into the model. The main challenge is how to compress the input without losing important information. Current methods mainly achieve this through compression based on information entropy, rule-based templates, or model-driven approaches.
\end{tcolorbox}

\hi{RAG Knowledge Compression}
The retrieved RAG knowledge can be compressed by a model to make small texts carry more information. Techniques like RECOMP~\cite{xu2023recomp}, CompAct~\cite{shi2024compressing}, and FAVICOMP~\cite{jung2024familiarity} adopt rule-based RAG context compression schemes, where predefined rules or templates explicitly guide the model to extract key information and remove redundant content. Alternatively, methods like xRAG~\cite{cheng2024xrag} and COCOM~\cite{rau2024context} use soft prompt-based RAG context compression schemes, where learnable parameters (such as the modality projector W in xRAG or the overall model training in COCOM) enable implicit vector learning. These implicit vectors dynamically adjust attention weights when the model processes input, allowing the model to adaptively optimize context representations under context compression.

\hi{Prompt Compression.} 
Prompt compression means that after the retrieved knowledge is put into the Prompt, the entire Prompt will be compressed.

\noindent\underline{(1) Metric-Based Compression.} Some studies~\cite{jiang2023llmlingua,jiang2023longllmlingua}, based on the hypothesis that a vast amount of knowledge is stored in the model parameters, have proposed methods to compress prompts while minimizing information loss.  
LLMLingua~\cite{jiang2023llmlingua} uses a perplexity criterion to remove redundant tokens from the original prompt. By quantifying the negative logarithmic probability (perplexity) of each token through a small model, LLMLingua identifies and removes tokens that can be predicted from the model's inherent knowledge, thereby shortening the prompt while retaining essential context. 

LLMLingua's extended version, LongLLMLingua~\cite{jiang2023longllmlingua}, uses a dual-granularity compression strategy: 
$(i)$ Coarse-grained compression initially filters key information at the document level to provide more focused content for fine-grained compression; 
$(ii)$ Fine-grained compression further optimizes at the token level to precisely retain key information. 
These two strategies work together to improve the quality of the prompt and model performance. LongLLMLingua also assigns different ``compression budgets'' to documents based on their importance, aiming to achieve the best global compression effect. 

\noindent\underline{(2) Finetuned-Model-Based Compression.}
Unlike the aforementioned methods that use a small model's perplexity for compression, some methods~\cite{pan2024llmlingua,mu2024learning,chevalier2023adapting} directly perform the compression task end-to-end by fine-tuning a model.
LLMLingua-2~\cite{pan2024llmlingua} defines prompt compression as a problem of classifying tokens and trains a dedicated model for compression. It uses a Transformer encoder to capture bidirectional contextual information, ensuring that the compressed prompt is faithful to the original. \cite{mu2024learning} proposes a technique called 'gisting', where a language model is trained to condense the prompt into a compact 'gist token'. These tokens encapsulate the core semantic content of the prompt and can be cached for later use. This method achieves a compression rate of up to 26 times. ~\cite{chevalier2023adapting} suggests a method to transform pre-trained language models into AutoCompressors. The AutoCompressor compresses long contexts into summary vectors, and training is performed on the model parameters using these summary vectors.

\subsubsection{Data Packing}

Data Packing aims to address the requirement for uniform sequence lengths in \llms' training inputs, which combines short texts in an appropriate way to enhance text coherence and reduce the number of padding tokens. In this way, we can avoid the excessive truncation caused by the drawbacks of simple concatenation and splitting methods~\cite{ding2024fewer}.

\hi{Short Sequence Insertion.} Some methods~\cite{ding2024fewer,liu2024bucket} involve inserting short sequences into long sequences to minimize padding. The Best-fit Packing~\cite{ding2024fewer} first splits long documents according to the model's context length, then sorts all document blocks in descending order of length. For each document block, it selects the training sequence set with the smallest remaining capacity that can accommodate it. \cite{liu2024bucket} prioritizes long documents and uses a greedy algorithm to fill remaining space with short document segments (sequences), reducing padding and minimizing document concatenation to lower contextual noise.

\begin{tcolorbox}[colback=gray!1,colframe=gray,lowerbox=visible]
  \textbf{Principles}
  \tcblower
Compared to traditional machine learning, \llms place higher demands on the semantic quality of training data. Additionally, due to the requirement for uniform input lengths, a key challenge is maintaining semantic integrity without excessive truncation. Existing techniques tackle this through short-sequence insertion, sequence concatenation, and semantic-aware composition. However, it remains crucial to account for the impact of these data packaging operations on overall training efficiency.
\end{tcolorbox}

\hi{Sequence Combination Optimization.} Some methods~\cite{krell2021efficient,pouransari2024dataset} optimize sequence combinations for efficient packing. \cite{krell2021efficient} proposes two efficient sequence packing algorithms: 
(1) The Shortest Pack First Histogram Packing (SPFHP) uses a sequence length histogram, sorts sequences from long to short, and applies a worst-fit algorithm to prioritize placing the histogram intervals into the remaining largest ``packs'', while limiting packing depth to avoid creating excessive small packs, thus improving space utilization. 
(2) The Non-Negative Least Squares Histogram Packing (NNLSHP) converts the packing problem into a non-negative least squares problem, using dynamic programming to enumerate reasonable sequence combination strategies, constructing a packing matrix to determine the strategy's repetition count. It also assigns small weights to short sequences' residuals to reduce long sequence leftovers, achieving efficient packing. \cite{pouransari2024dataset} splits documents into multiple fixed-length ``buckets'' based on their length, ensuring that each sequence comes from the same document to avoid cross-document attention issues. Additionally, by combining Variable Sequence Length Curriculum (VSL), different lengths of sequences are dynamically sampled during training to maintain a consistent total token count.

\hi{Semantic-Based Packing.} Some methods~\cite{staniszewski2023structured,shi2023context} improve data coherence through semantic-based data packing. \cite{shi2023context} reorders pretraining data by combining semantically related documents into coherent input contexts, allowing the \llm to read and reason across document boundaries. Similarly, SPLICE~\cite{staniszewski2023structured} randomly selects a document as the root document, and in a breadth-first manner, uses retrieval methods like BM25 and Contriever (trained from a mix of Wiki and CCNet data) to retrieve $k$ similar documents, adding them to the training sample until the maximum length is reached. Finally, the tree structure is flattened using a specific tree traversal strategy to generate the training example.

\subsubsection{Data Provenance}

Data Provenance is the process of tracking the sources, transformations, and lineage of data,  which is increasingly recognized critical in ensuring the reliability, transparency, and accountability of \llm data~\cite{DBLP:journals/jcs/AlamW21}.

\begin{tcolorbox}[colback=gray!1,colframe=gray,lowerbox=visible]
  \textbf{Principles}
  \tcblower
Compared with traditional machine-learning models, LLMs demand heightened safeguards for output security owing to their powerful generative capabilities. The central challenge is to preserve output integrity without degrading quality. Current solutions embed watermarks or deploy statistical-detection techniques to reveal any tampering.
\end{tcolorbox}

\hi{Embedding Markers.} Current data provenance methods~\cite{zhou2024bileve,christ2024undetectable,liu2023unforgeable,kirchenbauer2023watermark} generally modify the generation logic to embed covert markers into the text. This is done in a way that does not disrupt the text itself, thereby providing a medium for tracing the origin of the data. 

Bileve~\cite{zhou2024bileve} enhances the traceability and integrity of text by embedding two distinct levels of signals:  \bfit{(1) Statistical signal} embedded globally to detect whether the text originates from a specific model. \bfit{(2) Content-related signature} embedded within each generation unit to verify if the text has been tampered with. During detection, the validity of the signature is first verified; if the signature is invalid, a statistical test is then used to determine whether the text comes from the target model. 

Unlike Bileve that emphasizes strict traceability after text tampering, \cite{christ2024undetectable} focuses on embedding watermarks in a way that preserves the quality of the generated output. It embeds hidden markers that can only be detected by individuals possessing a specific key, while remaining imperceptible to others that the text has been altered. Specifically, the method employs a pseudo-random function (PRF, used to generate seemingly random numbers) to determine the shuffling of each output word, ensuring that the generated text is statistically indistinguishable from the original model's output. During detection, the presence of hidden markers is ascertained by calculating a score for each word in the text (based on the numbers generated by the pseudo-random function). 

Unlike previous approaches, UPV ~\cite{liu2023unforgeable} introduces a watermarking method that enables detection without requiring access to the key used during generation, thereby eliminating the risk of key leakage. It employs two independent neural networks for watermarking. During text generation, the watermark generation network utilizes an embedding module and a fully connected classifier to predict watermark signals based on token information within a sliding window, and accordingly adjusts the language model's output distribution. For detection, an LSTM-based network takes the text sequence as input and identifies the watermark, leveraging shared token embedding parameters with the generation network.

Compared to methods that require specific keys for detection, ~\cite{fairoze2023publicly} embeds a special type of watermark into text generated by language models, which can be detected by anyone without the need for any secret information. It selects specific lexical combinations (rejection sampling, ensuring that the embedding of the marker does not affect the naturalness of the text) during text generation, in conjunction with an error correction mechanism (error-correcting codes, allowing the marker to be recovered even after partial modification of the text), to embed an encrypted signature (public key signature, ensuring the non-forgeability of the marker) into the text. During detection, one only needs to extract these specific {\it lexical combinations} from the text and verify the validity of the signature to determine whether the text contains the marker.

\hi{Statistical Provenance.} Unlike the aforementioned methods that rely on detecting special markers for tracing the origin, ~\cite{kirchenbauer2023watermark} achieve data provenance through the statistical information of the vocabulary. Specifically, before generating each word, the model randomly divides the vocabulary into two parts (green-listed and red-listed tokens) and tends to favor the shuffling of green-listed tokens during the generation process (green-listed tokens are a randomly selected subset of the vocabulary). By employing statistical tests (a mathematical method used to determine whether text adheres to specific rules), it is possible to detect whether the proportion of green-listed tokens in the text is abnormal, thereby ascertaining if the text is machine-generated.

\section{LLM for Data Management}
After preparing the \llms with carefully processed / stored / served data, we next introduce the \llm techniques that can be utilized to enhance data management tasks, including data manipulation, data analysis, and data system optimization.



\begin{figure}[!t]
    \centering
    \includegraphics[width=1\linewidth]{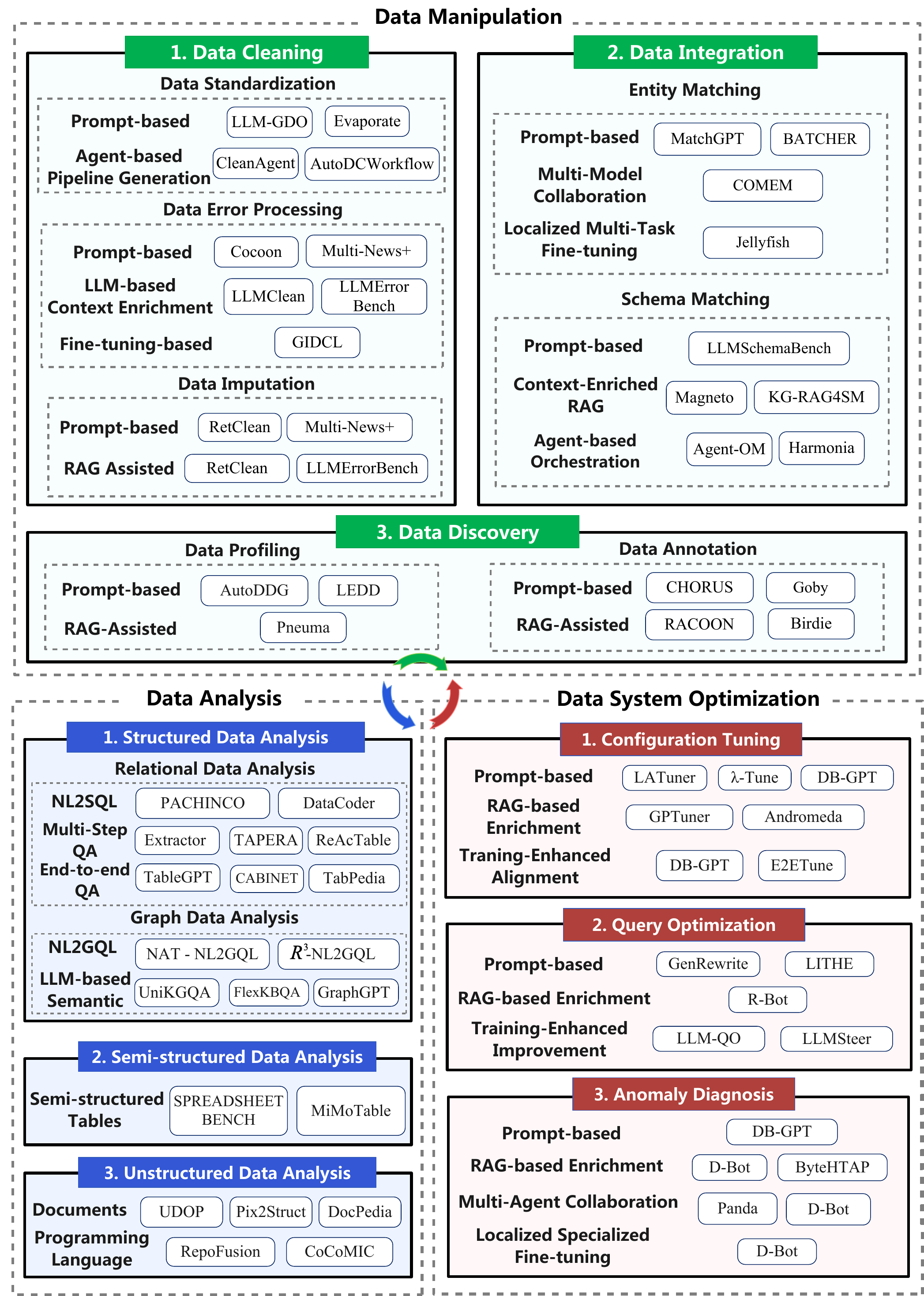}
    \vspace{-.5cm}
    \caption{Overview of \llmdata Techniques.}
    \label{fig:frameworkLLM4Data}
    \vspace{-.5cm}
\end{figure}

\subsection{LLM for Data Manipulation}
\label{subsec:manipulation}

\llm can be employed to explore and prepare appropriate data for non-\llm-oriented tasks, such as data cleaning for classification tasks, data integration for extracting well-structured tables from unstructured sources, and data discovery for identifying relevant datasets.
Unlike data preparation pipelines designed specifically for \llm applications, these methods focus on enhancing the quality and utility of data for downstream analytical or machine learning tasks.




\subsubsection{\llm for Data Cleaning}
\label{subsubsection:cleaning}

Data cleaning focuses on transforming corrupted or low-quality data into a reliable form suitable for downstream applications (e.g., statistical analysis or training machine learning models).
It encompasses a range of tasks such as handling missing values, correcting typos, resolving formatting inconsistencies, and addressing dependency violations.
These tasks are typically categorized into data standardization, error detection and correction, and data imputation.

Traditional data cleaning methods depend on rigid rules and constraints (e.g., zip code validation), demanding substantial manual effort and domain expertise (e.g., schema knowledge in financial data)~\cite{AutoDCWorkflow, GIDCL}.
Additionally, they often require domain-specific training, which restricts their generalizability~\cite{Evaporate}.
Recent studies show that large language models (\llms) can address these limitations by offering natural language interfaces that reduce manual and programming effort, eliminate the need for complex runtime environments, and support seamless integration of domain knowledge. These methods primarily target the following tasks.





\noindent \textbf{Data Standardization.}
Data standardization involves converting diverse, inconsistent, or non-conforming values into a consistent format to ensure reliable analysis and effective downstream processing.
Existing methods use either structured \llm prompting for specific cleaning operations or \llm agents for automated pipeline generation.

\noindent \bfit{(1) Prompt Based End-to-End Standardization.}
The first approach constructs well-structured prompts with explicit standardization instructions and employs advanced prompting techniques (e.g., Chain-of-Thought) to improve the effectiveness of \llm-based standardization methods.
For example, LLM-GDO~\cite{LLMGDO} utilizes user-defined prompts (UDPs), including in-context learning examples, to implement \llm-based operators that replace traditional user-defined functions (UDFs) across various standardization tasks (e.g., normalizing numerical values).
This method simplifies logic implementation and facilitates the seamless integration of domain knowledge.
Evaporate~\cite{Evaporate} employs \llms to transform semi-structured documents into structured views through two main strategies: (i) \textit{Evaporate-Direct}, which prompts the \llm to extract values directly, and (ii) \textit{Evaporate-Code}, which guides the \llm to synthesize extraction code and ensembles multiple candidate functions using weak supervision to improve output quality while maintaining low cost.




\noindent \bfit{(2) Agent Based Operation and Pipeline Generation.}
To address the inefficiencies of \llm-based solutions, such as the reliance on multi-turn prompts and expert-level prompt engineering, the second method employs \llm agents to automatically generate cleaning operations and orchestrate end-to-end pipelines.
For instance, CleanAgent~\cite{CleanAgent} integrates domain-specific APIs with autonomous agents to execute a standardization pipeline that includes API call generation (e.g., \texttt{clean\_date(df, ``Admission Date'', ``MM/DD/YYYY'')}) and iterative code execution.
Similarly, AutoDCWorkflow~\cite{AutoDCWorkflow} adopts \llm agents to construct pipelines for resolving duplicates and inconsistent formats.
The agent performs step-by-step reasoning to identify relevant columns, evaluate data quality, and generate appropriate operations (e.g., \texttt{upper()} and \texttt{trim()}), while leveraging tools such as OpenRefine for execution and feedback.

\noindent \textbf{Data Error Processing.}
Given a data entry, error processing typically involves two steps: detecting erroneous values and correcting these values.
Typical errors include typos, invalid formats, type mismatches, numeric outliers, and dependency violations.
Existing methods generally fall into two categories: employing \llms for direct end-to-end error processing, or enhancing context models to better guide the detection and correction process.




\noindent \bfit{(1) Prompt Based End-to-End Error Processing.}
To support end-to-end data error processing, the first approach employs prompting techniques to either directly handle data errors or generate the corresponding processing functions.
For instance, $\text{Multi-News}^{+}$~\cite{MultiNews} employs Chain-of-Thought (CoT) prompting, majority voting inspired by human annotation practices, and self-consistency checks to enhance classification accuracy and transparency when processing noisy documents.
Similarly, Cocoon~\cite{Cocoon} constructs semantic detection prompts and divides datasets into batches, allowing the \llm to analyze sampled values (e.g., 1,000 entries per column) and identify typos or inconsistencies (e.g., ``mapping English'' $\rightarrow$ ``eng''), thereby supporting {batch-wise} data cleaning.
GIDCL~\cite{GIDCL} adopts a creator-critic framework in which the \llm iteratively refines lightweight error detection models and generates pseudo-labeled data using handcrafted prompts and in-context examples to produce both detection and correction functions, further enhanced by structural correlation learning with Graph Neural Networks (GNNs).


\noindent \bfit{(2) \llm Based Cleaning Context Enrichment.}
To address the inefficiencies and limited scalability of manual cleaning context model construction in dynamic environments, the second approach leverages \llms to enrich data cleaning context models and more effectively capture semantic relationships within the data.
For example, LLMClean~\cite{LLMClean} proposes an automated \llm-based method for generating context models by extracting ontological functional dependencies (OFDs) using both prompt ensembling and fine-tuned \llms (e.g., Llama-2).
The extracted OFDs are then used to identify data errors (e.g., value inconsistencies) and guide \llm-based repairs through iterative feedback from integrated correction tools such as Baran.
LLMErrorBench~\cite{LLMErrorBench} employs \llm agents equipped with Python (via IPython) and prompted with task-specific instructions and contextual hints (e.g., error locations) to explore, modify, and repair datasets iteratively.
Corrections (e.g., value replacement, missing data handling) are guided by performance feedback from pre-defined code execution and evaluation pipelines.

\noindent \bfit{(3) Fine-tuning Based End-to-End Error Processing.}
To improve error correction accuracy while preserving computational efficiency and model adaptability, the third approach fine-tunes \llms to capture dataset-specific patterns and dependencies that are typically difficult to model through prompting alone.
For example, GIDCL~\cite{GIDCL} fine-tunes a local \llm (e.g., Mistral-7B) using Low-Rank Adaptation (LoRA) to optimize error correction, constructing training data from labeled tuples and pseudo-labeled tuples generated via \llm-based augmentation, with each training instance formatted as a context-enriched prompt comprising: (i) an instruction (e.g., ``Correct the ProviderID to a valid numeric format''), (ii) a serialized erroneous cell with row and column context (e.g., ``\texttt{<COL>ProviderID<VAL>1x1303...}''), (iii) in-context learning demonstrations (e.g., ``bxrmxngham $\rightarrow$ birmingham''), and (iv) retrieval-augmented examples from the same cluster (e.g., clean tuples via k-means).

\noindent \textbf{Data Imputation.}
Given a data entry with missing attribute values (e.g., {NULL}), data imputation aims to infer the missing values using available contextual information accurately.
Existing methods either (i) use structured prompts to convey contextual hints to \llm, or (ii) apply retrieval-augmented generation (RAG) to integrate relevant external data.

\noindent \bfit{(1) Prompt Based End-to-End Imputation.}
To incorporate contextual information for imputing missing values, the first approach constructs structured prompts.
For example, RetClean~\cite{RetClean} enhances \llm effectiveness by serializing each tuple into a formatted representation (e.g., ``[Name: John; Age: 25; Gender: NULL]'') and pairing it with a targeted question such as ``What is the correct value for Gender?''.
This prompt design enables the \llm to generate accurate, context-aware missing values.

\noindent \bfit{(2) RAG Assisted Localized Imputation.}
To enable online \llms in handling unseen, domain-specific, or private datasets, the second approach adopts the retrieval-augmented generation (RAG) paradigm.
For example, RetClean~\cite{RetClean} introduces a retrieval-based data cleaning framework that indexes a data lake using both syntactic (Elasticsearch) and semantic (Faiss/Qdrant) methods.
It retrieves the top-$k$ relevant tuples, reranks them (e.g., using ColBERT), and then leverages an \llm to infer missing values, while maintaining lineage tracking for transparency and traceability.

\subsubsection{\llm for Data Integration}
\label{subsubsection:integration}

Data integration aims to align elements across heterogeneous datasets to enable unified access, analysis, and knowledge extraction.
For instance, it includes identifying tables or records that correspond to the same real-world entity.
Moreover, it facilitates downstream tasks such as data augmentation by establishing semantic relationships across sources.

Traditional integration methods often struggle with semantic ambiguities and conflicts, particularly in complex integration scenarios without domain-specific knowledge~\cite{KGRAG4SM}.
Furthermore, classical models (e.g., pretrained models) generally require large amounts of task-specific training data and tend to degrade in performance when encountering out-of-distribution entities~\cite{MatchGPT}.
In contrast, recent studies have shown that \llms possess strong semantic understanding, enabling them to uncover correlations across datasets and incorporate domain-specific knowledge, thereby offering robust generalization across diverse integration tasks.




\noindent \textbf{Entity Matching.}
The goal of entity matching is to determine whether two entries refer to the same real-world entity.
Existing methods leverage \llms through well-structured prompts and advanced reasoning mechanisms, incorporate multiple models for collaborative matching, and apply multi-task fine-tuning to further enhance performance.

\noindent \bfit{(1) Prompt Based End-to-End Matching.}
To improve \llm's effectiveness on matching tasks, the first approach crafts well-structured prompts and integrates auxiliary mechanisms to strengthen the robustness of the reasoning process.

\noindent $\bullet$ \emph{Manually-Crafted Prompt.}
This method incorporates detailed instructions and illustrative examples into the prompts to guide \llm in performing entity matching more effectively.
For example, MatchGPT~\cite{MatchGPT} evaluates the performance of both open-source and closed-source \llms (e.g., Llama 3.1 and GPT-4o mini) with (i) different prompt designs, (ii) the selection of in-context demonstrations, (iii) automatic generation of matching rules, and (iv) fine-tuning \llms using a shared pool of training data.
To reduce inference costs, BATCHER~\cite{BATCHER} introduces a batch prompting method that allows multiple entity pairs to be processed simultaneously.
It optimizes in-context learning by (i) grouping entity pairs into a single prompt and (ii) applying a greedy cover-based strategy to select demonstrations such that each query in the batch is semantically close to at least one example.

\noindent $\bullet$ \emph{Pseudo-Code Guided Reasoning.}
To mitigate hallucinations arising from over-reliance on an \llm's internal knowledge, this method integrates external formalized representations to enhance the robustness and reliability of the reasoning process.
For example, KcMF~\cite{KcMF} guides \llms using expert-designed pseudo-code instructions structured as a sequence of if-then-else logical conditions, combined with external domain knowledge (e.g., datasets and examples).
It further adopts an ensemble strategy by generating outputs from different knowledge sources (e.g., Wikidata and domain-specific datasets) and applies a voting mechanism to aggregate results, improving consistency and accuracy.



\noindent \bfit{(2) End-to-End Matching with Multi-Model Collaboration.}
To leverage the strengths of different models across tasks, the second approach employs collaborative entity matching using models of varying sizes.
For example, COMEM~\cite{COMEM} introduces a compound entity matching framework that combines multiple strategies with \llm collaboration to address global consistency, which is often ignored in binary matching.
It employs (i) a local strategy using a medium-sized \llm (3B-11B) as a matcher or comparator to rank top-$k$ candidates via bubble sort, reducing position bias and context length dependency; and (ii) a global selection strategy using a stronger \llm (e.g., GPT-4o) to refine top-$k$ candidates by modeling inter-record interactions.

\noindent \bfit{(3) Localized \llm Fine-tuning of Multi-Task Learning.}
To enhance the generalization capability of local \llms, the last approach integrates multiple task-specific datasets within a unified multi-task instruction tuning framework.
For example, Jellyfish~\cite{Jellyfish} applies parameter-efficient instruction tuning to locally deployed \llms (7B-13B) across diverse data processing tasks.
It employs techniques such as chain-of-thought prompting over task-specific serialized data and reasoning data distillation, using explanation traces generated by a larger mixture-of-experts model (Mixtral-8x7B-Instruct) to guide the learning process.



\noindent \textbf{Schema Matching.}
The objective of schema matching is to identify correspondences between elements of different database schemas (e.g., matching attribute names ``employee ID'' and ``staff number'').
Existing approaches directly apply prompting techniques to enable \llms to perform end-to-end matching, utilize retrieval-augmented generation (RAG) to enhance contextual understanding, and employ \llm agents to orchestrate the overall matching workflow.

\noindent \bfit{(1) Prompt Based End-to-End Matching.}
To facilitate schema matching without requiring rigid code implementations, the first method employs various prompting techniques to guide \llm in identifying the desired mappings.
For example, LLMSchemaBench~\cite{LLMSchemaBench} applies prompt engineering techniques to interact with \llms, defining four task scopes that differ in the level of contextual information included in the prompts.
The prompts are constructed using established design patterns: the persona pattern (e.g., instructing the \llm to act as a schema matcher), meta language creation (e.g., explicitly defining valid match criteria), Chain-of-Thought reasoning, and the output automater (e.g., generating structured JSON outputs for downstream automation).


\noindent \bfit{(2) End-to-End Matching via Context-Enriched RAG.}
To enrich the matching context and improve accuracy, the second method integrates retrieval-augmented generation (RAG) with various strategies.
For example, Magneto~\cite{Magneto} employs a retrieve-rerank framework that combines small pre-trained language models (SLMs) with \llms to deliver cost-effective and generalizable schema matching.
SLMs serve as candidate retrievers, generating an initial ranked list of potential matches from the target table for each input column, which is then refined by \llms acting as rerankers to improve accuracy. 
KG-RAG4SM~\cite{KGRAG4SM} incorporates multiple retrieval strategies, including vector-based, graph traversal-based, and query-based, to extract relevant subgraphs from knowledge graphs (KGs).
These subgraphs are further refined through ranking mechanisms and used to augment \llm prompts, thereby improving schema matching performance through enriched contextual input.

\noindent \bfit{(3) Agent-Based Matching Workflow Orchestration.}
To address complex matching patterns, the final approach leverages \llm-based agents to orchestrate the end-to-end matching workflow.
For example, Agent-OM~\cite{AgentOM} employs two \llm agents (i.e., Retrieval Agent and Matching Agent) to control the workflow by decomposing tasks via Chain-of-Thought (CoT) prompting, invoking specialized tools (e.g., syntactic/lexical/semantic retrievers and matchers), and accessing a hybrid database (relational + vector) for memory storage and retrieval.
Harmonia~\cite{Harmonia} leverages \llm-based agents to orchestrate data harmonization tasks, combining predefined data integration primitives (e.g., schema matching, value matching) with on-demand code generation when the primitives are insufficient.
In addition, it employs techniques like ReAct for reasoning and action planning, interactive user feedback for error correction, and declarative pipeline specifications for reproducibility.

\subsubsection{\llm for Data Discovery}
\label{subsubsection:discovery}

Data discovery focuses on identifying relationships within datasets through tasks like data annotation (e.g., column type classification) and profiling (e.g., metadata generation).
Unlike data analysis, which emphasizes statistical computations or factual answer generation, data discovery enables deeper semantic understanding critical for downstream applications such as integration, search, and recommendation.

Existing data discovery methods face two limitations.
First, they typically consider limited interaction between queries and tables~\cite{Birdie}.
Second, many of these approaches rely heavily on large training datasets, struggle with distribution shifts, and fail to generalize to rare or domain-specific data~\cite{ArcheType, LLMCTA}.
Recent studies have shown that \llms can effectively address these challenges by generating high-quality metadata, enriching dataset context, and supporting natural language interfaces for data discovery tasks.








\noindent \textbf{Data Profiling.}
Data profiling typically involves characterizing a given dataset by generating additional information (e.g., dataset descriptions).
Recent methods often employ prompting techniques to guide \llm in generating such metadata by leveraging their pretrained knowledge and contextual understanding.

\noindent \bfit{(1) Manually Crafted Profiling Prompt Engineering.}
To profile different aspects of a dataset without extensive manual effort or code implementation, the first approach relies on a set of manually crafted profiling prompts.
For example, AutoDDG~\cite{AutoDDG} utilizes \llm with carefully designed prompts to generate two types of descriptions (i.e., User-Focused Descriptions (UFDs) for readability and Search-Focused Descriptions (SFDs) for search optimization) tailored to the dataset's content and intended usage.
LEDD~\cite{LEDD} employs carefully crafted prompts to support core data discovery tasks in data lakes.
For hierarchical cataloging, prompts instruct \llm to summarize data clusters into semantically meaningful categories.
For semantic search, prompts refine natural language queries before embedding and retrieval.
For real-time relation analysis, prompts guide \llm in comparing expanded graph nodes and describing inter-table relationships.

\noindent \bfit{(2) RAG Assisted Context Enrichment.}
To enhance retrieval effectiveness across diverse query types, the second method adopts a hybrid approach that integrates diverse retrieval techniques.
For example, Pneuma~\cite{Pneuma} adopts a RAG framework to retrieve relevant tables from databases, data lakes, or repositories based on natural language queries.
It combines \llms with traditional retrieval techniques, such as full-text and vector search, using \llms for both schema narration (i.e., generating meaningful column descriptions) and as judges to refine and rerank retrieved results.

\noindent \textbf{Data Annotation.}
Data annotation involves assigning semantic or structural labels to data elements, such as identifying column types (e.g., \textit{Manufacturer} or \textit{birthDate} from the DBPedia ontology).
Recent methods leveraging \llm typically design prompts with task-specific annotation instructions.
Additionally, some approaches employ retrieval-augmented generation (RAG) techniques and the contextual reasoning capabilities of \llms to further enrich the annotation context and improve performance.




\noindent \bfit{(1) Task-Specific Annotation Prompt Engineering.}
To flexibly support diverse annotation tasks, the first approach encodes task-specific instructions and requirements within carefully crafted prompt templates.
For example, CHORUS~\cite{CHORUS} integrates \llms into the annotation pipeline using task-specific prompts that incorporate instructions, demonstrations, data samples, metadata, domain knowledge, and output formatting guidance.
Goby~\cite{kayali2025goby} explores the use of \llms for semantic column type annotation in a domain-specific enterprise setting by crafting a set of tailored prompts.
It proposes several techniques to improve performance, including tree serialization (providing the full ontology as prompt context), grammar-constrained decoding (enforcing hierarchical structure during generation), and step-by-step prompting (Chain-of-Thought strategy to guide ontology navigation).
LLMCTA~\cite{LLMCTA} evaluates diverse \llms for generating and refining label definitions by employing methods like knowledge generation prompting (e.g., producing initial demonstrations), self-refinement (error-based definition improvement), and self-correction (two-step pipeline featuring a reviewer model).




\noindent \bfit{(2) RAG Assisted Annotation Context Enrichment.}
To supply \llm with relevant annotation context, the second approach utilizes diverse retrieval strategies within retrieval-augmented generation (RAG) frameworks to enrich the input.

\noindent $\bullet$ \emph{Classical Retrieval Technique.}
To mitigate the shortcomings of vanilla \llm-based annotation, such as outdated knowledge, this method augments the context with retrieved external knowledge.
For example, RACOON~\cite{RACOON} performs semantic type annotation by leveraging a Knowledge Graph (KG) to retrieve entity-related information (e.g., labels and triples) associated with column cells.
This information is then processed into concise contextual representations and incorporated into \llm prompts to improve annotation accuracy.

\noindent $\bullet$ \emph{\llm Based Generation.}
To fully leverage \llm's internal knowledge, this method relies on the model itself to generate relevant contextual information.
For example, Birdie~\cite{Birdie} leverages \llms to automatically generate natural language queries for training a differentiable search index (DSI), which facilitates linking relational tables to queryable knowledge by enriching them with contextual semantics.
It supports scalable structured data annotation, using prompts composed of structured markdown tables comprising captions, headers, and sample rows alongside explicit task instructions.

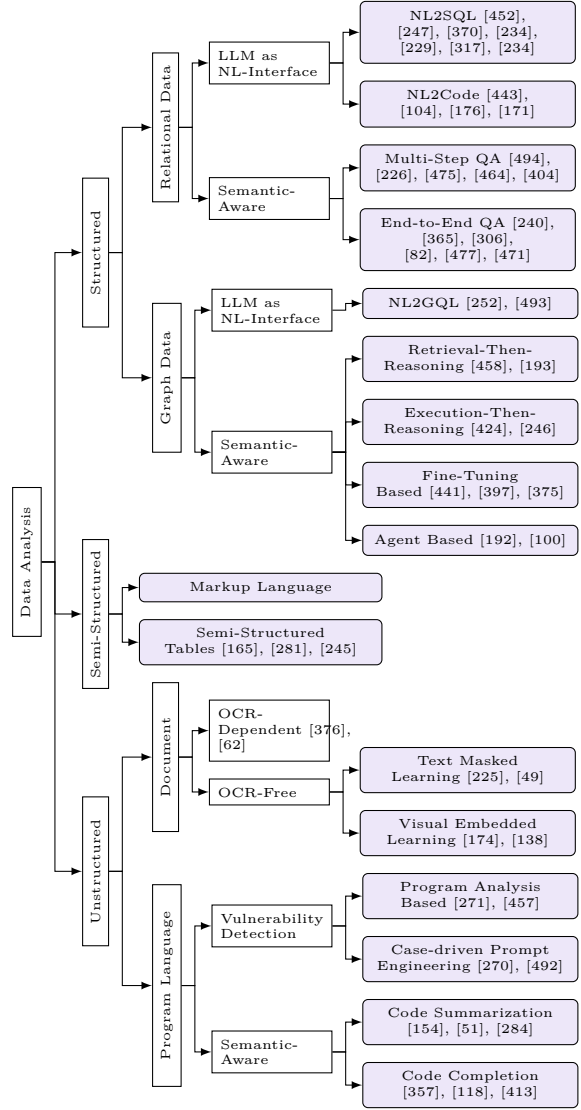
\begin{figure}[t]
\label{fig:data_analysis}
\centering
\tikzset{
    basic/.style  = {draw, text width=26pt, font=\sffamily, rectangle, font=\tiny},
    ver/.style    = {basic, rotate=90, child anchor=north, parent anchor=south, anchor=center},
    root/.style   = {basic, thin, align=center, text width=50pt},
    data_type_node/.style = {basic, thin, text width=50pt, align=center},
    uns_child/.style = {basic, text width=50pt, align=center},
    str_child/.style = {basic, text width=50pt, align=center},
    uns_child_child/.style = {basic, text width=39pt},
    str_child_child/.style = {basic, text width=39pt},
    leaf/.style = {basic, fill=pink!10!blue!80!red!10, rounded corners=2pt, align=center},
    sem_leaf/.style = {leaf, thin, text width=3cm},
    uns_leaf/.style = {leaf, thin, text width=75pt},
    str_leaf/.style = {leaf, thin, text width=75pt},
    edge from parent/.style={draw=black, edge from parent fork right}
}

\begin{forest} for tree={
    grow=east,
    reversed=true,
    parent anchor=east,
    child anchor=west,
    edge path={\noexpand\path[\forestoption{edge},->, >={latex}] 
         (!u.parent anchor) -- +(5pt,0pt) |- (.child anchor)
         \forestoption{edge label};}
}
[Data Analysis, root, rotate=90, parent anchor=south, child anchor=north
    [Structured, data_type_node, rotate=90, parent anchor=south, child anchor=north
        [Relational Data, str_child, rotate=90, parent anchor=south, child anchor=north
            [LLM as \\ NL-Interface, str_child_child
                [NL2SQL~\cite{zhang2024finsql,petsql,talaei2024chess,li2024codes,SQL360,DINSQL,li2024codes}, str_leaf]
                [NL2Code~\cite{yin2022code,chowdhery2022palm,huang2024datacoder,hong2024datainterpreter}, str_leaf] ]
            [Semantic-Aware, str_child_child
                [Multi-Step QA~\cite{tatllm,lei2023s3hqa,zhao2024tapera,zhang2023reactable,wang2024chainoftable}, str_leaf
                ]
                [End-to-End QA~\cite{li2023tablegpt,su2024tablegpt2,patnaik2024cabinet,cao2025tablemaster,zheng2024multimodal,zhao2024tabpedia}, str_leaf
                ] ] ]
        [Graph Data, str_child, rotate=90, parent anchor=south, child anchor=north
            [LLM as \\ NL-Interface, str_child_child
                [NL2GQL~\cite{liang2024natnl2gql,zhou2024r3}, str_leaf] ] 
            [Semantic-Aware, str_child_child
                [Retrieval-Then-Reasoning~\cite{zhang2022subgraph,jiang2023unikgqa}, str_leaf]
                [Execution-Then-Reasoning~\cite{xiong2024interactivekbqa,li2024flexkbqa}, str_leaf]
                [Fine-Tuning Based~\cite{ye2024instructglm,wang2024instructgraph,tang2024graphgpt}, str_leaf]
                [Agent Based~\cite{jiang2023structgpt,cheng2024necessary}, str_leaf] ] ]
    ]
    [Semi-Structured, data_type_node, rotate=90, parent anchor=south, child anchor=north
        [Markup Language, sem_leaf]
        [Semi-Structured Tables~\cite{gupta2023temptabqa,ma2024spreadsheet,li2024mimotable}, sem_leaf] 
    ]
    [Unstructured, data_type_node, rotate=90, parent anchor=south, child anchor=north
        [Document, uns_child, rotate=90, parent anchor=south, child anchor=north
            [OCR-Dependent~\cite{tang2023unifying,appalaraju2023docformerv2}, uns_child_child]
            [OCR-Free, uns_child_child
                [Text Masked Learning~\cite{lee2023pix2struct,aggarwal2023dublin}, uns_leaf]
                [Visual Embedded Learning~\cite{hu2024mplugdocowl15,feng2024docpedia}, uns_leaf] ] ],
        [Program Language, uns_child, rotate=90, parent anchor=south, child anchor=north, text width=70pt
            [Vulnerability Detection, uns_child_child
                [Program Analysis Based~\cite{liu2024pdbert,zhang2023bulnerability}, uns_leaf]
                [Case-driven Prompt Engineering~\cite{liu2023gptvulnerability,zhou2024llmvulnerability}, uns_leaf] ]
            [Semantic-Aware, uns_child_child
                [Code Summarization\\ \cite{geng2023llmfewshot,ahmed2024code,mao2024scla}, uns_leaf]
                [Code Completion\\ \cite{repofusion,cocomic,repoformer}, uns_leaf] ] ]
    ]
]
\end{forest}

\caption{Overview of LLM for Data Analysis.}
\end{figure}

\subsection{\llm for Data Analysis}
\label{subsec:analysis}

Apart from data manipulation, \llms hold the potential to revolutionize traditional data analysis paradigms by supporting natural language interfaces and enabling advanced, semantic-aware analysis tasks that typically require human involvement. In this section, we discuss the challenges and techniques of LLM-based data analysis, including structured data analysis, semi-structured data analysis, and unstructured data analysis.


\subsubsection{\llm for Structured Data Analysis}
Structured data refers to data with well-defined schemas like relational (tabular) data~\cite{relationaldata} and graph data~\cite{graphdata}.


\subsubsection*{3.2.1.1 Relational Data Analysis} 


\hi{LLM for Natural Language Interfaces.} Basic analysis jobs for relational data are typically characterized by well-defined operations. These include basic calculations (e.g., summation, averaging, counting, ranking), statistical analysis (e.g., regression, K-means clustering), and data quality assurance processes (e.g., constraint validation, outlier detection). Such tasks can generally be supported by tools like SQL or Python libraries (e.g., Pandas). 

\noindent\underline{(1) NL2SQL.} With the help of \llm, users can directly perform operations using natural language. NL2SQL focuses on translating natural language queries into SQL commands by leveraging techniques such as $(i)$ schema linking, which aligns user intents with database schema to resolve ambiguities~\cite{zhang2024finsql,petsql}, $(ii)$ content retrieval, which dynamically extracts relevant information from the database to refine query generation~\cite{talaei2024chess,li2024codes}, and $(iii)$ SQL generation strategies such as multi-step generation, intermediate SQL representation, and different decoding strategies~\cite{SQL360,DINSQL,li2024codes, zhou2025cracksql, zhou2025cracksqldemo}.

\noindent\underline{(2) NL2Code.} Different from NL2SQL, NL2Code approaches emphasize enhancing relational data analysis through generating Python code (e.g., Pandas, NumPy), which includes a vast number of library APIs characterized by high variability and complexity, and often requiring the handling of complex chain operations. Recent advancements address these issues to some extent. 

\noindent$\bullet$ \emph{Model Finetuning:}  PACHINCO~\cite{yin2022code} fine-tunes a 62B parameter PALM~\cite{chowdhery2022palm} model in two stages (i.e., separately using  a Python source code corpus with 64B tokens and a Jupyter notebook corpus with 9.6B tokens) so as to improve model performance on analysis-related tasks (e.g., calculate the amount of games added in each year for each month). DataCoder~\cite{huang2024datacoder} utilizes different types of contexts (e.g., code, text, and data) by employing dual encoders (e.g., data encoder and code + text encoder) and one general decoder to generate code in notebooks.

\noindent$\bullet$ \emph{\llm Based Analysis Agent:}  
Data Interpreter~\cite{hong2024datainterpreter}, on the other hand, leverages LLMs through APIs to generate task and action graphs. Specifically, they utilize \llm's semantic reasoning ability to accurately decompose complex user queries into subproblems (e.g., correlation analysis, data exploration, and anomaly detection), and refine and verify each subproblem to improve code generation results for data science tasks.

\begin{figure}[!t]
    \centering
    \includegraphics[width=1\linewidth]{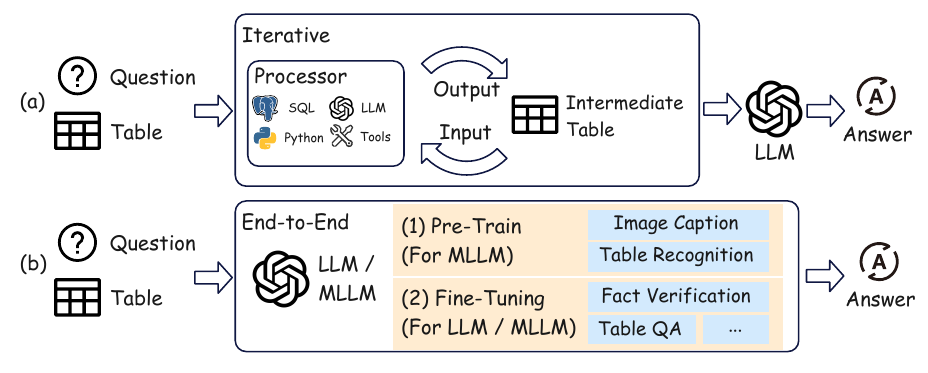}
    \caption{\textbf{General Workflows} - (a) Multi-Step Relational Data QA. (b) End-to-End Relational Data QA.}
    \label{fig:frameworkLLM4Data}
\end{figure}

\hi{LLM for Semantic Analysis.} Moreover, some jobs require LLM-based analysis, such as those that involve semantic understanding or demand outputs in natural language format (e.g., table  summarization). These challenges call for methodologies like (1) multi-step question answering (QA) with diverse decomposition strategies and (2) end-to-end QA leveraging specifically optimized LLMs.

\noindent$\bullet$ \bfit{Multi-Step QA.} Multi-step question answering (QA) refers to decomposing complex queries into a sequence of sub-questions to facilitate step-by-step reasoning. According to the question decomposition mechanisms, existing methods can be categorized into two types: (1) static decomposition, which follows predefined and fixed processing steps (e.g., retrieve-select-reason), and (2) LLM-driven iterative decomposition, in which the \llm dynamically determines the next operation based on the contextual history of the reasoning process.

\noindent\underline{(1) Static Decomposition.} The static decomposition includes Retriever-Selector-Reasoner frameworks and the variants, which partition tasks into modular components for better multi-step inference and enhanced interpretability. The Extractor-Reasoner-Executor paradigm~\cite{tatllm} extracts the relevant segments from the context, generates the logic rules or equations, and performs the rules or executes the equations to get the final answer through LLM prompting. S3HQA~\cite{lei2023s3hqa} trains a retriever which aims to perform initial filtering of heterogeneous resources, utilizes a selector to select the most relevant factual knowledge, and a generation-based reasoner to obtain final answers.


\noindent\underline{(2) Iterative Decomposition.} However, static decomposition paradigm performs poorly on multi-hop queries, while LLM-driven iterative decomposition, which dynamically refines subtasks through recursive reasoning, could effectively address the issue.  

TAPERA~\cite{zhao2024tapera} introduces the query decomposition step into the question answering process by adopting the LLM-driven approach. The Planner decomposes the query into sub-queries, forming an initial plan. The Reasoner then generates executable programs for each sub-query, while the Answer Generator derives answers based on the program outputs to fulfill the plan. Finally, the Planner updates or finalizes the plan as needed. 

Similarly, ReAcTable~\cite{zhang2023reactable} and CHAIN-OF-TABLE~\cite{wang2024chainoftable} iteratively generate operations and update the table to present a reasoning chain as a proxy for intermediate thoughts through prompting \llms and in-context learning.

\noindent$\bullet$ \bfit{End-to-End QA.} End-to-End Question Answering (QA) refers to approaches in which the answer-generating \llm directly produces the final response without intermediate steps or iterative refinement. Based on the data representation and processing mechanisms, the relevant methods can be classified into table-specific \llm fine-tuning, table content retrieval, and table-as-image analysis.

\noindent\underline{(1) Table-Specific \llm Fine-Tuning.} Fine-tuning \llms on task-specific table datasets enables them to internalize analytical knowledge directly within their parameters. TableGPT~\cite{li2023tablegpt} fine-tunes \llms like GPT-3.5 using a diverse set of table tasks synthesized from real-world tables. Building on Qwen2.5~\cite{qwen2025qwen25}, TableGPT2~\cite{su2024tablegpt2} introduces a table encoder to generate a hybrid table representation, an adapter to generate query representations, and a \llm decoder generates an agent workflow (i.e., the tool execution pipeline) to derive the final answer. The TableGPT2 model is pre-trained on 593.8K tables and fine-tuned 2.36M question-answer pairs.

\noindent\underline{(2) Table Content Retrieval.} Instead of embedding the whole table, table content retrieval enhances model performance by eliminating noisy parts of the table while retaining information relevant to question answering. CABINET~\cite{patnaik2024cabinet} employs a weakly supervised component to produce a parsing statement that defines the criteria for selecting relevant rows and columns, emphasizing the corresponding table cell content. TableMaster~\cite{cao2025tablemaster} constructs a refined subtable through row and column lookup. By leveraging carefully designed \llm prompts (e.g., provide objective, table definition, table information, question, instructions, and response format), it ranks all candidate columns, selects a relevant subset based on the query, and then instructs the LLM to generate an SQL query for extracting the most relevant rows.

\noindent\underline{(3) Table-As-Image Analysis.} Due to the limitations of (text-only) LLMs in understanding table structures, the Table-as-Image approach has been proposed, converting tables into images for analysis using multimodal LLMs. Table-LLaVA~\cite{zheng2024multimodal} applies incremental pretraining to LLaVA-7B~\cite{liu2024llava} on 150K table recognition samples (e.g., input a table image and output table representations in HTML, Markdown, or LaTeX), enabling the model to align table structures and elements with textual modality. It is further fine-tuned on 232K samples on question answering, text generation, fact verification, and structure understanding tasks to enhance its instruction-following ability. To enable a single model to perform various analytical tasks, TabPedia~\cite{zhao2024tabpedia} introduces the concept synergy mechanism, abstracting all table analysis tasks into concepts. Built on Vicuna-7B~\cite{zheng2023judging}, it appends meditative tokens to the input of the \llm decoder, which adaptively activates different regions of visual tokens and helps the model interpret the intent behind specific task questions. However, such methods face limitations when processing twisted or distorted tables, and their performance degrades significantly when directly handling document images.

\subsubsection*{3.2.1.2 Graph Data Analysis} 
Different from relational data, graph data represents entities (vertices) and their inter-dependencies (relationships) to explicit model of complex network semantics (e.g., social networks and knowledge graphs) beyond rigid tabular schema, which presents unique challenges due to the vast search space and complex path reasoning in multi-hop queries~\cite{defgraphdata}. Compared with relational data analysis, graph data analysis involves more complex jobs like summarization based on the multi-hop relations across the graph vertices and reasoning over text-attributed graphs whose nodes and edges are associated with text~\cite{liang2024natnl2gql,zhou2024r3}. Graph data can not only be stored in relational databases, but also be stored and queried in knowledge graphs and accessed through SPARQL in RDF databases (e.g., Blazegraph~\cite{blazegraph} and GraphDB~\cite{graphdb}) or Cypher in Neo4j~\cite{neo4j}.


{Traditional} graph analysis (e.g., statistical methods, graph neural network (GNN) based methods) encompasses a spectrum of tasks, including {\it node classification} (e.g., categorizing academic papers into research domains), {\it graph classification} (e.g., predicting node properties over molecular graphs), {\it link prediction} (i.e., inferring latent relationships between graph nodes), {\it community detection} (i.e., identifying densely connected subgraphs), {\it anomaly detection} (i.e., identifying deviations from expected patterns), {\it graph clustering}, and etc. However, these methods have their own limitations. Statistics-based methods fail to handle complex semantic information (e.g., query can be extremely complex and requires human expertise), while graph neural networks (GNNs) exhibit limited generalization capabilities, necessitating task-specific retraining on different tasks.


In contrast, the advent of \llms offers transformative potential by leveraging their advanced reasoning capacities and cross-domain generalization abilities, which can (1) simplify the query writing costs (e.g., NL interfaces) and (2) achieve semantic-aware analysis unsupported in traditional ones.

\hi{Natural Language To Graph Analysis Query.} {Different from NL2SQL, the syntax of graph query language generation is more complex (i.e., MATCH, LOOKUP, GET and other operations unique to graph data manipulation) and there exist two operation objects (i.e., vertex and edge)~\cite{zhou2024r3}.} 
By integrating natural language interfaces with graph data, LLMs facilitate flexible and efficient query generation without the need for specialized model architectures.

To enhance LLMs' comprehension of the complex syntax of Graph Query Language (GQL), $R^3$-NL2GQL~\cite{zhou2024r3} proposes a hybrid approach leveraging relatively small \llm (e.g., LLaMA3-7B) as a selector and GQL rewriter, while employing a larger LLM (e.g., GPT-4) as a reasoner. The selector identifies the necessary CRUD functions, clauses, and schema, while the rewriter refines the query by aligning it with the relevant graph data retrieved by minimum edit distance and semantic similarity calculation. The LLM then synthesizes the aligned question, selected operations, and schema to generate the final GQL query.

To address the limitations of LLMs in planning and collaborating with other LLMs, NAT-NL2GQL~\cite{liang2024natnl2gql} introduces a three-agent framework. The Preprocessor agent constructs context information, including query rewriting, path linking, and the extraction of query-relevant schemas. The Generator agent, an LLM fine-tuned with NL-GQL data, generates GQL statements based on the rewritten queries and extracted schemas. The Refiner agent iteratively enhances the GQL or contextual information by leveraging error feedback from GQL execution results. 

Note that, within the context of AI for Science (AI4Science), the integration of LLMs with graph data analysis has also shown significant potential and wide-ranging applications (e.g., treat polymers as graphs and predict their properties~\cite{li2024graph,pei2024bio}), which is not the primary focus of this survey.


\hi{LLM-based Semantic Analysis.} Furthermore, certain jobs necessitate semantic-aware analysis, such as summarizing textual paragraphs embedded within graph nodes. 
Based on the adopted \llm strategies, we classify the relevant methods into retrieval-then-reasoning methods, execution-then-reasoning methods, graph task based fine-tuning methods, and agent based methods.

\noindent$\bullet$ \bfit{Retrieval-Then-Reasoning.} 
Retrieval-then-reasoning first extracts a question-specific subgraph from the graph to identify the most relevant entities and then generates answers using \llms.
To address the challenge of a vast search space, \cite{zhang2022subgraph} introduces a two-stage approach. First, a trainable and decoupled subgraph retriever selects a relevant subgraph based on the query. Then, reasoning is performed over the retrieved subgraph to derive the final answer. UniKGQA~\cite{jiang2023unikgqa} integrates retrieval and reasoning within a unified model architecture. It comprises a semantic matching module, leveraging a pre-trained RoBERTa~\cite{RoBERTa} for the semantic alignment between questions and relations in graphs, and a matching information propagation module that propagates matching signals along directed edges in graphs.

\noindent$\bullet$ \bfit{Execution-Then-Reasoning.} Execution-then-reasoning refers to the process of parsing natural language queries into executable logical forms (e.g., SPARQL) that align with the graph data, followed by reasoning based on the output of the executed program.
Interactive-KBQA~\cite{xiong2024interactivekbqa} introduces an interactive \llm QA framework with a unified SPARQL-based toolset (e.g., entity search, graph pattern search, SPARQL execution, etc.) designed to address complex queries. 
FlexKBQA ~\cite{li2024flexkbqa} addresses the challenge of lacking high-quality annotated data in real-world scenarios. By prompting \llms as program translators, it samples program-answer pairs from the knowledge base and generates corresponding natural language questions. The synthetic question-program-answer dataset is used to train lightweight models through execution-guided self-training, which are subsequently employed to annotate real user queries. This approach addresses the distribution shifts between synthetic and actual data, leading to significant improvements in few-shot learning scenarios.

\noindent$\bullet$ \bfit{Graph Task Based Fine-tuning Methods.}
InstructGLM~\cite{ye2024instructglm} enables generative graph learning by fine-tuning an \llm and leveraging natural language descriptions of graph structures (e.g., offer the first node and the 1-/2-/3-hop neighbors' information).
InstructGraph~\cite{wang2024instructgraph} introduces a stricter code-like graph representation format which constructs entities and triples in the form of list, whose backbone \llm (LLaMA2-7B) is fine-tuned on a graph-centric corpus comprising 1.6 million instances. To mitigate the issue of hallucination, it incorporates Direct Preference Optimization (DPO) algorithm~\cite{rafailov2024dpo} for preference alignment.
GraphGPT~\cite{tang2024graphgpt} enhances model performance in zero-shot scenarios by incorporating a structural information encoding module based on Graph-SAGE~\cite{10.5555/3294771.3294869} and GCN~\cite{kipf2017gcn}. It fine-tunes the projector bridging the graph encoder and the LLM decoder to align the language capabilities of the foundation LLM (Vicuna-7B) with the graph learning tasks.

\noindent$\bullet$ \bfit{Agent Based Methods.}
Agent-based methods involve leveraging LLM-based agents with predefined tools (e.g., human-written interfaces or graph processing library APIs) that iteratively interact with the graph data to retrieve, refine, and operate information.
StructGPT~\cite{jiang2023structgpt} introduces an iterative reading-then-reasoning framework, leveraging specialized interfaces to operate on graph data. It repeatedly applies an invoke-linearize-generate procedure to derive query results. 
Another approach is to generate an entire reasoning path based on the query and refine it only when necessary. Readi~\cite{cheng2024necessary} initially constructs a reasoning path and instantiates it on the graph. When execution errors occur, it collects error messages and invokes an LLM to revise the path. The final answer is inferred from the instantiated graphs.

\subsubsection{\llm for Semi-Structured Data Analysis}
Semi-structured data refers to data that are neither with strictly predefined schema like relational models nor raw data (e.g., plain text or images)~\cite{abiteboul1997querying}. Meanwhile, they still maintain part of organizational properties (e.g., tags, headers) and have hierarchical or nested representation (e.g., \textit{County} - \textit{Province} - \textit{City} in a nested JSON).

\subsubsection*{3.2.2.1 Markup Language}
Markup languages (e.g., XML, JSON, and HTML) are widely used for structuring and exchanging data across systems. Traditional approaches for processing these formats typically involve transforming them into structured tables or representing them as hierarchical tree structures. Leveraging the reasoning capabilities of LLMs, it becomes possible to directly extract and interpret hierarchical relationships, attributes, and nested structures from data without the need for intermediate transformations.

\subsubsection*{3.2.2.2 Semi-Structured Tables} 
Compared to structured relational data, semi-structured tables exhibit a more complex structural organization characterized by merged cells. This inherent complexity presents a significant challenge in aligning queries with the table content and structure in query answering tasks. The lack of efficient tools (usually using the openpyxl library) and representation methods (usually stored in Excel or HTML files) for handling semi-structured tables makes it more difficult to process such data.

Although research on semi-structured table analysis is limited, several studies have compiled various semi-structured table reasoning datasets, providing valuable data support.
TEMPTABQA~\cite{gupta2023temptabqa} consists of 11,454 question-answer pairs focused on temporal queries, while SPREADSHEETBENCH~\cite{ma2024spreadsheet} presents a challenging benchmark for spreadsheet manipulation, with 912 questions derived from real-world scenarios. MiMoTable~\cite{li2024mimotable} incorporates reasoning across multiple sheets and files, containing 1,719 queries within 428 spreadsheets. Evaluation results on these benchmarks highlight a significant performance gap (ranging from 20\% to 50\%) between state-of-the-art models and human performance, calling for further exploration in this area.

\subsubsection{\llm for Unstructured Data Analysis}
Unstructured data refers to data that lacks explicit structure, as it does not adhere to a predefined schema. Additionally, it exhibits high variability in format, length, and modality, which further complicates its processing and analysis.

\subsubsection*{3.2.3.1 Documents} 
Documents exhibit complex layouts and styles with diverse elements, including a hybrid of images, tables, charts, plain text, and formulas.

\noindent$\bullet$ \bfit{OCR-Dependent Methods.}
OCR-based methods refer to approaches that involve performing Optical Character Recognition on document images, followed by the integration of textual, layout, and visual features for reasoning. 
UDOP~\cite{tang2023unifying} integrates text and layout modalities within a unified encoder, dynamically fusing image patch tokens and text tokens based on their spatial information. Specifically, when the center of a text token's bounding box falls within an image patch, the corresponding image patch embedding is added to the text token embedding, enabling a more cohesive representation of document structure.
DocFormerV2~\cite{appalaraju2023docformerv2} preserves the integrity of layout information by employing a visual encoder. Image patches and text bounding box positions are embedded through a linear layer and added to the corresponding token embeddings as input to the T5~\cite{C4-T5} encoder. To achieve local feature semantic alignment, the model undergoes pretraining on token-to-line (i.e., predict whether a key-value pair is on the same line or adjacent lines) and token-to-grid (i.e., predict each token located in which image grid) tasks. The T5 decoder is then incorporated to fine-tune the whole model on downstream tasks.

\noindent$\bullet$ \bfit{OCR Free Methods.}
However, the OCR step often introduces semantic errors, resulting in suboptimal performance. To fill this gap, OCR-free methods have emerged, directly generating the target token sequences with end-to-end multimodal \llms~\cite{liu2024fox,wei2024got}. Based on different approaches to enhancing model understanding of textual semantics, related works can be categorized into text masked learning and visual embedded learning.

\noindent\underline{(1) Text Masked Learning.}
Text Masked Learning involves masking textual content within a document and training the model to predict the missing text. 
Pix2Struct~\cite{lee2023pix2struct} is a typical vision-encoder-text-decoder pre-trained image-to-text model designed for visual language understanding based on ViT~\cite{dosovitskiy2021image}. It is pretrained to parse masked web pages into simplified HTML. The model introduces a variable-resolution input representation, rescaling input images to maximize the number of patches that can fit within the given sequence length, to prevent aspect ratio distortion.
DUBLIN~\cite{aggarwal2023dublin} designed multiple fine-tuning tasks (i.e., bounding box prediction based on given text, text prediction based on given bounding box, masked text generation, and query answering) to improve the generalization ability. 

\noindent\underline{(2) Visual Embedded Learning.}
In Visual Embedded Learning, there are no specially designed training objectives. Instead, the model is directly fine-tuned on downstream tasks to enhance its understanding of textual content within images.
mPLUG-DocOwl1.5~\cite{hu2024mplugdocowl15} introduces a spatial-aware vision-to-text module designed for representing high-resolution, text-rich images. This module preserves structural information while reducing the length of visual features. It consists of a convolution layer to shorten the sequence length and a fully connected layer that projects visual features into the language embedding space.
Unlike most methods that crop or resize the initial image before feeding it into a vision encoder, DocPedia~\cite{feng2024docpedia} directly processes visual input in the frequency domain. It utilizes JPEG DCT~\cite{wallace1992dct} extraction to obtain DCT coefficients, which are then processed using a frequency adapter before being input into the vision encoder. This approach allows the model to capture more visual and textual information while using a limited number of tokens. The performance improvement observed in the experiment suggests that this method offers a novel approach for processing high-resolution images.


\subsubsection*{3.2.3.2 Program Language Analysis} 
Programming language analysis involves multiple levels of abstraction, including lexical analysis, parsing, and semantic analysis, each requiring distinct techniques to process source code effectively. Additionally, it must handle both local and global information, such as variable scopes, function call chains, and complex dependencies, which pose significant challenges for accurate program understanding.

\hi{LLM as Program Vulnerability Detection Tools.}
Recent advancements in LLMs have opened new avenues for improving vulnerability detection tools. Training LLMs based on program analysis techniques enhances their ability to understand programs at both the lexical and syntactic levels. Leveraging in-context learning through case-driven prompt engineering enhances the model's accuracy by providing relevant examples.

\noindent$\bullet$ \bfit{Program Analysis based Training.}
Static and dynamic program analysis are commonly used methods for detecting vulnerabilities in programs. By assisting these processes, LLMs improve the accuracy of vulnerability detection.
PDBER~\cite{liu2024pdbert} is a model fine-tuned on CodeBERT~\cite{feng2020codebert} through three tasks (i.e., Predicting Masked Tokens, Predicting Statement-Level Control Dependencies, and Predicting Token-Level Data Dependencies). This enables more fine-grained vulnerability analysis at the statement level.
To reduce the impact of irrelevant information, \cite{zhang2023bulnerability} decomposes the control flow graph (CFG) into multiple execution paths from the entry node to the exit node. CodeBERT and a CNN are employed to capture intra-path and inter-path representations, respectively. The extracted feature vectors are then combined as a unified program representation, which serves as input to a MLP classifier for vulnerability detection.

\noindent$\bullet$ \bfit{Case-driven Prompt Engineering.}
Leveraging the in-context learning and few-shot learning capabilities of LLMs can significantly improve their accuracy in vulnerability detection. 
VUL-GPT~\cite{liu2023gptvulnerability} uses GPT-3.5 to generate analysis content (i.e., the program interpretation) for the input code and retrieves similar code snippets and corresponding vulnerability information through BM25~\cite{bm25} or TF-IDF. The retrieved information, along with the original code and analysis, is then input into GPT to detect vulnerabilities.
~\cite{zhou2024llmvulnerability} designs various prompts, such as random code samples and retrieve-based code samples, and demonstrates that GPT-4 outperforms state-of-the-art models in vulnerability detection.

\hi{LLM-based Semantic-aware Analysis.}
Traditional semantic-aware tasks convert programs into ASTs~\cite{song2024improvingcs} or graph structures~\cite{gao2021sgtrans} and train Seq2Seq models to learn program syntax, dependencies, and semantics. However, these approaches lack general knowledge, leading to limited generalization ability. By leveraging the world knowledge and few-shot learning capabilities of LLMs, the performance of tasks such as code summarization and code completion has been significantly improved.

\noindent$\bullet$ \bfit{LLM as Code Summarizer.}
Recent advancements in LLM-powered code summarization focus on retrieving similar code snippets and leverage LLMs' few-shot learning capability to enhance performance. 
\cite{geng2023llmfewshot} retrieves similar code examples by measuring token overlap and the cosine distance between embedding vectors of code snippets. In contrast, ~\cite{ahmed2024code} employs the BM25 algorithm and incorporates repository information, data flow information, and variable information to construct three-shot prompts.
SCLA~\cite{mao2024scla} further enhances code semantics in LLM prompts by preprocessing the code sample pool to extract semantic information. By simultaneously leveraging few-shot learning, it achieves state-of-the-art performance based on Gemini-1.5-Pro.

\noindent$\bullet$ \bfit{LLM as Repository-Level Code Completer.}
Repository context (e.g., imports, related classes, etc.) plays a crucial role in code completion. Given the strong semantic understanding and generative capabilities of LLMs, how to integrate contextual information into code completion has become a key research focus. 
RepoFusion~\cite{repofusion} appends the surrounding text of the target code to the repository context retrieved based on BM25, encoding and concatenating them as input to the decoder for code generation. This approach enables the model to produce context-aware code completions by leveraging both local and repository-level information. 
CoCoMIC~\cite{cocomic} proposes a more robust retrieval method based on program dependency graphs. Given an incomplete program, it retrieves the most relevant context by analyzing file imports within the constructed graph. By defining the relevant context as files within a two-hop neighborhood, this approach mitigates the risk of excluding vital dependencies while avoiding the inclusion of irrelevant information.
However, some researchers have found that simple retrieval methods fail to improve performance in up to 80\% of cases and may even degrade performance due to the inclusion of irrelevant information~\cite{repoformer}. As a result, Repoformer introduces a self-supervised learning approach to enable the model to accurately judge whether retrieval can improve its output quality. A new \textit{$<$eof$>$} token is introduced to guide the model in determining whether context retrieval is necessary. Based on the output after \textit{$<$eof$>$} token, it decides whether to generate the output directly or to perform retrieval first.

\subsection{LLM for Data System Optimization}
\label{subsec:optimization}

This section presents the application of \llm to optimize the performance of different data systems across three key tasks:
\emph{(1) Configuration Tuning:} selecting effective system configurations, such as database knobs and indexes;
\emph{(2) Query Optimization:} accelerating input SQL queries through logical rewrites and physical plan selection;
\emph{(3) Anomaly Diagnosis:} addressing system anomalies, such as spikes in the usage of specific system resources.



\subsubsection{\llm for Configuration Tuning}
Configuration tuning aims to identify effective configurations, such as database knobs~\cite{li2019qtune, TuningSurvey} and indexes~\cite{BID, TRAP, vita}, to optimize the system performance. 
Traditional tuning approaches, including rule-based methods and learning-based techniques with classical machine learning models, often require extensive explorations without a promising starting point~\cite{li2019qtune}.
Furthermore, they might result in sub-optimal configurations, despite using advanced techniques such as transfer learning~\cite{knobTransfer, cathpo}.

A key limitation of these methods is the failure to incorporate extensive domain knowledge (e.g., information from system manuals and public forum discussions) into the tuning process, relying solely on runtime feedback from benchmark evaluations to guide optimization.
To address this issue, recent approaches utilize \llm with large-scale domain knowledge to enhance the tuning process via the following methods.



\noindent \textbf{Tuning Task-Aware Prompt Engineering.}
The first method manually designs prompts with informative details (e.g., system status) to assist \llm~in configuration tuning (e.g., database knobs and indexes).
Some approaches further enhance this by introducing automatic prompt generation techniques or by formulating it as an optimization problem.

\noindent \bfit{(1) Manually-Crafted Tuning Prompt.}
Existing methods design prompts that incorporate essential details (e.g., system status) tailored to the characteristics of specific tasks. 
In particular, the constructed prompts typically consist of the following components.

\noindent $\bullet$ \bfit{Configuration Task Instruction.}
To convey the overall tuning objective, existing methods specify task instructions in the prompts using chain-of-thought (CoT) and role-play-based guidance.
For instance, LLMBench~\cite{LLMBench} explicitly defines the goals of three key subtasks in knob tuning: (i) knob pruning to retain the most influential knobs, (ii) model initialization to select promising knobs for warm-starting bayesian optimization, and (iii) knob recommendation to return optimal configurations for specific workloads.
Similarly, LATuner~\cite{LATuner} instructs \llm to identify critical knobs for warm-starting the tuning process and select promising knobs as training samples for boosting the sampling procedure.

\noindent $\bullet$ \bfit{Input Tuning Context.}
To enable \llm to effectively support the tuning process for specific workloads, existing methods enrich the tuning context with detailed information. Specifically, prompts are carefully structured to include: (i) Configuration Specifications: list of tunable knobs (e.g., names and allowable value ranges) and usage descriptions, including fixed-task demonstrations (e.g., LLMBench~\cite{LLMBench}, LATuner~\cite{LATuner}); 
(ii) Environment Information: covering workload and database characteristics (e.g., compressed SQL snippets with join conditions in $\lambda\text{-}Tune$~\cite{lambdaTune}), as well as hardware settings (e.g., memory size and CPU core count).


\noindent $\bullet$ \bfit{Output Tuning Requirement.}
To ensure accurate parsing and interpretation of configurations generated by \llm, output formats are explicitly specified in the prompt. For instance, LLMBench~\cite{LLMBench} requires that recommended knob values be returned in JSON format, while LATuner~\cite{LATuner} enforces constraints such as excluding the use of the “None” value in the configuration output.

\noindent \bfit{(2) Automatic Tuning Prompt Generation.}
To improve the efficiency of prompt generation for different workloads, existing methods propose the following techniques to automate the process of identifying effective prompts.

\noindent $\bullet$ \bfit{Input Specific Prompt Generation.}
To identify the most suitable prompts for varying tasks, existing methods automatically tailor prompt generation based on specific inputs.
For example, DB-GPT~\cite{DBGPT} introduces an automatic prompt generation framework that leverages \llm to produce multiple instruction candidates, selecting the optimal ones using scoring functions associated with the performance improvement.
Additionally, DB-GPT~\cite{DBGPT} and LLMIdxAdvis~\cite{LLMIdxAdvis} select demonstration examples in the prompts based on semantic similarity between candidate examples and input queries, as computed by a model-based encoder.

\noindent $\bullet$ \bfit{Optimization Problem Formulation.}
To reduce token usage and convey the most relevant context to the \llm, some methods formulate prompt generation as a cost-based optimization problem.
For instance, $\lambda\text{-}Tune$~\cite{lambdaTune} compresses workload representations by modeling the selection of join conditions as an integer linear programming problem, introducing binary decision variables to capture the positional relationships of different columns.


\noindent \textbf{RAG Based Tuning Experience Enrichment.}
The second method builds an offline knowledge base from diverse external sources and performs online retrieval to provide \llm with context-specific knowledge (e.g., similar historical tuning cases). This approach addresses the limitations of direct prompting, which often yields overly generic responses lacking concrete commands and effective configurations~\cite{Andromeda}. 

\noindent \bfit{(1) \llm Based Tuning Experience Preparation.}
Given that existing tuning knowledge is distributed across heterogeneous formats, \llms are employed to construct a knowledge base by processing and integrating multi-source external experience in an offline manner.
For example, GPTuner~\cite{GPTuner} prompts \llm to extract implicit knowledge, remove noisy content, and summarize relevant information from multiple sources. Additionally, it introduces a prompt ensemble algorithm that generates multiple prompts by varying the demonstration examples, aiming to mitigate hallucination issues.

\noindent \bfit{(2) Semantic Based Tuning Experience Retrieval.}
To improve the accuracy of relevant experience retrieval, existing methods employ model-based encoders to capture semantic relationships (e.g., documents conveying similar meanings with different expressions).
For instance, Andromeda~\cite{Andromeda} utilizes a Sentence-BERT encoder trained with contrastive learning to generate embeddings, which are then used to perform similarity searches across various sources, including historical queries and troubleshooting manuals.


\noindent \textbf{Training Enhanced Tuning Goal Alignment.}
The third method introduces additional training to further refine \llms, improving their alignment with tuning objectives.
For example, DB-GPT~\cite{DBGPT} proposes techniques to facilitate effective fine-tuning, including: (i) heuristic statistical data embedding, (ii) \llm-assisted annotation of high-quality samples, (iii) contrastive learning of supplementary training data generation, and (iv) delta tuning to minimize trainable parameters while maintaining performance.
Similarly, E2ETune~\cite{E2ETune} fine-tunes \llms (e.g., Mistral-7B) using training data comprising ``\emph{(workload) $\rightarrow$ (configuration)}'' pairs, where diverse workloads are generated via GPT-4 prompting and optimal configurations are identified using the HEBO algorithm~\cite{hebo}.


\subsubsection{\llm for Query Optimization}
Query optimization aims to accelerate SQL execution through logical (e.g., query rewriting) and physical (e.g., join order and plan selection) enhancements.
Traditional logical optimization relies on predefined rewrite rules or learning-based approaches to determine rule application order, while physical optimization employs heuristic algorithms using statistical data or learning-based techniques leveraging query plan features. However, these approaches often overlook external SQL optimization knowledge, limiting their effectiveness and generalizability across diverse SQL patterns.

To address these limitations, recent studies investigate the use of \llm to directly rewrite input SQL queries or determine optimal rule application sequences for logical optimization.
They also explore leveraging \llm to select optimal query execution plans for physical optimization, drawing on the extensive SQL optimization knowledge encoded within the model.
These methods can be broadly categorized as follows.

\noindent \textbf{Optimization-Aware Prompt Engineering.}
The first method directly employs \llms to perform query optimization using well-structured prompts composed of two key components: (i) manually crafted templates enriched with task-specific details (e.g., explicit task instructions), and (ii) relevant optimization examples automatically selected to more effectively guide the optimization process.

\noindent \bfit{(1) Manually-Crafted Optimization Prompt.}
Existing methods construct prompts with the following components to facilitate the query optimization task.

\noindent $\bullet$ \bfit{Optimization Task Instruction.}
To clarify the optimization objective and guide \llms to produce specific optimization actions, detailed task instructions are included in the prompts.
For logical query optimization, some methods instruct \llms to directly generate equivalent rewritten queries with improved performance (e.g., DB-GPT~\cite{DBGPT}, GenRewrite~\cite{GenRewrite}, and LITHE~\cite{LITHE}), while others ask them to determine the optimal sequence of rewrite rule applications for a given query (e.g., $\llm\text{-}R^2$\cite{LLMR2} and R-Bot\cite{RBot}).
For physical query optimization, some approaches prompt \llms to generate complete query plans with specified operators and join orders (e.g., LLM-QO~\cite{LLMQO}), while others instruct \llms to generate optimization hints or select the most effective plan from a set of candidates (e.g., LLMOpt~\cite{LLMOpt}).


\noindent $\bullet$ \bfit{Input Optimization Context.}
To enable effective query optimization for specific workloads, existing methods augment prompts with additional contextual information to better inform \llms. This includes:
(i) Database Statistics: column selectivity~\cite{LITHE}, histograms, distinct value counts, and estimated cardinalities~\cite{LLMQO};
(ii) Rule Specifications: a list of applicable rewrite rules accompanied by usage descriptions (e.g., GenRewrite~\cite{GenRewrite} presents natural language hints as the rules) and illustrative examples~\cite{LLMR2}.


\noindent $\bullet$ \bfit{Output Optimization Requirement.}
To ensure that the optimizations produced by \llms are valid and easily processed for downstream use, some methods explicitly define output formatting requirements within the prompts.
For example, $\llm\text{-}R^2$ enforces that selected rewrite rules be returned in the format ``rules selected: [rule names]''~\cite{LLMR2}, while LLM-QO specifies that the generated query plan should follow the ``join operator(table1, table2)'' format~\cite{LLMQO}.

\noindent \bfit{(2) In-Context Learning with Optimization Example.}
Rather than relying on fixed examples to illustrate how \llm should perform optimization, some methods automatically retrieve examples that are semantically similar to the input query to provide more effective guidance.
For instance, $\llm\text{-}R^2$~\cite{LLMR2} introduces a contrastive representation model to encode query plans based on features such as operators, cardinalities, and costs, and retrieves a set of high-quality demonstrations, i.e., successfully optimized rewritten queries.

\noindent \textbf{RAG Based Optimization Experience Enrichment.}
The second method adopts the retrieval-augmented generation (RAG) paradigm to equip \llm with relevant contextual information for targeted optimization of specific queries.
It constructs and retrieves optimization knowledge from multiple sources that are semantically related to the input query.

\noindent \bfit{(1) \llm Based Optimization Experience Preparation.}
To consolidate optimization experience from multiple sources, existing methods introduce an offline preparation pipeline that leverages \llm to process and integrate data into a unified format.
For example, R-Bot~\cite{RBot} employs \llm to generate rewrite rule specifications by (i) summarizing rule code within a hierarchical structure and (ii) extracting information from structured documentation blocks.
It further uses \llm to standardize the resulting specifications, explicitly outlining application conditions and detailed rewrite transformations.

\noindent \bfit{(2) Hybrid Optimization Experience Retrieval.}
To more accurately identify relevant optimization experiences, both structural and semantic characteristics of the input queries are considered during similarity search.
For instance, R-Bot~\cite{RBot} introduces a hybrid retrieval approach that computes similarity using concatenated embeddings capturing structural features (e.g., rewrite rule explanations) and semantic representations (e.g., query template structures).
Based on the retrieved experience, R-Bot employs a step-by-step \llm-driven rewrite process, further enhanced through a self-reflection mechanism to improve rewrite quality.

\noindent \textbf{Training Enhanced Optimization Improvement.}
The third method either uses \llm outputs to train smaller models or fine-tunes \llms on task-specific data to support various query optimization tasks (e.g., query plan generation).
For instance, LLMSteer~\cite{LLMSteer} uses \llm-generated embeddings to train a classifier for selecting optimal hints of the input SQL.
LLM-QO~\cite{LLMQO} fine-tunes \llms to generate execution plans directly through a two-stage pipeline: (i) Query Instruction Tuning (QIT) for producing valid plans; (ii) Query Direct Preference Optimization (QDPO) for distinguishing high-quality plans.
The fine-tuning data is structured as ``\emph{(query, task instruction, auxiliary information such as schema and statistics, demonstration)}'' paired with the corresponding efficient execution plan.
LLMOpt~\cite{LLMOpt} fine-tunes two models: (i) LLMOpt(G), which generates candidate hints, and (ii) LLMOpt(S), which selects the optimal hint as a list-wise cost model.
The fine-tuning data is structured as ``\emph{(query, statistics such as histograms) $\rightarrow$ (optimal hint)}'' for LLMOpt(G) and ``\emph{(query, statistics such as histograms, candidate hints) $\rightarrow$ (index of optimal hint)}'' for LLMOpt(S).



\subsubsection{\llm for Anomaly Diagnosis}

Anomaly diagnosis focuses on analyzing root causes and identifying recovery solutions for anomalies (e.g., spikes in system resource usage) during the system runtime, such as databases.
Traditional rule-based methods often fail to accurately identify root causes across diverse scenarios, while classical machine learning models (e.g., random forests) cannot generate comprehensive reports with detailed recovery solutions.

Recent studies demonstrate that \llms, with their advanced textual understanding and reasoning capabilities, can effectively pinpoint root causes and generate detailed diagnosis reports with recovery solutions in various formats.
These \llm-based approaches can be categorized as follows.

\noindent \textbf{Manually Crafted Prompts for Anomaly Diagnosis.}
The first method emulates the reasoning process of a human DBA, which involves referencing essential statistical information and conducting an in-depth analysis during diagnosis.
The information is incorporated into well-structured prompts to enhance diagnosis accuracy.
For example, DBG-PT~\cite{DBGPTDebugger} utilizes \llm to detect query execution slowdowns caused by changes in query plans, using prompts that include: (i) a summary of plan differences, (ii) a request for feasible configuration recommendations, and (iii) a specification of the reasoning process with output formatted in JSON format.

\noindent \textbf{RAG Based Diagnosis Experience Enrichment.}
The second method adopts retrieval-augmented generation (RAG) paradigm to provide \llm with relevant diagnosis knowledge, leveraging two key components: a knowledge base and a retriever.
For instance, D-Bot~\cite{DBot, LLMasDBA} enhances database anomaly diagnosis by preparing a corpus of documents and tools considering the hierarchical document structure, then using a fine-tuned Sentence-BERT encoder to retrieve relevant materials and guide \llm via prompts enriched with the retrieved content.
ByteHTAP~\cite{ByteHTAP} supports \llm-based diagnosis of query performance regressions in HTAP systems by first constructing a knowledge base of historical queries and their associated performance explanations.
It then employs an enhanced tree-CNN classifier to encode and retrieve relevant plan pairs.
The retrieved information is incorporated into prompts that include: (i) background information (e.g., key differences among HTAP system engines), (ii) a task description (e.g., retrieved diagnosis knowledge with explicit input-output specifications), and (iii) additional user-provided context (e.g., recent index changes).

\noindent \textbf{Multi-Agent Mechanism for Collaborative Diagnosis.}
The third method adopts an agent-based diagnosis framework, where specialized agents with distinct responsibilities collaborate to improve diagnosis accuracy and efficiency.
For example, D-Bot~\cite{DBot, LLMasDBA} orchestrates multiple domain-specific \llm agents, each aligned with a cluster of preprocessed diagnosis knowledge, to support precise anomaly diagnosis in databases.
These agents, coordinated by a chief agent, conduct multi-step root cause analysis via a tree-search algorithm.
Similarly, Panda~\cite{Panda} emulates experienced database engineers by leveraging \llm agents across five functional components: (i) question verification to eliminate irrelevant queries, (ii) grounding to provide necessary input query context, (iii) verification to ensure diagnosis accuracy and source attribution, (iv) feedback integration to incorporate user input, and (v) affordance assessment to estimate the performance impact of generated solutions.

\noindent \textbf{Localized LLM Enhancement via Specialized Fine-Tuning.}
The last method employs specialized fine-tuning strategies for localized \llms of modest scale (e.g., 6B-14B), leveraging distilled knowledge to approximate the outputs of larger models while achieving comparable performance.
For instance, D-Bot~\cite{DBot} applies multi-task fine-tuning to improve the diagnosis capabilities of localized \llms.
Specifically, three models (i.e., Llama2-13B, CodeLlama-13B, and Baichuan2-13B) are fine-tuned to replicate the diagnosis results generated by the GPT-4-powered D-Bot.
The fine-tuning dataset consists of samples covering D-Bot diagnosis workflows across five sub-tasks (e.g., tool invocation), along with associated prompts and historical dialogue messages.

\begin{tcolorbox}[colback=gray!1,colframe=gray,lowerbox=visible] \textbf{Practices of LLMs for Data Management} \tcblower Alibaba Cloud~\cite{xiyangbi} has integrated Text-to-SQL features into its BI platform, facilitating NL queries over structured datasets. Amazon Nova~\cite{awsnova} employs automated document processing to extract structured information from diverse unstructured sources. In terms of data systems, PawSQL~\cite{pawsql}, an advanced query optimization platform, offers both SQL rewriting and index recommendation capabilities, adopted by over 10,000 professionals. Database diagnosis also thrives on a robust ecosystem. For instance, DBDoctor~\cite{dbdoctor}, compatible with mainstream databases, delivers kernel-level performance diagnostics for comprehensive system analysis and optimization. \end{tcolorbox}

\section{Challenges and Future Directions}
\label{sec:futurework}

\subsection{Data Management for LLM}

\subsection*{4.1.1 Task-Specific Data Selection for Efficient Pretraining}

In \llm pre-training, vast amounts of general data are typically used, but much of this data may not be relevant to the target task. The inclusion of irrelevant data not only increases training time but also impedes the model’s adaptability to specific tasks. For instance, when training a model for the medical domain, unrelated data sources such as news articles and social media posts may hinder the learning of domain-specific knowledge. Consequently, the challenge lies in automatically selecting task-relevant data while discarding irrelevant information during pretraining. Currently, most approaches rely on hand-crafted filtering rules or fixed labeled datasets for data selection, lacking dynamic strategies that adapt to the model’s evolving task-specific needs. Exploring methods to automatically select relevant data and discard irrelevant data during pre-training represents a promising avenue for improving task adaptability and training efficiency.

\subsection*{4.1.2 Optimizing Data Processing Pipelines}

Currently, the construction of data processing pipelines for \llms relies heavily on experience and experimentation. For instance, in building the FineWeb dataset, decisions such as whether to use the WET or WARC format for text extraction from CommonCrawl, or whether to apply a global MinHash approach for deduplication or perform it separately for each snapshot, are made only after training models and benchmarking their performance. However, this experimental methodology is resource-intensive. In the case of FineWeb, over 70 models with 1 billion parameters were trained, consuming a total of 80,000 H100 GPU hours. To improve the efficiency of these pipelines, future research should focus on developing data-driven methods that can predict optimal {preprocessing configurations}. in advance, reducing the reliance on costly trial-and-error approaches. This would not only minimize computational costs but also accelerate the development of high-quality datasets for LLMs.

\subsection*{4.1.3 \llm Knowledge Update and Version Control}

In fast-evolving domains (e.g., healthcare, finance, law), knowledge is constantly updated. To ensure the reliability of \llms, the data used for training and fine-tuning must be up-to-date. Delays in incorporating the latest knowledge can result in outdated or harmful outputs, particularly in fields like medicine where guidelines frequently change. While there have been various approaches to data synthesis and augmentation, little attention has been given to efficiently managing rapid knowledge updates or resolving contradictions when new information conflicts with older data. Existing systems often rely on static datasets, which are problematic in dynamic sectors. Although platforms like ChatGPT and Deepseek allow LLMs to search the web, this approach may not always guarantee accuracy or relevance, leading to suboptimal results. A more effective solution would involve a platform that facilitates the creation, sharing, and version control of datasets with real-time knowledge updates. By leveraging community-driven contributions, this platform could enable users to synthesize and share datasets using customizable methods, such as LLM-generated prompts from documents or websites, offering continuous, high-quality updates and improving the overall accuracy and reliability of LLMs.




\subsection*{4.1.4 Comprehensive Dataset Evaluation}

The performance enhancement of models is closely tied to the use of 'high-quality' datasets. However, determining what constitutes a high-quality dataset remains a challenge. Typically, the quality of a dataset can only be inferred after training and evaluating a model, which makes the process indirect and resource-intensive. When a dataset's quality is subpar, it can lead to significant computational overhead and inefficiencies. While existing research ~\cite{wang2022statisticaldatasetevaluationreliability} has proposed a model-agnostic method for evaluating datasets across three aspects: reliability, difficulty, and validity. These dimensions alone do not fully capture a dataset's quality. The current framework falls short of providing a comprehensive evaluation that aligns with the model's capabilities and performance improvements. Therefore, a promising direction for future research is the development of a robust dataset evaluation system that does not rely on model training. This system should provide consistent quality scores that directly correlate with model performance enhancements, enabling more efficient dataset selection and use without the need for exhaustive training cycles.

\subsection*{4.1.5 Hybrid RAG Indexing and Retrieval}
Currently, there lacks a single database that integrates full-text, vector, knowledge graph, and structured search interfaces into a cohesive indexing and retrieval engine for Retrieval-Augmented Generation (RAG) training. While systems like Elasticsearch~\cite{elasticsearch} excel in full-text and vector search, and LightRAG~\cite{guo2024lightrag} has introduced advanced vector and graph processing, these solutions remain siloed. They lack a unified platform designed specifically for hybrid RAG, where multiple indexing and search mechanisms coexist to support efficient downstream applications. Although emerging platforms like AutoRAG~\cite{kim2024autoragautomatedframeworkoptimization} provide frameworks for constructing RAG pipelines, they focus on workflow management, model integration, and automation rather than offering a fully integrated database with indexing and retrieval engines. A promising direction for future RAG data serving is the development of an integrated platform that provides seamless indexing and retrieval for diverse data types, while also integrating data serving features such as knowledge filtering and re-ranking~\cite{abdallah2025asrank}, thereby improving the efficiency and flexibility of RAG applications.

\subsection{LLM for Data Management}

\subsection*{4.2.1 Unified Data Analysis System}
One of the major challenges in LLM for Data Analysis is the absence of a unified system capable of handling diverse data types. Currently, analyzing different data formats often requires designing task-specific models separately. The most straightforward approach to enabling a system to process all types of data is to integrate these models into a single framework. However, this leads to prohibitively high deployment and maintenance costs due to the need to manage multiple models simultaneously.
A more promising direction is to develop a model that can flexibly accommodate various data inputs and user requirements while supporting the analysis of structured, semi-structured, and unstructured data. Such a system would establish a paradigm for LLM for Data Analysis at the system level and offer a generalized capability for analyzing data across different structural types, thereby facilitating data automation.

\subsection*{4.2.2 Data Analysis with Private Domain Knowledge}
Another challenge in leveraging LLMs for data analysis is the effective utilization of private domain knowledge. Current approaches primarily rely on RAG to retrieve relevant knowledge or fine-tune models on domain-specific datasets. However, these methods struggle when dealing with novel or highly complex domain knowledge. For example, in Text-to-SQL tasks involving large-scale databases with 10,000 columns and 1,000,000 rows, where each column is associated with specific domain knowledge, existing techniques often fail to generalize effectively. The lack of datasets that explicitly incorporate domain knowledge further exacerbates this issue, making it difficult to meet the demands of real-world industrial applications. Consequently, developing more advanced mechanisms for integrating domain knowledge into LLMs remains a critical open research problem.

\subsection*{4.2.3 Representing Non-Sequential and Non-Textual Data}

Current \llm-based approaches typically transform non-sequential and non-textual data into serialized textual formats to align with the input requirements of \llms~\cite{RetClean, LLMQO, LLMOpt}.
While this enables basic compatibility, it overlooks the original structural semantics of the data and can lead to significant information loss in downstream tasks.
For instance, in data manipulation and analysis, relational tables (originally structured as two-dimensional matrices) are typically flattened into multiple serialized sequences, obscuring inherent row-column relationships~\cite{LLMClean, LLMErrorBench, CleanAgent}.
Similarly, in system optimization tasks, crucial statistical signals such as column selectivities and histograms are either omitted or naively encoded as plain texts, limiting their utility in guiding optimization decisions~\cite{lambdaTune, LATuner}.
Consequently, a promising future direction is to develop more expressive and task-aware representations that preserve the structural and statistical integrity of such data.
This includes leveraging multi-modal \llms or designing tailored encoding strategies that maintain the uniqueness of these data types, thereby enabling more effective and semantically informed \llm applications.

\subsection*{4.2.4 Efficient \llm Utilization Under Budget Constraints}

While \llms have shown strong potential across data manipulation, analysis, and system optimization tasks, their high computational cost and latency pose challenges for real-time or large-scale applications~\cite{LLMQO, LLMSteer}.
For example, relying solely on \llms is impractical for processing tens of millions of rows in relational table analysis due to prohibitive resource demands~\cite{GIDCL, LLMSchemaBench}. 
Similarly, current \llm-based query optimizers often require minutes per query, far exceeding the millisecond-level efficiency of traditional statistical methods~\cite{RBot, LLMR2}.
Therefore, a promising direction is to develop hybrid strategies that integrate \llms with traditional techniques or to devise scheduling mechanisms that allocate tasks across multiple \llms based on cost-performance trade-offs.
Such approaches can enhance the practicality and scalability of \llm-based systems under real-world budget constraints.

\section{Conclusion}
\label{sec:conclusion}

In this paper, we summarize the recent techniques on \datallm and \llmdata. The former focuses on  utilizing data processing, storage, serving techniques to address the data problems in different \llm stages. The latter focuses on using \llm capabilities to reduce the complexity of conducting data management, e.g., data manipulation, data analysis, and data system optimization. We also provide some research challenges and open problems in \datallm, 
 \llmdata, and hybrid data and \llm optimization.

\balance
\scriptsize
\bibliographystyle{abbrv}
\bibliography{ref/DA}

\end{document}